\documentclass[oneside,english]{elsart}
\usepackage[T1]{fontenc}
\usepackage[latin1]{inputenc}
\usepackage{amsmath}
\usepackage{graphicx}
\usepackage{amssymb}

\makeatletter
\usepackage{babel}
\makeatother
\begin{document}
\begin{frontmatter}

\title{Gaussian matrix elements in a cylindrical harmonic oscillator basis}

\author{W. Younes}

\ead{younes1@llnl.gov}

\address{Lawrence Livermore National Laboratory, Livermore, CA 94551}

\begin{abstract}
We derive a formalism, the separation method, for the efficient and
accurate calculation of two-body matrix elements for a Gaussian potential
in the cylindrical harmonic-oscillator basis. This formalism is of
critical importance for Hartree-Fock and Hartree-Fock-Bogoliubov calculations
in deformed nuclei using realistic, finite-range effective interactions
between nucleons. The results given here are also relevant for microscopic
many-body calculations in atomic and molecular physics, as the formalism
can be applied to other types of interactions beyond the Gaussian
form. The derivation is presented in great detail to emphasize the
methodology, which relies on generating functions. The resulting analytical
expressions for the Gaussian matrix elements are checked for speed
and accuracy as a function of the number of oscillator shells and
against direct numerical integration.
\end{abstract}
\end{frontmatter}

\section{Introduction}

Gaussian interactions play an important role in the microscopic description
of molecular and nuclear processes \cite{brink67}. The Gaussian form
represents a relatively simple two-body potential with a finite range,
which is needed in many realistic descriptions of many-body systems.
In nuclear physics for example, the Gogny interaction \cite{decharge80}

\begin{eqnarray}
V\left(\vec{r}_{1},\vec{r}_{2}\right) & = & \sum_{i=1}^{2}\left(W_{i}+B_{i}\hat{P}_{\sigma}-H_{i}\hat{P}_{\tau}-M_{i}\hat{P}_{\sigma}\hat{P}_{\tau}\right)e^{-\left(\vec{r}_{1}-\vec{r}_{2}\right)^{2}/\mu_{i}^{2}}\nonumber \\
 &  & +iW_{LS}\left(\overleftarrow{\nabla}_{1}-\overleftarrow{\nabla}_{2}\right)\times\delta\left(\vec{r}_{1}-\vec{r}_{2}\right)\left(\overrightarrow{\nabla}_{1}-\overrightarrow{\nabla}_{2}\right)\cdot\left(\vec{\sigma}_{1}+\vec{\sigma}_{2}\right)\nonumber \\
 &  & +t_{0}\left(1+x_{0}\hat{P}_{\sigma}\right)\delta\left(\vec{r}_{1}-\vec{r}_{2}\right)\rho^{\gamma}\left(\frac{\vec{r}_{1}+\vec{r}_{2}}{2}\right)+V_{\textrm{Coul}}\label{eq:gogny-force}\end{eqnarray}
where $\hat{P}_{\sigma}$ and $\hat{P}_{\tau}$ are spin- and isospin-exchange
operators and $\rho$ is the total nuclear density, gives the effective
(in-medium) potential between nucleons. Two Gaussian terms appear
explicitly with range parameters $\mu_{1}$ and $\mu_{2}$. A spin-orbit
term with strength $W_{LS}$ uses a Dirac-delta function, but extensions
of the Gogny force have been proposed \cite{gogny70} that introduce
a Gaussian form for this term. Finally the Coulomb interaction $V_{\textrm{Coul}}\sim1/\left|\vec{r}_{1}-\vec{r}_{2}\right|$
between protons is clearly not of Gaussian form, but the mathematical
framework presented in this paper can be applied equally well to a
Coulomb potential.

For the calculation of matrix elements in molecular, atomic, and nuclear
physics, harmonic-oscillator functions provide a convenient and popular
orthogonal basis. The calculation of Gaussian matrix elements in a
harmonic-oscillator basis, however, poses definite technical challenges
in accuracy as well as execution time. In previous work \cite{gogny75},
the separation method was introduced as a way of calculating the Gaussian
matrix elements efficiently and accurately for systems with spherical
symmetry. In the separation method, two-body matrix elements are expressed
as a more manageable finite sum of products of one-body matrix elements.
In this paper, we derive the separation method for a wider class of
systems that exhibit axial symmetry. These results are crucial, for
example, in microscopic calculations of nuclear fission using the
Gogny force, where the nucleus elongates along a symmetry axis, until
scission occurs.

Fission calculations in particular bring to the fore many of the technical
difficulties involved in the computation of Gaussian matrix elements.
On the other hand, microscopic calculations of fission using the interaction
in Eq. (\ref{eq:gogny-force}) have had considerable success in recent
years \cite{berger84,goutte05,dubray08}, and are therefore of great
interest. In the microscopic description of fission, the matrix elements
of the nucleon-nucleon interaction are typically used in a Hartree-Fock-Bogoliubov
(HFB) procedure to construct a Slater-determinant wave function for
the nucleus. Scission configurations are then found by driving the
nucleus to such exotic shapes that the delicate balance between its
surface tension and the Coulomb repulsion between the nascent fission
fragments is broken. The proper identification of scission configurations
and the calculation of their properties depend sensitively on accurate
calculations of the matrix elements of the effective interaction.
Fission also implies the evolution of the nucleus through a variety
of exotic shapes leading to scission. Therefore many sets of matrix
elements need to be calculated, each set corresponding to a harmonic-oscillator
basis optimized for a particular nuclear shape, and each set requiring
a large number of oscillator shells. The resulting large-scale computations
can become very time-consuming and are prone to errors in accuracy.
Thus microscopic fission calculations must rely on fast and accurate
algorithms to evaluate the two-body matrix elements, such as the separation
method. The separation method is especially well-suited to the HFB
algorithm, because the coefficients needed to calculate the two-body
matrix elements derived in this paper can be calculated quickly once
and for all, and stored with relatively little computer memory.

The goal of this paper is to derive the separation-method formalism
for Gaussian matrix elements in a cylindrical harmonic-oscillator
basis, with particular emphasis placed on the details of the derivation
because of its relevance to other types of interactions, and other
applications involving the harmonic-oscillator basis. In particular,
we rely heavily on the power and versatility of generating functions
to derive many of the present results. We also present the derivations
in great detail because they are rather involved, and although the
same results may be arrived at by alternate approaches, the formulas
will tend to be much more cumbersome and less computationally efficient
than the ones obtained by the generating-function methods outlined
here. Because of the lengthy and detailed derivations involved, many
of the intermediary results have been placed in the appendices. These
intermediary results are important in their own right, as they provide
useful properties of harmonic-oscillator functions in a cylindrical
basis, and the mapping between cylindrical and Cartesian harmonic-oscillator
bases.

In section \ref{sec:Theory}, the basic formalism for the calculation
of both radial and axial components of the Gaussian matrix elements
by the separation method are derived. In section \ref{sec:Discussion},
the accuracy of the method is examined both relative to direct numerical
integration, and as a function of the number of shells in the oscillator
basis. The execution times for the separation method are also compared
to those of the numerical integration. The mapping between harmonic-oscillator
function in polar and Cartesian coordinates, needed in the development
of the separation-method formalism, is derived in appendix \ref{sec:Mapping-from-Cartesian}.
In appendix \ref{sec:Decomposition-of-two-body}, the Gaussian two-body
potential, $V\left(\vec{r}_{1},\vec{r}_{2}\right)$, is written in
separated form with respect to $\vec{r}_{1}$ and $\vec{r}_{2}$.
Formulas reducing the products of harmonic-oscillator functions are
derived in appendix \ref{sec:Product-of-harmonic-oscillator}, and
provide a powerful tool in the evaluation of integrals involving those
functions. In appendix \ref{sec:Formalism-for-large}, the result
quoted in \cite{egido97} for the separation-method formalism in the
case of large oscillator-shell numbers is derived in detail. Finally,
in appendix \ref{sec:Angular-integral}, we obtain a series expansion
for the direct angular integral of the Gaussian potential, which we
use in the numerical integration of the potential in section \ref{sec:Discussion}.

\section{Theory\label{sec:Theory}}

\subsection{General formalism}

We wish to calculate matrix elements of the two-body potential function\begin{eqnarray}
V\left(\vec{r}_{1},\vec{r}_{2}\right) & = & e^{-\left(\vec{r}_{1}-\vec{r}_{2}\right)^{2}/\mu^{2}}\label{eq:vgauss}\end{eqnarray}
in the cylindrical harmonic-oscillator basis. We will write the matrix
elements as\begin{eqnarray}
V_{ijkl} & \equiv & \left\langle ij\left|V\right|kl\right\rangle \nonumber \\
 & = & \int d^{3}r_{1}\int d^{3}r_{2}\Phi_{n_{r}^{(i)},\Lambda^{(i)},n_{z}^{(i)}}^{*}\left(\vec{r}_{1};b_{\perp},b_{z}\right)\Phi_{n_{r}^{(j)},\Lambda^{(j)},n_{z}^{(j)}}^{*}\left(\vec{r}_{2};b_{\perp},b_{z}\right)\nonumber \\
 &  & \times V\left(\vec{r}_{1},\vec{r}_{2}\right)\Phi_{n_{r}^{(k)},\Lambda^{(k)},n_{z}^{(k)}}\left(\vec{r}_{1};b_{\perp},b_{z}\right)\Phi_{n_{r}^{(l)},\Lambda^{(l)},n_{z}^{(l)}}\left(\vec{r}_{2};b_{\perp},b_{z}\right)\label{eq:Vijkl-integral}\end{eqnarray}
where we have introduced the stretched harmonic-oscillator basis functions
in the cylindrical coordinates $\left(\rho,\varphi,z\right)$%
\footnote{We will drop the qualifier {}``stretched'' when referring to the
deformed harmonic-oscillator function in subsequent discussion for
the sake of brevity.%
}\begin{eqnarray}
\Phi_{n_{r},\Lambda,n_{z}}\left(\vec{r};b_{\perp},b_{z}\right) & = & \Phi_{n_{r},\Lambda}\left(\rho,\varphi;b_{\perp}\right)\Phi_{n_{z}}\left(z;b_{z}\right)\nonumber \\
 & = & \Phi_{n_{r},\left|\Lambda\right|}\left(\rho;b_{\perp}\right)\frac{e^{\mathrm{i}\Lambda\varphi}}{\sqrt{2\pi}}\Phi_{n_{z}}\left(z;b_{z}\right)\label{eq:phipol-def}\end{eqnarray}
with the radial-component function\begin{eqnarray}
\Phi_{n_{r},\left|\Lambda\right|}\left(\rho;b_{\perp}\right) & = & \mathcal{N}_{n_{r}}^{\left|\Lambda\right|}\eta^{\left|\Lambda\right|/2}e^{-\eta/2}L_{n_{r}}^{\left|\Lambda\right|}\left(\eta\right)\label{eq:phirad-def}\end{eqnarray}
defined in terms of associated Laguerre polynomials $L_{n_{r}}^{\left|\Lambda\right|}\left(\eta\right)$
as a function of\begin{eqnarray*}
\eta & \equiv & \rho^{2}/b_{\perp}^{2}\end{eqnarray*}
and with a normalization constant given by\begin{eqnarray}
\mathcal{N}_{n_{r},\left|\Lambda\right|} & \equiv & \frac{1}{b_{\perp}}\left[\frac{2n_{r}!}{\left(n_{r}+\left|\Lambda\right|\right)!}\right]^{1/2}\label{eq:normpol-def}\end{eqnarray}
The Cartesian, z-axis-component function in Eq. (\ref{eq:phipol-def}),\begin{eqnarray}
\Phi_{n_{z}}\left(z;b_{z}\right) & = & \mathcal{N}_{n_{z}}e^{-\xi^{2}/2}H_{n_{z}}\left(\xi\right)\label{eq:phicar-def}\end{eqnarray}
is expressed in terms of Hermite polynomials $H_{n_{z}}\left(\xi\right)$
with\begin{eqnarray*}
\xi & \equiv & z/b_{z}\end{eqnarray*}
and normalization constant\begin{eqnarray*}
\mathcal{N}_{n_{z}} & \equiv & \frac{1}{\left(b_{z}\sqrt{\pi}2^{n_{z}}n_{z}!\right)^{1/2}}\end{eqnarray*}
The harmonic-oscillator functions defined in Eqs. (\ref{eq:phipol-def})
and (\ref{eq:phicar-def}) satisfy the orthonormalization conditions\begin{eqnarray*}
\int_{0}^{\infty}\rho d\rho\int_{0}^{2\pi}d\varphi\,\Phi_{n_{r},\Lambda}^{*}\left(\rho,\varphi;b_{\perp}\right)\Phi_{n_{r}^{\prime},\Lambda^{\prime}}\left(\rho,\varphi;b_{\perp}\right) & = & \delta_{n_{r},n_{r}^{\prime}}\delta_{\Lambda,\Lambda^{\prime}}\end{eqnarray*}
\begin{eqnarray*}
\int_{-\infty}^{\infty}dz\,\Phi_{n_{z}}\left(z;b_{z}\right)\Phi_{n_{z}^{\prime}}\left(z;b_{z}\right) & = & \delta_{n_{z},n_{z}^{\prime}}\end{eqnarray*}
The parameters $b_{\perp}$ and $b_{z}$ appearing in the harmonic-oscillator
function definitions are usually treated as variational parameters
in HFB calculations, and chosen to minimize the energy. 

The central idea in this paper is to express the two-body potential
as a sum of products of one-body potential functions\begin{eqnarray*}
e^{-\left(\vec{r}_{1}-\vec{r}_{2}\right)^{2}/\mu^{2}} & = & \sum_{n_{r},\Lambda,n_{z}}f_{n_{r},\Lambda,n_{z}}\left(\vec{r}_{1};b_{\perp},b_{z}\right)\hat{\Phi}_{n_{r},\Lambda,n_{z}}\left(\vec{r}_{2};b_{\perp},b_{z}\right)\end{eqnarray*}
Then the two-body matrix elements can be written in terms of one-body
matrix elements\begin{eqnarray}
V_{ijkl} & = & \sum_{n_{r},\Lambda,n_{z}}\left\langle i\left|f_{n_{r},\Lambda,n_{z}}\right|k\right\rangle \left\langle j\left|\hat{\Phi}_{n_{r},\Lambda,n_{z}}\right|l\right\rangle \label{eq:vijkl-z-original}\end{eqnarray}

where we will show that this last sum is limited to a finite number
of terms. It will be useful to separate the radial and Cartesian components
in each one-body matrix element to write\begin{eqnarray*}
\left\langle i\left|f_{n_{r},\Lambda,n_{z}}\right|k\right\rangle  & = & \int d^{3}r\,\Phi_{n_{r}^{(i)},\Lambda^{(i)},n_{z}^{(i)}}^{*}\left(\vec{r};b_{\perp},b_{z}\right)f_{n_{r},\Lambda,n_{z}}\left(\vec{r};b_{\perp},b_{z}\right)\\
 &  & \times\Phi_{n_{r}^{(k)},\Lambda^{(k)},n_{z}^{(k)}}\left(\vec{r};b_{\perp},b_{z}\right)\\
 & = & \int_{0}^{\infty}\rho d\rho\int_{0}^{2\pi}d\varphi\,\Phi_{n_{r}^{(i)},\Lambda^{(i)}}\left(\rho,\varphi;b_{\perp}\right)f_{n_{r},\Lambda}\left(\rho,\varphi;b_{\perp}\right)\\
 &  & \times\Phi_{n_{r}^{(k)},\Lambda^{(k)}}\left(\rho,\varphi;b_{\perp}\right)\\
 &  & \times\int_{-\infty}^{\infty}dz\,\Phi_{n_{z}^{(i)}}\left(z;b_{z}\right)f_{n_{z}}\left(z;b_{z}\right)\Phi_{n_{z}^{(k)}}\left(z;b_{z}\right)\\
 & \equiv & \left\langle i\left|f_{n_{r},\Lambda}\right|k\right\rangle \left\langle i\left|f_{n_{z}}\right|k\right\rangle \end{eqnarray*}
and, similarly,\begin{eqnarray*}
\left\langle j\left|\hat{\Phi}_{n_{r},\Lambda,n_{z}}\right|l\right\rangle  & = & \int_{0}^{\infty}\rho d\rho\int_{0}^{2\pi}d\varphi\,\Phi_{n_{r}^{(j)},\Lambda^{(j)}}\left(\rho,\varphi;b_{\perp}\right)\hat{\Phi}_{n_{r},\Lambda}\left(\rho,\varphi;b_{\perp}\right)\\
 &  & \times\Phi_{n_{r}^{(l)},\Lambda^{(l)}}\left(\rho,\varphi;b_{\perp}\right)\\
 &  & \times\int_{-\infty}^{\infty}dz\,\Phi_{n_{z}^{(j)}}\left(z;b_{z}\right)\hat{\Phi}_{n_{z}}\left(z;b_{z}\right)\Phi_{n_{z}^{(l)}}\left(z;b_{z}\right)\\
 & \equiv & \left\langle j\left|\hat{\Phi}_{n_{r},\Lambda}\right|l\right\rangle \left\langle j\left|\hat{\Phi}_{n_{z}}\right|l\right\rangle \end{eqnarray*}
so that we can write Eq. (\ref{eq:vijkl-z-original}) as\begin{eqnarray}
V_{ijkl} & = & \left[\sum_{n_{r},\Lambda}\left\langle i\left|f_{n_{r},\Lambda}\right|k\right\rangle \left\langle j\left|\hat{\Phi}_{n_{r},\Lambda}\right|l\right\rangle \right]\left[\sum_{n_{z}}\left\langle i\left|f_{n_{z}}\right|k\right\rangle \left\langle j\left|\hat{\Phi}_{n_{z}}\right|l\right\rangle \right]\nonumber \\
 & \equiv & V_{ijkl}^{\left(r\right)}V_{ijkl}^{\left(z\right)}\label{eq:vijkl-separated}\end{eqnarray}
In the remainder of this section we calculate the explicit expressions
needed to evaluate the matrix elements $V_{ijkl}$.

\subsection{Cartesian component}

Here we derive an expression for the Cartesian component,$V_{ijkl}^{\left(z\right)}$
, in Eq. (\ref{eq:vijkl-separated}). We will show that\begin{eqnarray}
V_{ijkl}^{\left(z\right)} & = & \sqrt{\frac{G_{z}-1}{G_{z}+1}}\sum_{m_{z}=\left|n_{z}^{(i)}-n_{z}^{(k)}\right|,2}^{n_{z}^{(i)}+n_{z}^{(k)}}\sum_{n_{z}=\left|n_{z}^{(j)}-n_{z}^{(l)}\right|,2}^{n_{z}^{(j)}+n_{z}^{(l)}}T_{n_{z}^{(i)},n_{z}^{(k)}}^{m_{z}}T_{n_{z}^{(j)},n_{z}^{(l)}}^{n_{z}}\bar{I}\left(m_{z},n_{z}\right)\label{eq:Vijkl-z}\end{eqnarray}
where $G_{z}$ is defined by Eq. (\ref{eq:Gz-def}), the $T_{n_{1},n_{2}}^{n_{3}}$
coefficients by Eq. (\ref{eq:carprod-coeff}), and the $\bar{I}\left(m_{z},n_{z}\right)$
coefficients by Eq. (\ref{eq:Ibar_mn}).

We start by evaluating\begin{eqnarray*}
\left\langle j\left|\hat{\Phi}_{n_{z}}\right|l\right\rangle  & = & \int_{-\infty}^{\infty}dz\,\Phi_{n_{z}^{(j)}}\left(z;b_{z}\right)\hat{\Phi}_{n_{z}}\left(z;b_{z}\right)\Phi_{n_{z}^{(l)}}\left(z;b_{z}\right)\end{eqnarray*}
Using Eqs. (\ref{eq:phihatcar}) which gives the explicit form of
$\hat{\Phi}_{n_{z}}\left(z;b_{z}\right)$ and Eq. (\ref{eq:carprod-expr})
to reduce the product of harmonic-oscillator functions,\begin{eqnarray*}
\left\langle j\left|\hat{\Phi}_{n_{z}}\right|l\right\rangle  & = & \frac{1}{\sqrt{b_{z}\sqrt{\pi}}}\sum_{m_{z}=\left|n_{z}^{(j)}-n_{z}^{(l)}\right|,2}^{n_{z}^{(j)}+n_{z}^{(l)}}T_{n_{z}^{(j)},n_{z}^{(l)}}^{m_{z}}\int_{-\infty}^{\infty}dz\,\Phi_{m_{z}}\left(z;b_{z}\right)\Phi_{n_{z}}\left(z;b_{z}\right)\end{eqnarray*}
By orthogonality of the harmonic-oscillator functions this is simply\begin{eqnarray}
\left\langle j\left|\hat{\Phi}_{n_{z}}\right|l\right\rangle  & = & \frac{1}{\sqrt{b_{z}\sqrt{\pi}}}T_{n_{z}^{(j)},n_{z}^{(l)}}^{n_{z}}\label{eq:phihatcar_jl}\end{eqnarray}
where we must have $\left|n_{z}^{(j)}-n_{z}^{(l)}\right|\leq n_{z}\leq n_{z}^{(j)}+n_{z}^{(l)}$
for the $T_{n_{z}^{(j)},n_{z}^{(l)}}^{n_{z}}$ coefficient to be non-zero.
Next, we use the explicit form of $f_{n_{z}}\left(z;b_{z}\right)$
from Eq. (\ref{eq:fcar}) to write\begin{eqnarray}
\left\langle i\left|f_{n_{z}}\right|k\right\rangle  & = & \int_{-\infty}^{\infty}dz\,\Phi_{n_{z}^{(i)}}\left(z;b_{z}\right)f_{n_{z}}\left(z;b_{z}\right)\Phi_{n_{z}^{(k)}}\left(z;b_{z}\right)\nonumber \\
 & = & K_{z}^{1/2}\lambda_{n_{z}}\int_{-\infty}^{\infty}dz\,\Phi_{n_{z}^{(i)}}\left(z;b_{z}\right)e^{-z^{2}/\left(2G_{z}b_{z}^{2}\right)}\nonumber \\
 &  & \times\Phi_{n_{z}}\left(z;G_{z}^{1/2}b_{z}\right)\Phi_{n_{z}^{(k)}}\left(z;b_{z}\right)\label{eq:fcar-ik-def}\end{eqnarray}
Two of the harmonic-oscillator functions can be replaced with a single
one, thanks to Eq. (\ref{eq:carprod-expr}),\begin{eqnarray}
\left\langle i\left|f_{n_{z}}\right|k\right\rangle  & = & \frac{K_{z}^{1/2}\lambda_{n_{z}}}{\sqrt{b_{z}\sqrt{\pi}}}\sum_{m_{z}=\left|n_{z}^{(i)}-n_{z}^{(k)}\right|,2}^{n_{z}^{(i)}+n_{z}^{(k)}}T_{n_{z}^{(i)},n_{z}^{(k)}}^{m_{z}}\int_{-\infty}^{\infty}dz\, e^{-z^{2}/\left(2b_{z}^{2}\right)-z^{2}/\left(2G_{z}b_{z}^{2}\right)}\nonumber \\
 &  & \times\Phi_{m_{z}}\left(z;b_{z}\right)\Phi_{n_{z}}\left(z;G_{z}^{1/2}b_{z}\right)\label{eq:fcar-ik-integral}\end{eqnarray}
The remaining integral, which we write in terms of the function\begin{eqnarray*}
I\left(m,n\right) & \equiv & \int_{-\infty}^{\infty}dz\, e^{-z^{2}/\left(2b_{z}^{2}\right)-z^{2}/\left(2B_{z}^{2}\right)}\Phi_{m}\left(z;b_{z}\right)\Phi_{n}\left(z;B_{z}\right)\end{eqnarray*}
where $B_{z}\equiv G_{z}^{1/2}b_{z}$, can be calculated with the
help of generating functions. Indeed, using Eq. (\ref{eq:carho-genf})
to form the product of the harmonic-oscillator functions, we have
for any $t_{1}$ and $t_{2}$\begin{eqnarray*}
e^{-t_{1}^{2}+2t_{1}z/b_{z}-z^{2}/\left(2b_{z}^{2}\right)}e^{-t_{2}^{2}+2t_{2}z/B_{z}-z^{2}/\left(2B_{z}^{2}\right)} & = & \sqrt{b_{z}B_{z}\pi}\sum_{m=0}^{\infty}\sum_{n=0}^{\infty}\frac{2^{\left(m+n\right)/2}}{\sqrt{m!n!}}\\
 &  & \times t_{1}^{m}t_{2}^{n}\Phi_{m}\left(z;b_{z}\right)\Phi_{n}\left(z;B_{z}\right)\end{eqnarray*}
from which, multiplying by the Gaussian factors in the definition
of $I\left(m,n\right)$ and integrating both sides of the equation,\begin{eqnarray}
 &  & e^{-t_{1}^{2}-t_{2}^{2}}\int_{-\infty}^{\infty}dz\, e^{2t_{1}z/b_{z}-z^{2}/b_{z}^{2}+2t_{2}z/B_{z}-z^{2}/B_{z}^{2}}\nonumber \\
 & = & \sqrt{b_{z}B_{z}\pi}\sum_{m=0}^{\infty}\sum_{n=0}^{\infty}\frac{2^{\left(m+n\right)/2}}{\sqrt{m!n!}}t_{1}^{m}t_{2}^{n}I\left(m,n\right)\label{eq:Imn-genf}\end{eqnarray}
The integral on the left-hand side can be evaluated by completing
the square,\begin{eqnarray*}
\int_{-\infty}^{\infty}dz\, e^{2t_{1}z/b_{z}-z^{2}/b_{z}^{2}+2t_{2}z/B_{z}-z^{2}/B_{z}^{2}} & = & e^{t^{2}/\nu}\int_{-\infty}^{\infty}dz\, e^{-\left(\sqrt{\nu}z-t/\sqrt{\nu}\right)^{2}}\\
 & = & \sqrt{\frac{\pi}{\nu}}e^{t^{2}/\nu}\end{eqnarray*}
where we have defined\begin{eqnarray*}
\nu & \equiv & \frac{1}{b_{z}^{2}}+\frac{1}{B_{z}^{2}}\\
t & \equiv & \frac{t_{1}}{b_{z}}+\frac{t_{2}}{B_{z}}\end{eqnarray*}
Thus, the left-hand side of Eq. (\ref{eq:Imn-genf}) becomes\begin{eqnarray*}
LHS & = & \sqrt{\frac{\pi}{\nu}}e^{t^{2}/\nu-t_{1}^{2}-t_{2}^{2}}\\
 & = & \sqrt{\frac{\pi}{\nu}}e^{-\left(b_{z}t_{1}-B_{z}t_{2}\right)^{2}/\left(\nu b_{z}^{2}B_{z}^{2}\right)}\end{eqnarray*}
which can be expanded as\begin{eqnarray*}
LHS & = & \sqrt{\frac{\pi}{\nu}}\sum_{p=0}^{\infty}\frac{\left(-1\right)^{p}\left(b_{z}t_{1}-B_{z}t_{2}\right)^{2p}}{p!\left(\nu b_{z}^{2}B_{z}^{2}\right)^{p}}\\
 & = & \sqrt{\frac{\pi}{\nu}}\sum_{p=0}^{\infty}\sum_{q=0}^{2p}\left(\begin{array}{c}
2p\\
q\end{array}\right)\frac{\left(-1\right)^{p+q}}{p!\nu^{p}b_{z}^{q}B_{z}^{2p-q}}t_{1}^{2p-q}t_{2}^{q}\end{eqnarray*}
Comparing with the right-hand side of Eq. (\ref{eq:Imn-genf}), we
see that we must make the identifications $m=2p-q$ and $n=q$ in
order for the equation to hold for any $t_{1}$ and $t_{2}$. Then,\begin{eqnarray*}
LHS & = & \sqrt{\frac{\pi}{\nu}}\sum_{m=0}^{\infty}\sum_{n=0}^{\infty}\left(\begin{array}{c}
2p\\
q\end{array}\right)\frac{\left(-1\right)^{p+q}}{p!\nu^{p}b_{z}^{q}B_{z}^{2p-q}}t_{1}^{2p-q}t_{2}^{q}\delta_{n,q}\delta_{m,2p-q}\end{eqnarray*}
and the comparison with the right-hand side of Eq. (\ref{eq:Imn-genf})
yields\begin{eqnarray*}
I\left(m,n\right) & = & \frac{\left(-1\right)^{\left(m+n\right)/2+n}\sqrt{m!n!}}{\left(\frac{m+n}{2}\right)!\left(2\nu\right)^{\left(m+n\right)/2}b_{z}^{n}B_{z}^{m}\sqrt{b_{z}B_{z}\nu}}\left(\begin{array}{c}
m+n\\
n\end{array}\right)\end{eqnarray*}
Note that $m+n$ must be even. We simplify this form further by noting
that\begin{eqnarray}
b_{z}B_{z}\nu & = & G_{z}^{1/2}+G_{z}^{-1/2}\label{eq:bBnu}\\
B_{z}^{2}\nu & = & 1+G_{z}\label{eq:BBnu}\\
b_{z}^{2}\nu & = & 1+G_{z}^{-1}\label{eq:bbnu}\end{eqnarray}
where $G_{z}$ is defined in Eq. (\ref{eq:Gz-def}). This leads us
to write\begin{eqnarray*}
I\left(m,n\right) & = & \frac{G_{z}^{1/4}G_{z}^{n/2}}{\sqrt{1+G_{z}}}\sqrt{\frac{m!n!}{2^{m+n}}}\frac{\left(-1\right)^{\left(m-n\right)/2}}{\left(\frac{m+n}{2}\right)!\left(1+G_{z}\right)^{\left(m+n\right)/2}}\left(\begin{array}{c}
m+n\\
n\end{array}\right)\end{eqnarray*}
Some of the constants can be factored out by defining the coefficient\begin{eqnarray}
\bar{I}\left(m,n\right) & \equiv & \frac{\sqrt{1+G_{z}}}{G_{z}^{1/4}G_{z}^{n/2}}I\left(m,n\right)\nonumber \\
 & = & \sqrt{\frac{m!n!}{2^{m+n}}}\frac{\left(-1\right)^{\left(m-n\right)/2}}{\left(\frac{m+n}{2}\right)!\left(1+G_{z}\right)^{\left(m+n\right)/2}}\left(\begin{array}{c}
m+n\\
n\end{array}\right)\label{eq:Ibar_mn}\end{eqnarray}
Then, returning to Eq. (\ref{eq:fcar-ik-integral}), we obtain after
some simplification\begin{eqnarray}
\left\langle i\left|f_{n_{z}}\right|k\right\rangle  & = & \frac{K_{z}^{1/2}\lambda_{n_{z}}}{\sqrt{b_{z}\sqrt{\pi}}}\frac{G_{z}^{1/4}G_{z}^{n_{z}/2}}{\sqrt{1+G_{z}}}\sum_{m_{z}=\left|n_{z}^{(i)}-n_{z}^{(k)}\right|,2}^{n_{z}^{(i)}+n_{z}^{(k)}}T_{n_{z}^{(i)},n_{z}^{(k)}}^{m_{z}}\bar{I}\left(m_{z},n_{z}\right)\label{eq:fcar-ik}\end{eqnarray}
Having derived the explicit forms in Eqs. (\ref{eq:phihatcar_jl})
and (\ref{eq:fcar-ik}), we can express the Cartesian component in
Eq. (\ref{eq:vijkl-separated}) as\begin{eqnarray*}
V_{ijkl}^{\left(z\right)} & \equiv & \sum_{n_{z}}\left\langle i\left|f_{n_{z}}\right|k\right\rangle \left\langle j\left|\hat{\Phi}_{n_{z}}\right|l\right\rangle \\
 & = & \sqrt{\frac{G_{z}-1}{G_{z}+1}}\sum_{m_{z}=\left|n_{z}^{(i)}-n_{z}^{(k)}\right|,2}^{n_{z}^{(i)}+n_{z}^{(k)}}\sum_{n_{z}=\left|n_{z}^{(j)}-n_{z}^{(l)}\right|,2}^{n_{z}^{(j)}+n_{z}^{(l)}}T_{n_{z}^{(i)},n_{z}^{(k)}}^{m_{z}}T_{n_{z}^{(j)},n_{z}^{(l)}}^{n_{z}}\bar{I}\left(m_{z},n_{z}\right)\end{eqnarray*}
where $\bar{I}\left(m_{z},n_{z}\right)$ is given by Eq. (\ref{eq:Ibar_mn}),
and the $T$ coefficients are given by Eq. (\ref{eq:carprod-coeff}).
An alternate form of $V_{ijkl}^{\left(z\right)}$ was proposed by
Egido et al. \cite{egido97} which yields more accurate results for
large oscillator shell numbers, and is derived as Eq. (\ref{eq:vijkl-z-big})
in appendix \ref{sec:Formalism-for-large}.

\subsection{Radial component}

A formula similar to Eq. (\ref{eq:Vijkl-z}) can be derived for the
radial component, $V_{ijkl}^{\left(r\right)}$, in Eq. (\ref{eq:vijkl-separated}).
We will show that\begin{eqnarray}
V_{ijkl}^{\left(r\right)} & = & \frac{G_{\perp}-1}{G_{\perp}+1}\sum_{n_{r}=0}^{n_{\bar{j},l}}\sum_{n=0}^{n_{\bar{i},k}}T_{n_{r}^{(i)},-\Lambda^{(i)};n_{r}^{(k)},\Lambda^{(k)}}^{n,-\Lambda^{(i)}+\Lambda^{(k)}}T_{n_{r}^{(j)},-\Lambda^{(j)};n_{r}^{(l)},\Lambda^{(l)}}^{n_{r},-\Lambda^{(j)}+\Lambda^{(l)}}\nonumber \\
 &  & \times\bar{I}\left(n_{r},-\Lambda^{(j)}+\Lambda^{(l)};n,-\Lambda^{(i)}+\Lambda^{(k)}\right)\label{eq:vijkl-r}\end{eqnarray}
where $G_{\perp}$ is defined by Eq. (\ref{eq:Gp-def}), the $T$
coefficients by Eq. (\ref{eq:polprod-coeff1}), and the $\bar{I}$
coefficients by Eq. (\ref{eq:Ibar-pol}). The indices $n_{\bar{j},l}$
and $n_{\bar{i},k}$ are given by Eq. (\ref{eq:n12-def}), where the
bar indicates that $-\Lambda^{(j)}$ and $-\Lambda^{(i)}$ , respectively,
should be used in that definition due to the complex conjugation in
Eq. (\ref{eq:Vijkl-integral}).

Using Eqs. (\ref{eq:phihatpol}) for the explicit form of $\hat{\Phi}_{n_{r},\Lambda}\left(\rho,\varphi;b_{\perp}\right)$,
Eq. (\ref{eq:polprod-expr}) to reduce the product of harmonic-oscillator
functions, and the orthogonality of harmonic-oscillator functions\begin{eqnarray*}
\left\langle j\left|\hat{\Phi}_{n_{r},\Lambda}\right|l\right\rangle  & = & \int_{0}^{\infty}\rho d\rho\int_{0}^{2\pi}d\varphi\,\Phi_{n_{r}^{(j)},\Lambda^{(j)}}^{*}\left(\rho,\varphi;b_{\perp}\right)\hat{\Phi}_{n_{r},\Lambda}\left(\rho,\varphi;b_{\perp}\right)\\
 &  & \times\Phi_{n_{r}^{(l)},\Lambda^{(l)}}\left(\rho,\varphi;b_{\perp}\right)\\
 & = & \frac{1}{\sqrt{\pi}b_{\perp}}\sum_{n=0}^{n_{\bar{j},l}}T_{n_{r}^{(j)},-\Lambda^{(j)};n_{r}^{(l)},\Lambda^{(l)}}^{n,-\Lambda^{(j)}+\Lambda^{(l)}}\\
 &  & \times\int_{0}^{\infty}\rho d\rho\int_{0}^{2\pi}d\varphi\,\Phi_{n,\Lambda^{(j)}+\Lambda^{(l)}}^{*}\left(\rho,\varphi;b_{\perp}\right)\Phi_{n_{r},\Lambda}\left(\rho,\varphi;b_{\perp}\right)\\
 & = & \frac{1}{\sqrt{\pi}b_{\perp}}T_{n_{r}^{(j)},-\Lambda^{(j)};n_{r}^{(l)},\Lambda^{(l)}}^{n_{r},-\Lambda^{(j)}+\Lambda^{(l)}}\delta_{n_{r}\leq n_{\bar{j},l}}\delta_{\Lambda,-\Lambda^{(j)}+\Lambda^{(l)}}\end{eqnarray*}
where the bar superscript in the $n_{\bar{j},l}$ symbol serves as
a reminder that we must use $-\Lambda^{(j)}$ in Eq. (\ref{eq:n12-def}),
because of the complex conjugation. The condition $\delta_{n_{r}\leq n_{\bar{j},l}}$
comes about from the definition of the $T$ coefficients in Eq. (\ref{eq:polprod-coeff1}).
The other matrix element in the radial component of Eq. (\ref{eq:vijkl-separated})
is written explicitly using the explicit form for $f_{n_{r},\Lambda}\left(\rho,\varphi;b_{\perp}\right)$
in Eq. (\ref{eq:fpol}) as\begin{eqnarray*}
\left\langle i\left|f_{n_{r},\Lambda}\right|k\right\rangle  & = & \int_{0}^{\infty}\rho d\rho\int_{0}^{2\pi}d\varphi\,\Phi_{n_{r}^{(i)},\Lambda^{(i)}}^{*}\left(\rho,\varphi;b_{\perp}\right)f_{n_{r},\Lambda}\left(\rho,\varphi;b_{\perp}\right)\\
 &  & \times\Phi_{n_{r}^{(k)},\Lambda^{(k)}}\left(\rho,\varphi;b_{\perp}\right)\\
 & = & K_{\bot}\lambda_{2n_{r}+\left|\Lambda\right|}\int_{0}^{\infty}\rho d\rho\int_{0}^{2\pi}d\varphi\, e^{-\rho^{2}/\left(2G_{\bot}b_{\bot}^{2}\right)}\Phi_{n_{r},\Lambda}\left(\rho,\varphi;G_{\bot}^{1/2}b_{\perp}\right)\\
 &  & \times\Phi_{n_{r}^{(i)},\Lambda^{(i)}}^{*}\left(\rho,\varphi;b_{\perp}\right)\Phi_{n_{r}^{(k)},\Lambda^{(k)}}\left(\rho,\varphi;b_{\perp}\right)\end{eqnarray*}
and using Eq. (\ref{eq:polprod-expr}), the product of harmonic-oscillator
functions can be reduced\begin{eqnarray*}
\left\langle i\left|f_{n_{r},\Lambda}\right|k\right\rangle  & = & \frac{K_{\bot}\lambda_{2n_{r}+\left|\Lambda\right|}}{\sqrt{\pi}b_{\perp}}\sum_{n=0}^{n_{\bar{i},k}}T_{n_{r}^{(i)},-\Lambda^{(i)};n_{r}^{(k)},\Lambda^{(k)}}^{n,-\Lambda^{(i)}+\Lambda^{(k)}}\\
 &  & \times\int_{0}^{\infty}\rho d\rho\int_{0}^{2\pi}d\varphi\, e^{-\rho^{2}/\left(2B_{\bot}^{2}\right)-\rho^{2}/\left(2b_{\bot}^{2}\right)}\\
 &  & \times\Phi_{n_{r},\Lambda}\left(\rho,\varphi;B_{\bot}\right)\Phi_{n,-\Lambda^{(i)}+\Lambda^{(k)}}\left(\rho,\varphi;b_{\perp}\right)\end{eqnarray*}
where $B_{\bot}\equiv G_{\bot}^{1/2}b_{\perp}$, and the $\bar{i}$
in $n_{\bar{i},k}$ is a reminder that we must use $-\Lambda^{(i)}$
in Eq. (\ref{eq:n12-def}). The remaining integral to be calculated
is\begin{eqnarray}
I\left(n_{1},k_{1};n_{2},k_{2}\right) & \equiv & \int_{0}^{\infty}\rho d\rho\int_{0}^{2\pi}d\varphi\, e^{-\rho^{2}/\left(2B_{\bot}^{2}\right)-\rho^{2}/\left(2b_{\bot}^{2}\right)}\nonumber \\
 &  & \times\Phi_{n_{1},k_{1}}\left(\rho,\varphi;B_{\bot}\right)\Phi_{n_{2},k_{2}}\left(\rho,\varphi;b_{\perp}\right)\label{eq:Ipol-def}\end{eqnarray}
and can be evaluated using the generating function in Eq. (\ref{eq:polho-genf-vec})
by writing, for arbitrary vectors $\vec{t}_{1}$ and $\vec{t}_{2}$,\begin{eqnarray}
 &  & e^{-\vec{t}_{1}^{2}+2\vec{\rho}\cdot\vec{t}_{1}/B_{\bot}-\rho^{2}/\left(2B_{\bot}^{2}\right)}e^{-\vec{t}_{2}^{2}+2\vec{\rho}\cdot\vec{t}_{2}/b_{\bot}-\rho^{2}/\left(2b_{\bot}^{2}\right)}\nonumber \\
 & = & B_{\bot}^{2}\sqrt{\frac{\pi}{2}}\sum_{k_{1}=-\infty}^{\infty}\sum_{n_{1}=0}^{\infty}\mathcal{N}_{n_{1},\left|k_{1}\right|}\left(B_{\bot}\right)\chi_{n_{1},k_{1}}\left(\vec{t}_{1}\right)\Phi_{n_{1},k_{1}}\left(\rho,\varphi;B_{\bot}\right)\nonumber \\
 &  & \times b_{\bot}^{2}\sqrt{\frac{\pi}{2}}\sum_{k_{2}=-\infty}^{\infty}\sum_{n_{2}=0}^{\infty}\mathcal{N}_{n_{2},\left|k_{2}\right|}\left(b_{\bot}\right)\chi_{n_{2},k_{2}}\left(\vec{t}_{2}\right)\Phi_{n_{2},k_{2}}\left(\rho,\varphi;b_{\bot}\right)\label{eq:Ipol-genf1}\end{eqnarray}
note that, for clarity, we have explicitly written the parameter dependence
for the normalization coefficients $\mathcal{N}_{n_{1},\left|k_{1}\right|}\left(B_{\bot}\right)$
and $\mathcal{N}_{n_{2},\left|k_{2}\right|}\left(b_{\bot}\right)$
given by Eq. (\ref{eq:normpol-def}). Multiplying both sides of Eq.
(\ref{eq:Ipol-genf1}) by the Gaussian factor that appears in Eq.
(\ref{eq:Ipol-def}) and integrating, we obtain on the left-hand side\begin{eqnarray}
LHS & = & e^{-\vec{t}_{1}^{2}-\vec{t}_{2}^{2}}\int_{0}^{\infty}\rho d\rho\int_{0}^{2\pi}d\varphi\, e^{-\rho^{2}/B_{\bot}^{2}-\rho^{2}/b_{\bot}^{2}}e^{2\vec{\rho}\cdot\vec{t}_{1}/B_{\bot}}e^{2\vec{\rho}\cdot\vec{t}_{2}/b_{\bot}}\label{eq:rad-comp-lhs}\end{eqnarray}
and on the right-hand side\begin{eqnarray}
RHS & = & \frac{\pi}{2}B_{\bot}^{2}b_{\bot}^{2}\sum_{k_{1}=-\infty}^{\infty}\sum_{n_{1}=0}^{\infty}\sum_{k_{2}=-\infty}^{\infty}\sum_{n_{2}=0}^{\infty}\mathcal{N}_{n_{1},\left|k_{1}\right|}\left(B_{\bot}\right)\mathcal{N}_{n_{2},\left|k_{2}\right|}\left(b_{\bot}\right)\nonumber \\
 &  & \times\chi_{n_{1},k_{1}}\left(\vec{t}_{1}\right)\chi_{n_{2},k_{2}}\left(\vec{t}_{2}\right)I\left(n_{1},k_{1};n_{2},k_{2}\right)\label{eq:Ipol-genf-rhs}\end{eqnarray}
which contains the desired coefficients $I\left(n_{1},k_{1};n_{2},k_{2}\right)$.
The integral in Eq. (\ref{eq:rad-comp-lhs}) can be evaluated by introducing\begin{eqnarray*}
\vec{t} & \equiv & \frac{\vec{t}_{1}}{B_{\bot}}+\frac{\vec{t}_{2}}{b_{\bot}}\\
\nu & \equiv & \frac{1}{B_{\bot}^{2}}+\frac{1}{b_{\bot}^{2}}\end{eqnarray*}
and completing the square,\begin{eqnarray*}
LHS & = & e^{t^{2}/\nu-t_{1}^{2}-t_{2}^{2}}\int_{0}^{\infty}\rho d\rho\int_{0}^{2\pi}d\varphi\,\exp\left[-\left(\sqrt{\nu}\vec{\rho}-\frac{\vec{t}}{\sqrt{\nu}}\right)^{2}\right]\\
 & = & \frac{\pi}{\nu}e^{t^{2}/\nu-t_{1}^{2}-t_{2}^{2}}\\
 & = & \frac{\pi}{\nu}e^{\left(2b_{\bot}B_{\bot}\vec{t}_{1}\cdot\vec{t}_{2}-B_{\bot}^{2}t_{1}^{2}-b_{\bot}^{2}t_{2}^{2}\right)/\left(B_{\bot}^{2}b_{\bot}^{2}\nu\right)}\end{eqnarray*}
using Eq. (\ref{eq:chiprops-expt1t2}) with $\vec{t}_{1}\rightarrow\vec{t}_{1}/\left(b_{\bot}\sqrt{\nu}\right)$
and $\vec{t}_{2}\rightarrow\vec{t}_{2}/\left(B_{\bot}\sqrt{\nu}\right)$,
this can be further expanded as\begin{eqnarray*}
LHS & = & \frac{\pi B_{\bot}^{2}b_{\bot}^{2}}{2}e^{-\left(B_{\bot}^{2}t_{1}^{2}+b_{\bot}^{2}t_{2}^{2}\right)/\left(B_{\bot}^{2}b_{\bot}^{2}\nu\right)}\sum_{n=0}^{\infty}\sum_{k=-\infty}^{\infty}\mathcal{N}_{n,\left|k\right|}^{2}\left(B_{\bot}b_{\bot}\sqrt{\nu}\right)\\
 &  & \times\chi_{n,k}^{*}\left(\frac{\vec{t}_{1}}{b_{\bot}\sqrt{\nu}}\right)\chi_{n,k}\left(\frac{\vec{t}_{2}}{B_{\bot}\sqrt{\nu}}\right)\end{eqnarray*}
Next, we use Eq. (\ref{eq:chiprops-exptransl}) to eliminate the remaining
exponential,\begin{eqnarray*}
LHS & = & \frac{\pi B_{\bot}^{2}b_{\bot}^{2}}{2}\sum_{n=0}^{\infty}\sum_{k=-\infty}^{\infty}\sum_{m_{1}=0}^{\infty}\sum_{m_{2}=0}^{\infty}\frac{\left(n+m_{1}\right)!\left(n+m_{2}\right)!}{m_{1}!m_{2}!\left(n!\right)^{2}}\mathcal{N}_{n,\left|k\right|}^{2}\left(B_{\bot}b_{\bot}\sqrt{\nu}\right)\\
 &  & \times\chi_{n+m_{1},k}^{*}\left(\frac{\vec{t}_{1}}{b_{\bot}\sqrt{\nu}}\right)\chi_{n+m_{2},k}\left(\frac{\vec{t}_{2}}{B_{\bot}\sqrt{\nu}}\right)\end{eqnarray*}
Using Eqs (\ref{eq:chiprops-complex}) to eliminate the complex conjugation,
and (\ref{eq:chiprops-scale}) to factor out the coefficients inside
the $\chi$ functions, this takes the form\begin{eqnarray*}
LHS & = & \frac{\pi B_{\bot}^{2}b_{\bot}^{2}}{2}\sum_{n=0}^{\infty}\sum_{k=-\infty}^{\infty}\sum_{m_{1}=0}^{\infty}\sum_{m_{2}=0}^{\infty}\frac{\left(n+m_{1}\right)!\left(n+m_{2}\right)!}{m_{1}!m_{2}!\left(n!\right)^{2}}\mathcal{N}_{n,\left|k\right|}^{2}\left(B_{\bot}b_{\bot}\sqrt{\nu}\right)\\
 &  & \times\left(b_{\bot}\sqrt{\nu}\right)^{-2\left(n+m_{1}\right)-\left|k\right|}\left(B_{\bot}\sqrt{\nu}\right)^{-2\left(n+m_{2}\right)-\left|k\right|}\\
 &  & \times\chi_{n+m_{1},-k}\left(\vec{t}_{1}\right)\chi_{n+m_{2},k}\left(\vec{t}_{2}\right)\end{eqnarray*}
Comparing this result for $LHS$ with $RHS$ in Eq. (\ref{eq:Ipol-genf-rhs})
for arbitrary vectors $\vec{t}_{1}$ and $\vec{t}_{2}$, we are led
to conclude that \begin{eqnarray}
I\left(n_{1},k_{1};n_{2},k_{2}\right) & = & 0\quad\textrm{if}\; k_{1}+k_{2}\neq0\label{eq:Ipol-cond}\end{eqnarray}
We are also led to make the identifications\begin{eqnarray*}
n+m_{1} & = & n_{1}\\
n+m_{2} & = & n_{2}\\
-k & = & k_{1}\\
k & = & k_{2}\end{eqnarray*}
which allow us to write\begin{eqnarray}
LHS & = & \frac{\pi B_{\bot}^{2}b_{\bot}^{2}}{2}\sum_{n=0}^{\infty}\sum_{k=-\infty}^{\infty}\sum_{n_{1}=0}^{\infty}\sum_{n_{2}=0}^{\infty}\frac{n_{1}!n_{2}!}{\left(n_{1}-n\right)!\left(n_{2}-n\right)!\left(n!\right)^{2}}\mathcal{N}_{n,\left|k\right|}^{2}\left(B_{\bot}b_{\bot}\sqrt{\nu}\right)\nonumber \\
 &  & \times\left(b_{\bot}\sqrt{\nu}\right)^{-2n_{1}-\left|k\right|}\left(B_{\bot}\sqrt{\nu}\right)^{-2n_{2}-\left|k\right|}\chi_{n_{1},-k}\left(\vec{t}_{1}\right)\chi_{n_{2},k}\left(\vec{t}_{2}\right)\label{eq:rad-comp-lhs2}\end{eqnarray}
and therefore, assuming $\left|k_{1}\right|=\left|k_{2}\right|\equiv\left|k\right|$
because of Eq. (\ref{eq:Ipol-cond}), the comparison between $LHS$
and $RHS$, in Eqs. (\ref{eq:rad-comp-lhs2}) and (\ref{eq:Ipol-genf-rhs})
respectively, yields\begin{eqnarray*}
I\left(n_{1},k_{1};n_{2},k_{2}\right) & = & \frac{\delta_{k_{1}+k_{2},0}\left(b_{\bot}\sqrt{\nu}\right)^{-2n_{1}-\left|k\right|}\left(B_{\bot}\sqrt{\nu}\right)^{-2n_{2}-\left|k\right|}n_{1}!n_{2}!}{\mathcal{N}_{n_{1},\left|k\right|}\left(B_{\bot}\right)\mathcal{N}_{n_{2},\left|k\right|}\left(b_{\bot}\right)}\\
 &  & \times\sum_{n=0}^{\infty}\frac{\mathcal{N}_{n,\left|k\right|}^{2}\left(B_{\bot}b_{\bot}\sqrt{\nu}\right)}{\left(n_{1}-n\right)!\left(n_{2}-n\right)!\left(n!\right)^{2}}\\
 & = & \delta_{k_{1}+k_{2},0}\frac{\left(b_{\bot}\sqrt{\nu}\right)^{-2n_{1}-\left|k\right|}\left(B_{\bot}\sqrt{\nu}\right)^{-2n_{2}-\left|k\right|}}{B_{\bot}b_{\bot}\nu}\\
 &  & \times\sqrt{n_{1}!\left(n_{1}+\left|k\right|\right)!n_{2}!\left(n_{2}+\left|k\right|\right)!}\\
 &  & \sum_{n=0}^{\infty}\frac{1}{\left(n_{1}-n\right)!\left(n_{2}-n\right)!n!\left(n+\left|k\right|\right)!}\end{eqnarray*}
Using Eqs. (\ref{eq:bBnu})-(\ref{eq:bbnu}) with $G_{\bot}$ instead
of $G_{z}$ we can simplify the factor outside the summation\begin{eqnarray*}
\frac{\left(b_{\bot}\sqrt{\nu}\right)^{-2n_{1}-\left|k\right|}\left(B_{\bot}\sqrt{\nu}\right)^{-2n_{2}-\left|k\right|}}{B_{\bot}b_{\bot}\nu} & = & \frac{\left(1+G_{\bot}^{-1}\right)^{-n_{1}-\left|k\right|/2}\left(1+G_{\bot}\right)^{-n_{2}-\left|k\right|/2}}{G_{\bot}^{1/2}+G_{\bot}^{-1/2}}\\
 & = & \frac{G_{\bot}^{\left(n_{1}-n_{2}\right)/2}}{\left(G_{\bot}^{1/2}+G_{\bot}^{-1/2}\right)^{n_{1}+n_{2}+\left|k\right|+1}}\end{eqnarray*}
and, for compactness of notation, we define\begin{eqnarray*}
\Xi\left(n_{1},n_{2},\left|k\right|\right) & \equiv & \sum_{n=0}^{\infty}\frac{1}{\left(n_{1}-n\right)!\left(n_{2}-n\right)!n!\left(n+\left|k\right|\right)!}\end{eqnarray*}
which, after some simplification can be written as\begin{eqnarray*}
\Xi\left(n_{1},n_{2},\left|k\right|\right) & = & \frac{1}{n_{1}!\left(n_{2}+\left|k\right|\right)!}\sum_{n=0}^{\infty}\left(\begin{array}{c}
n_{1}\\
n\end{array}\right)\left(\begin{array}{c}
n_{2}+\left|k\right|\\
n_{2}-n\end{array}\right)\\
 & = & \frac{1}{\left(n_{1}+n_{2}+\left|k\right|\right)!}\left(\begin{array}{c}
n_{1}+n_{2}+\left|k\right|\\
n_{1}\end{array}\right)\left(\begin{array}{c}
n_{1}+n_{2}+\left|k\right|\\
n_{2}\end{array}\right)\end{eqnarray*}
where Eq. 0.156(1) in \cite{gradshteyn79} was used to obtain the
second line. Therefore, we finally have\begin{eqnarray*}
I\left(n_{1},k_{1};n_{2},k_{2}\right) & = & \delta_{k_{1}+k_{2},0}\frac{G_{\bot}^{\left(n_{1}-n_{2}\right)/2}}{\left(G_{\bot}^{1/2}+G_{\bot}^{-1/2}\right)^{n_{1}+n_{2}+\left|k\right|+1}}\\
 &  & \times\sqrt{n_{1}!\left(n_{1}+\left|k\right|\right)!n_{2}!\left(n_{2}+\left|k\right|\right)!}\Xi\left(n_{1},n_{2},\left|k\right|\right)\end{eqnarray*}
As in Eq. (\ref{eq:Ibar_mn}), it will be convenient to factor out
some constant terms. Therefore we define\begin{eqnarray}
\bar{I}\left(n_{1},k_{1};n_{2},k_{2}\right) & \equiv & \frac{K_{\bot}\lambda_{2n_{1}+\left|k\right|}}{\pi b_{\perp}^{2}}\frac{G_{\perp}+1}{G_{\perp}-1}I\left(n_{1},k_{1};n_{2},k_{2}\right)\nonumber \\
 & = & \delta_{k_{1}+k_{2},0}\frac{\sqrt{n_{1}!\left(n_{1}+\left|k\right|\right)!n_{2}!\left(n_{2}+\left|k\right|\right)!}}{\left(G_{\perp}+1\right)^{n_{1}+n_{2}+\left|k\right|}}\Xi\left(n_{1},n_{2},\left|k\right|\right)\label{eq:Ibar-pol}\end{eqnarray}
and the radial component in Eq. (\ref{eq:vijkl-separated}) becomes\begin{eqnarray*}
V_{ijkl}^{\left(r\right)} & = & \sum_{n_{r},\Lambda}\left\langle i\left|f_{n_{r},\Lambda}\right|k\right\rangle \left\langle j\left|\hat{\Phi}_{n_{r},\Lambda}\right|l\right\rangle \\
 & = & \frac{G_{\perp}-1}{G_{\perp}+1}\sum_{n_{r}=0}^{\infty}\sum_{\Lambda=-\infty}^{\infty}\sum_{n=0}^{n_{\bar{i},k}}T_{n_{r}^{(i)},-\Lambda^{(i)};n_{r}^{(k)},\Lambda^{(k)}}^{n,-\Lambda^{(i)}+\Lambda^{(k)}}\bar{I}\left(n_{r},\Lambda;n,-\Lambda^{(i)}+\Lambda^{(k)}\right)\\
 &  & \times T_{n_{r}^{(j)},\Lambda^{(j)};n_{r}^{(l)},\Lambda^{(l)}}^{n_{r},\Lambda^{(j)}+\Lambda^{(l)}}\delta_{n_{r}\leq n_{j,l}}\delta_{\Lambda,-\Lambda^{(j)}+\Lambda^{(l)}}\\
 & = & \frac{G_{\perp}-1}{G_{\perp}+1}\sum_{n_{r}=0}^{n_{\bar{j},l}}\sum_{n=0}^{n_{\bar{i},k}}T_{n_{r}^{(i)},-\Lambda^{(i)};n_{r}^{(k)},\Lambda^{(k)}}^{n,-\Lambda^{(i)}+\Lambda^{(k)}}T_{n_{r}^{(j)},-\Lambda^{(j)};n_{r}^{(l)},\Lambda^{(l)}}^{n_{r},-\Lambda^{(j)}+\Lambda^{(l)}}\\
 &  & \times\bar{I}\left(n_{r},-\Lambda^{(j)}+\Lambda^{(l)};n,-\Lambda^{(i)}+\Lambda^{(k)}\right)\end{eqnarray*}
Thus , using Eqs. (\ref{eq:Vijkl-z}) or (\ref{eq:vijkl-z-big}) and
(\ref{eq:vijkl-r}), the full matrix element $V_{ijkl}$ in Eq. (\ref{eq:vijkl-separated})
can be calculated as an analytical expression. In the next section,
we will examine the computational merits of these results.

\section{Discussion\label{sec:Discussion}}

In this section, we will compare three different ways of evaluating
the Cartesian ($V_{ijkl}^{\left(z\right)}$) and radial ($V_{ijkl}^{\left(r\right)}$)
components of the Gaussian matrix elements in Eq. (\ref{eq:Vijkl-integral}):
1) direct numerical integration of Eq. (\ref{eq:Vijkl-integral}),
2) numerical evaluation of the separation-method equations (Eqs. (\ref{eq:Vijkl-z})
or (\ref{eq:vijkl-z-big}) for the Cartesian component, and Eq. (\ref{eq:vijkl-r})
for the radial component) in double-precision mode, and 3) exact evaluation
of the separation-method equations using the symbolic-algebra package
Mathematica \cite{wolfram03}. In principle, the first two methods--numerical
evaluation by either integration or the separation method--will give
the values of $V_{ijkl}^{\left(z\right)}$ and $V_{ijkl}^{\left(r\right)}$
to within the limits of machine accuracy and roundoff errors, whereas
the third--exact evaluation of the separation-method equations using
Mathematica--will produce these matrix elements to any desired accuracy
(even beyond machine accuracy) and will serve as a reference check
for numerical convergence of the integrals and roundoff errors.

We begin by comparing the relative merits of the separation-method
Eqs. (\ref{eq:Vijkl-z}) and (\ref{eq:vijkl-z-big}) for the Cartesian
component of the matrix element. The two equations are mathematically
equivalent, but Eq. (\ref{eq:vijkl-z-big}) was obtained from Eq.
(\ref{eq:Vijkl-z}) specifically to provide greater accuracy in numerical
calculations. For all quantitative applications in this work, we have
used\begin{eqnarray*}
\mu & = & 1.2\;\textrm{fm}\\
b_{z} & = & 3.3\;\textrm{fm}\\
b_{\bot} & = & 2\;\textrm{fm}\end{eqnarray*}
These values of $\mu$, $b_{z}$, $b_{\bot}$ are typical in HFB calculations
using the Gogny interaction for $^{240}\textrm{Pu}$ along the most
likely path to scission \cite{younes07}.

In practice, both Eqs. (\ref{eq:Vijkl-z}) and (\ref{eq:vijkl-z-big})
can be evaluated efficiently because the $T_{n_{1},n_{2}}^{n}$ and
$\bar{I}\left(m,n\right)$ or $\bar{F}_{n_{1},n_{2}}^{n}$ coefficients
can easily be calculated once and for all and stored with relatively
little memory, to be used in reconstructing the matrix elements $V_{ijkl}^{\left(z\right)}$
whenever they are needed. However, for large values of the quantum
numbers $n_{i}$, $n_{j}$, $n_{k}$, and $n_{l}$ the sums in Eq.
(\ref{eq:Vijkl-z}) rapidly lead to sizable numerical inaccuracies.
These inaccuracies arise because the $T$ coefficients grow progressively
larger with increasing values of the arguments, whereas the $\bar{I}$
coefficients decrease. The resulting sum of products of small and
large numbers in Eq. (\ref{eq:Vijkl-z}) becomes numerically unstable.
The formula obtained by Egido et al. in \cite{egido97}, and derived
as Eq. (\ref{eq:vijkl-z-big}) in the present work, avoids this problem. 

\begin{figure}
\includegraphics[%
  scale=0.5,
  angle=-90]{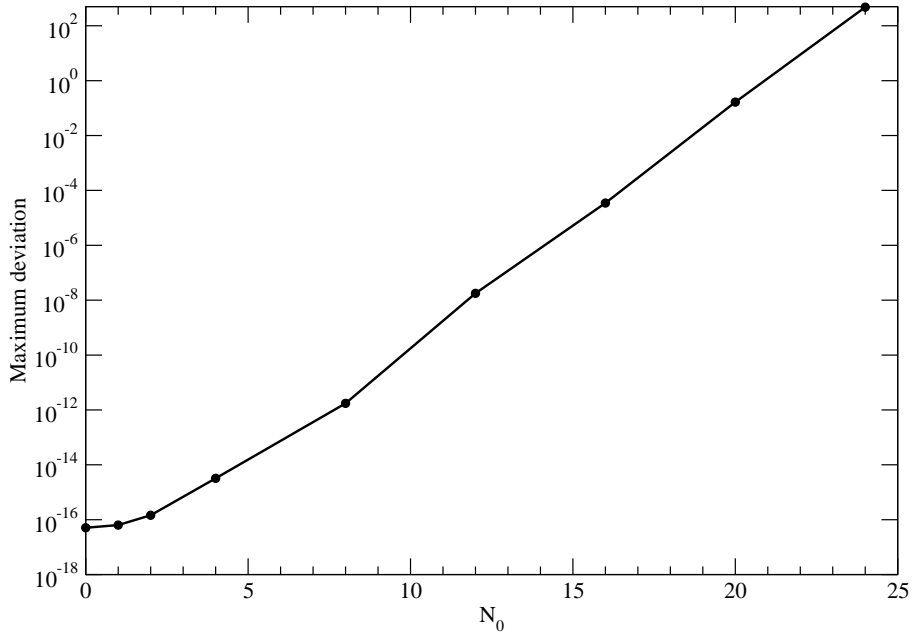}

\caption{\label{cap:zsep-vs-zsepbig}Maximum deviation between calculations
of the matrix elements $V_{ijkl}^{\left(z\right)}$ using the separation
method in Eq. (\ref{eq:Vijkl-z}) on one hand, and Eq. (\ref{eq:vijkl-z-big})
on the other, plotted as a function of basis size $N_{0}$.}
\end{figure}

Fig. \ref{cap:zsep-vs-zsepbig} gives the maximum deviation between
matrix elements calculated using numerical evaluations of Eqs. (\ref{eq:Vijkl-z})
and (\ref{eq:vijkl-z-big}). To generate the plot, the equations were
compared for calculations of $V_{ijkl}^{\left(z\right)}$ as a function
of the maximum harmonic-oscillator shell number $N_{0}$, i.e. for
all possible quantum numbers such that $0\leq n_{i},n_{j},n_{k},n_{l}\leq N_{0}$,
and the largest deviation was recorded for each point on the plot.
We will refer to $N_{0}$ as the size of the basis in the discussion
below. The deviations plotted in Fig. \ref{cap:zsep-vs-zsepbig} are
based on the dimensionless Gaussian function in Eq. (\ref{eq:vgauss}),
but with realistic interaction strengths for the Gogny force \cite{warda02},
a deviation as small as $10^{-2}$ on the plot, can correspond to
a discrepancy of the order of an MeV. Thus, for $N_{0}$ greater than
about 16, Eq. (\ref{eq:vijkl-z-big}) should certainly always be used
instead of Eq. (\ref{eq:Vijkl-z}), and in the remainder of this paper
we will use it consistently for all $N_{0}$ instead of Eq. (\ref{eq:Vijkl-z}).

\begin{figure}
\includegraphics[%
  scale=0.5,
  angle=-90]{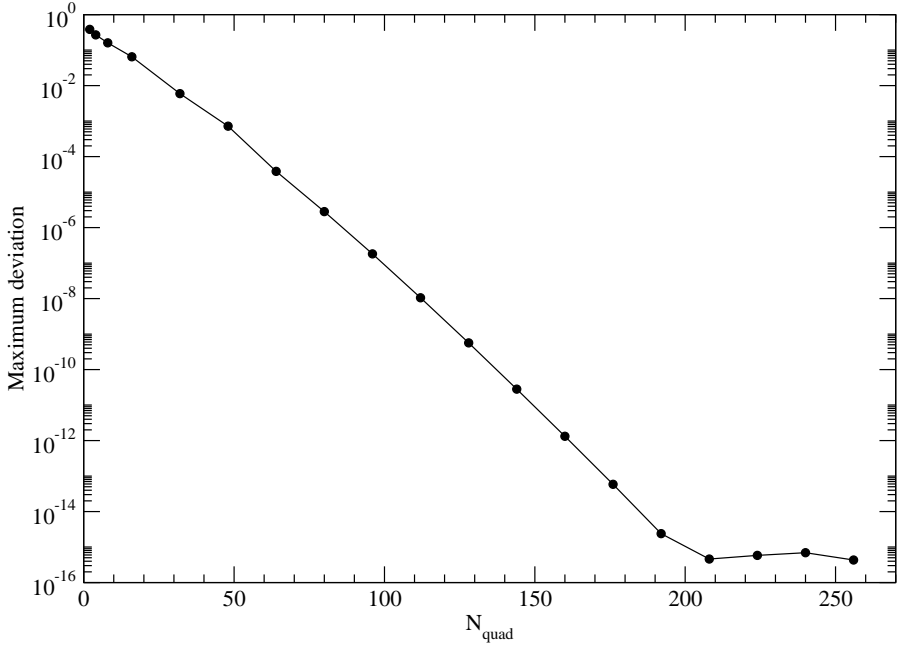}

\caption{\label{cap:vzint12-conv}Maximum deviation between the numerical
integration of the matrix elements $V_{ijkl}^{\left(z\right)}$ and
their exact evaluation using the separation method in Eq. (\ref{eq:vijkl-z-big})
with Mathematica for basis size $N_{0}=12$, plotted as a function
of the number of Gauss-Hermite quadrature points in the integral.}
\end{figure}

Next, we compare an exact evaluation of Eq. (\ref{eq:vijkl-z-big})
to the numerical integration of the Cartesian component in Eq. (\ref{eq:Vijkl-integral}).
We choose to compare the separation method to a numerical integral
of the potential because the latter is easily implemented, requires
very little computer memory, and can be made arbitrarily accurate.
The exact evaluation of Eq. (\ref{eq:vijkl-z-big}) was obtained using
the symbolic-algebra package Mathematica. Within Mathematica, the
expression in Eq. (\ref{eq:vijkl-z-big}) was first reduced by symbolic
manipulation to the exact algebraic form $a\sqrt{b}/c$, where $a$,
$b$, and $c$ are integers, for each choice of the quantum numbers
$n_{i}$, $n_{j}$, $n_{k}$, and $n_{l}$. That algebraic number
could then be evaluated numerically to any desired accuracy. The numerical
integration, on the other hand, was performed by Gauss-Hermite quadrature
in double-precision mode (i.e., with 16 significant figures). The
purpose of the comparison between the exact evaluation of Eq. (\ref{eq:vijkl-z-big})
and the numerical integration is to show that the numerical integration
can be made arbitrarily close (up to the limits of machine accuracy)
to the exact result, thereby validating Eq. (\ref{eq:vijkl-z-big}).
In Fig. \ref{cap:vzint12-conv}, the maximum deviation between the
exact calculation and numerical integration of the $V_{ijkl}^{\left(z\right)}$
values is plotted as a function of the number $N_{\textrm{quad}}$
of quadrature points for a basis size $N_{0}=12$. For $N_{\textrm{quad}}\geq208$,
the limits of machine accuracy are reached in the numerical integration,
and the maximum deviation between the two methods of calculating $V_{ijkl}^{\left(z\right)}$
matrix elements levels out slightly above $4.3\times10^{-16}$.

\begin{figure}
\includegraphics[%
  scale=0.5,
  angle=-90]{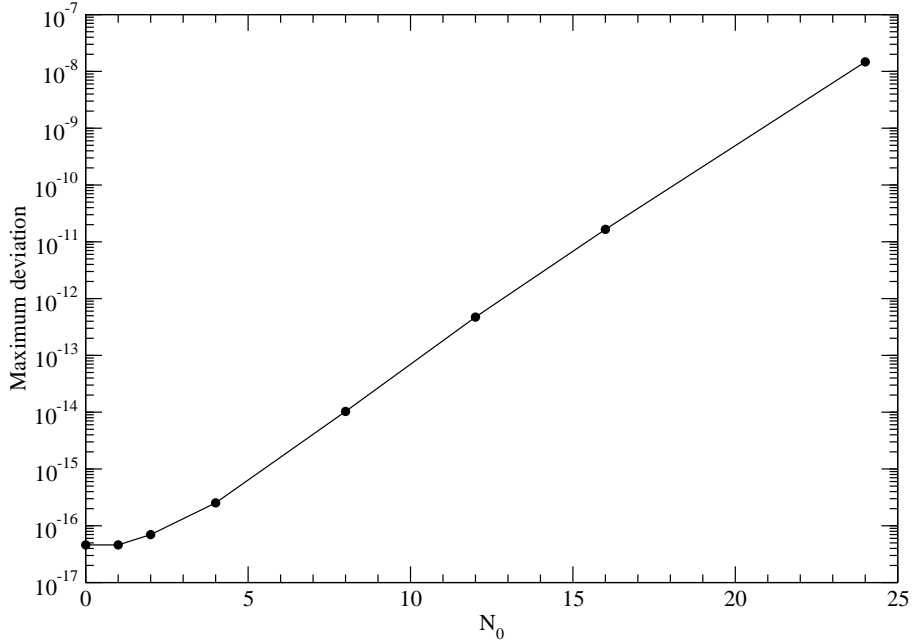}

\caption{\label{cap:vzbig-excomp}Maximum deviation between the numerical
calculation and exact Mathematica evaluation of the matrix elements
$V_{ijkl}^{\left(z\right)}$ using the separation method in Eq. (\ref{eq:vijkl-z-big}),
plotted as a function of basis size $N_{0}$.}
\end{figure}

In Fig. \ref{cap:vzbig-excomp}, we compare the exact evaluation of
Eq. (\ref{eq:vijkl-z-big}) using Mathematica to its numerical evaluation
in double-precision mode, as a function of basis size $N_{0}$. The
trend in Fig. \ref{cap:vzbig-excomp} shows the effect of roundoff
error in the numerical evaluation of Eq. (\ref{eq:vijkl-z-big}).
However, despite a clear decrease in accuracy with increasing basis
size, Fig. \ref{cap:vzbig-excomp} shows that a double-precision numerical
evaluation of Eq. (\ref{eq:vijkl-z-big}) still gives the value of
the matrix elements $V_{ijkl}^{\left(z\right)}$ to a very high level
of accuracy. Even for a basis size as large as $N_{0}=24$, the largest
deviation from the exact values is still only $1.5\times10^{-8}$.
For the remainder of this discussion, we will use the numerical evaluation
of Eq. (\ref{eq:vijkl-z-big}) in double-precision mode rather than
the exact Mathematica result, because the Mathematica calculations
are prohibitively time-consuming, and the accuracy of the numerical
evaluation of the separation-method formulas is more than sufficient
for most applications.

\begin{figure}
\includegraphics[%
  scale=0.5,
  angle=-90]{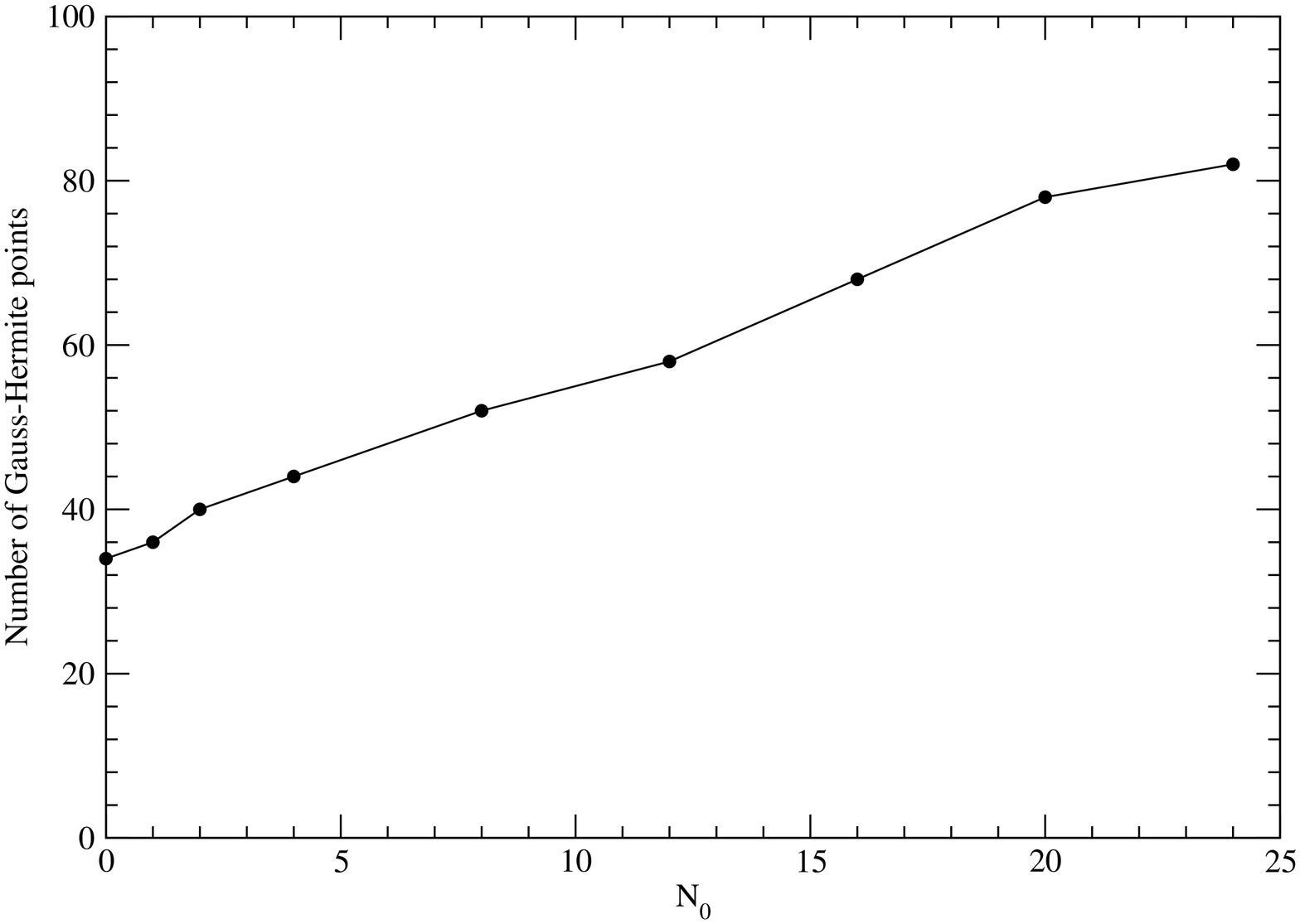}

\caption{\label{cap:sepgauss-zstats}Minimum number of Gauss-Hermite quadrature
points needed to achieve $10^{-4}$ or better agreement between the
numerical integration of the matrix elements $V_{ijkl}^{\left(z\right)}$
and their evaluation using the separation method in Eq. (\ref{eq:vijkl-z-big}),
plotted as a function of basis size $N_{0}$.}
\end{figure}

In Fig. \ref{cap:sepgauss-zstats}, we extract the number of Gauss-Hermite
quadrature points required by the numerical integration to obtain
values that are satisfactorily close (say, within a $10^{-4}$ discrepancy
at most) to the values given by a numerical evaluation of Eq. (\ref{eq:vijkl-z-big}).
The number of quadrature points plotted as a function of basis size
$N_{0}$ is moderately large, and increases steadily with $N_{0}$.
Further below we will gauge the cost in computational time incurred
by the numerical integration with these relatively large numbers of
quadrature points.

\begin{figure}
\includegraphics[%
  scale=0.5,
  angle=-90]{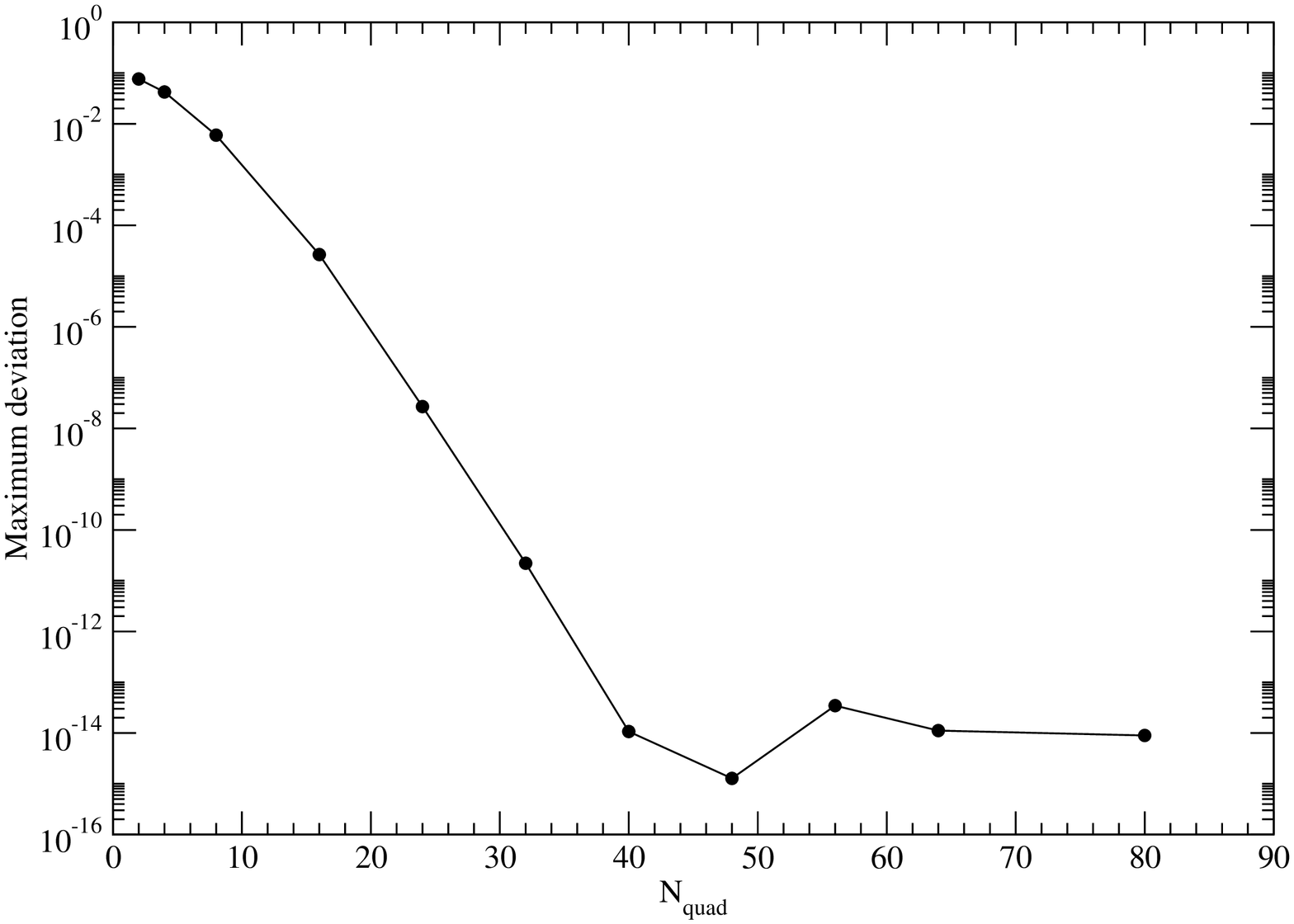}

\caption{\label{cap:vrint8-conv}Maximum deviation between the numerical integration
of the matrix elements $V_{ijkl}^{\left(r\right)}$ and their exact
evaluation using the separation method in Eq. (\ref{eq:vijkl-r})
with Mathematica for basis size $N_{0}=8$, plotted as a function
of the number of Gauss-Laguerre quadrature points in the integral.}
\end{figure}

We carry out a similar analysis for the radial component,$V_{ijkl}^{\left(r\right)}$,
of the matrix elements. In this case, for a given basis size $N_{0}$,
the quantum numbers for the radial matrix element $V_{ijkl}^{\left(r\right)}$
in Eq. (\ref{eq:Vijkl-integral}) take on all values such that $0\leq2n_{r}+\left|\Lambda\right|\leq N_{0}$
with $n_{r}\geq0$. As we did in Fig. \ref{cap:vzint12-conv} for
the Cartesian component, we compare in Fig. \ref{cap:vrint8-conv}
an exact (Mathematica) calculation of Eq. (\ref{eq:vijkl-r}) to a
numerical integration of the radial component in Eq. (\ref{eq:Vijkl-integral})
using double-precision Gauss-Laguerre quadrature, for a basis size
$N_{0}=8$. In Fig. \ref{cap:vrint8-conv}, the maximum deviation
between exact evaluation and numerical integration, plotted as a function
of the number $N_{\textrm{quad}}$ of quadrature points, is made arbitrarily
small with increasing $N_{\textrm{quad}}$ values until the limits
of machine accuracy and roundoff error are reached for $N_{\textrm{quad}}\geq48$,
where the maximum discrepancy settles above $1.3\times10^{-15}$.

\begin{figure}
\includegraphics[%
  scale=0.5,
  angle=-90]{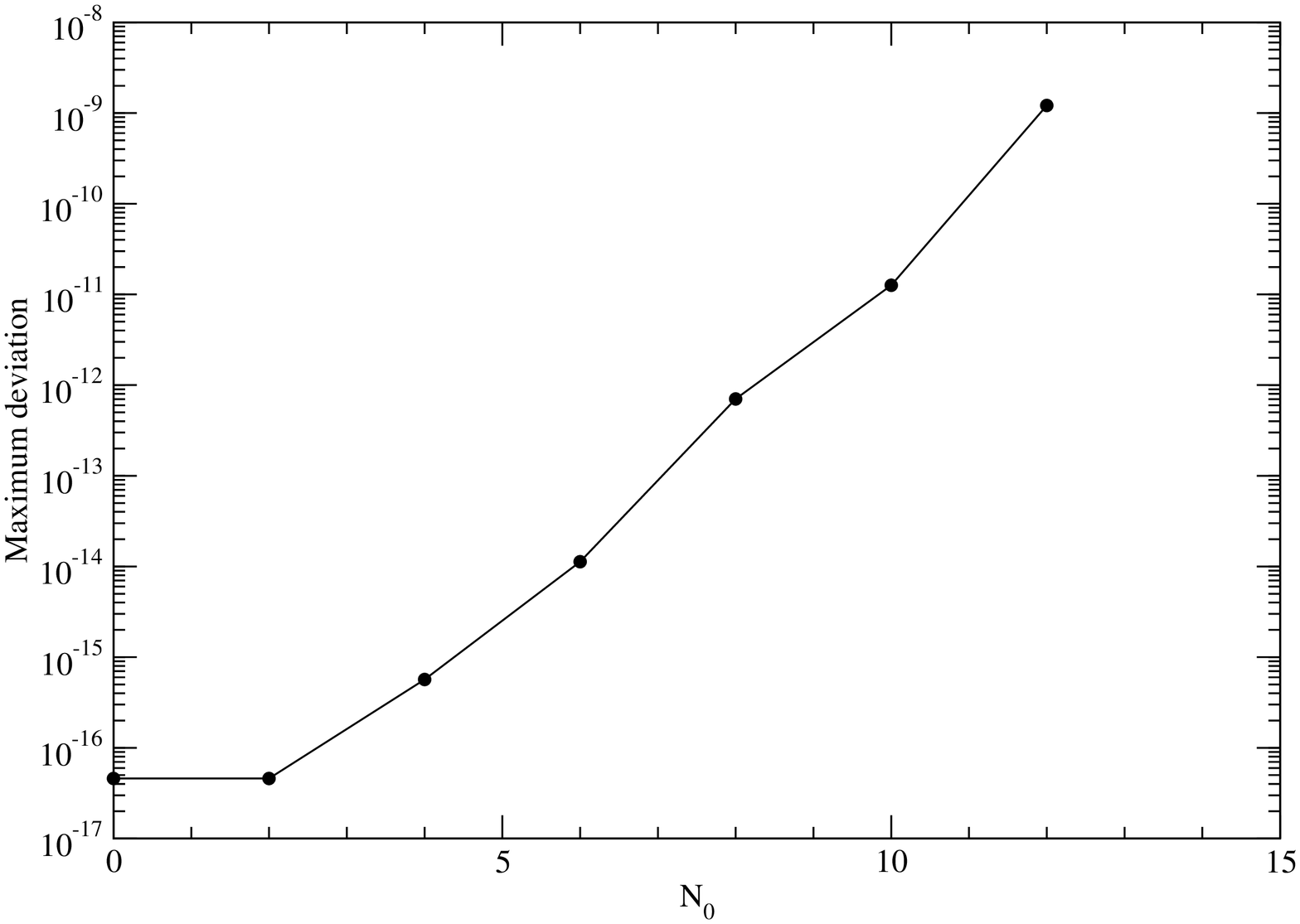}

\caption{\label{cap:vr-excomp}Maximum deviation between the numerical calculation
and exact Mathematica evaluation of the matrix elements $V_{ijkl}^{\left(r\right)}$
using the separation method in Eq. (\ref{eq:vijkl-r}), plotted as
a function of basis size $N_{0}$.}
\end{figure}

A comparison between exact (Mathematica) and double-precision numerical
evaluations of the separation-method result in Eq. (\ref{eq:vijkl-r})
is plotted in Fig. \ref{cap:vr-excomp} as a function of basis size
$N_{0}$. The accuracy of the numerical evaluation clearly deteriorates
with increasing basis size, but remains quite good nevertheless, reaching
only a $1.2\times10^{-9}$ maximum deviation for $N_{0}=12$. For
practical reasons, we will use the numerical evaluation of Eq. (\ref{eq:vijkl-r})
in the remainder of this discussion, rather than the exact--but much
slower--Mathematica calculation.

\begin{figure}
\includegraphics[%
  scale=0.5,
  angle=-90]{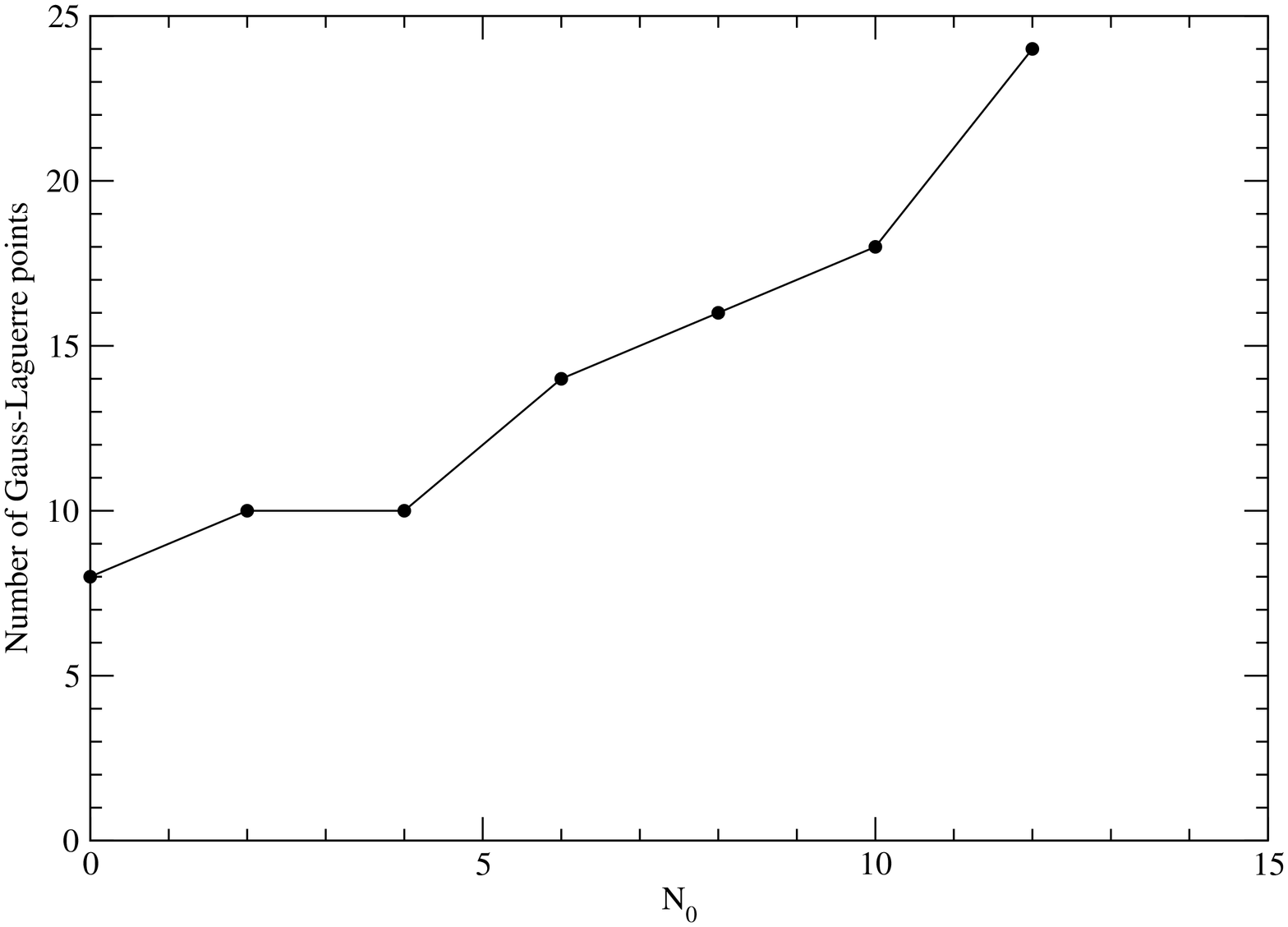}

\caption{\label{cap:sepgauss-rstats}Minimum number of Gauss-Laguerre quadrature
points needed to achieve $10^{-4}$ or better agreement between the
numerical integration of the matrix elements $V_{ijkl}^{\left(r\right)}$
and their evaluation using the separation method in Eq. (\ref{eq:vijkl-r}),
plotted as a function of basis size $N_{0}$.}
\end{figure}

The number of Gauss-Laguerre quadrature points needed to obtain a
discrepancy of $10^{-4}$ or less between the numerical integration
and numerical separation method for $V_{ijkl}^{\left(r\right)}$ matrix
elements is plotted in Fig. \ref{cap:sepgauss-rstats} as a function
of basis size. As in Fig. \ref{cap:sepgauss-zstats} for the Cartesian
matrix elements, the required number of quadrature points is moderate
and increases with basis size. The impact of these numbers of quadrature
points on execution time will be investigated next.

We now compare execution times for the numerical integration and numerical
separation methods. The numerical integrations for the Cartesian and
radial components are performed with the number of quadrature points
given in Figs. \ref{cap:sepgauss-zstats} and \ref{cap:sepgauss-rstats},
respectively, to ensure agreement to $10^{-4}$ or better with the
separation-method results. In order to speed up the numerical integrations,
the harmonic-oscillator functions are calculated at the appropriate
quadrature points and stored once and for all. A set of nested loops
then evaluate the multidimensional integrals by recalling the stored
values of the functions as the terms in the quadrature are summed.
Likewise, for the calculations by the separation method, the $T$,
$\bar{I}$, and $\bar{F}$ coefficients are calculated ahead of time
and recalled as needed in the evaluation of the matrix elements using
Eqs. (\ref{eq:vijkl-z-big}) and (\ref{eq:vijkl-r}).

\begin{figure}
\includegraphics[%
  scale=0.5,
  angle=-90]{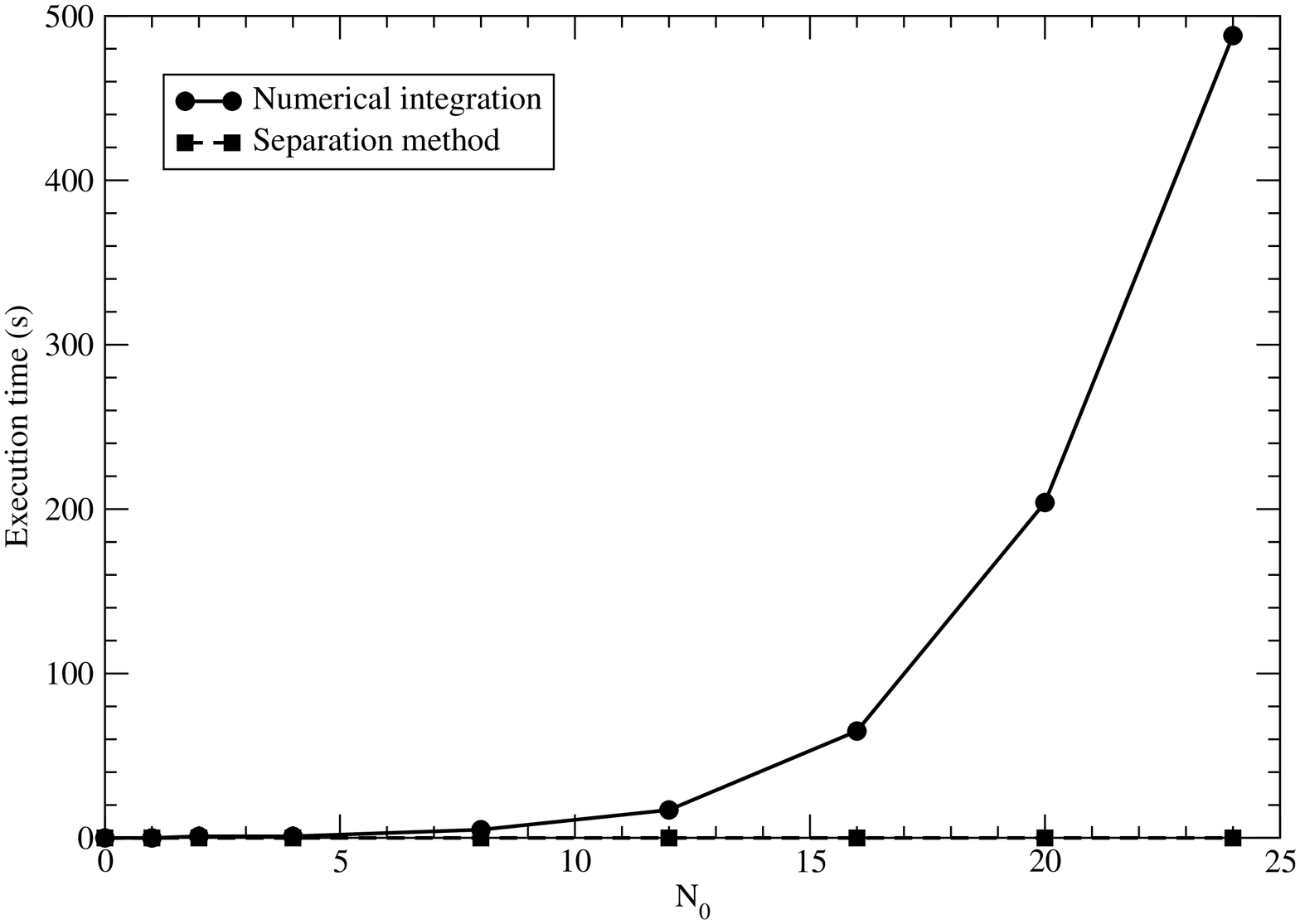}

\caption{\label{cap:ztime}Comparison of total execution times for the evaluation
of $V_{ijkl}^{\left(z\right)}$ by numerical integration and by the
separation method in Eq. (\ref{eq:vijkl-z-big}), as a function of
basis size $N_{0}$.}
\end{figure}

\begin{figure}
\includegraphics[%
  scale=0.5,
  angle=-90]{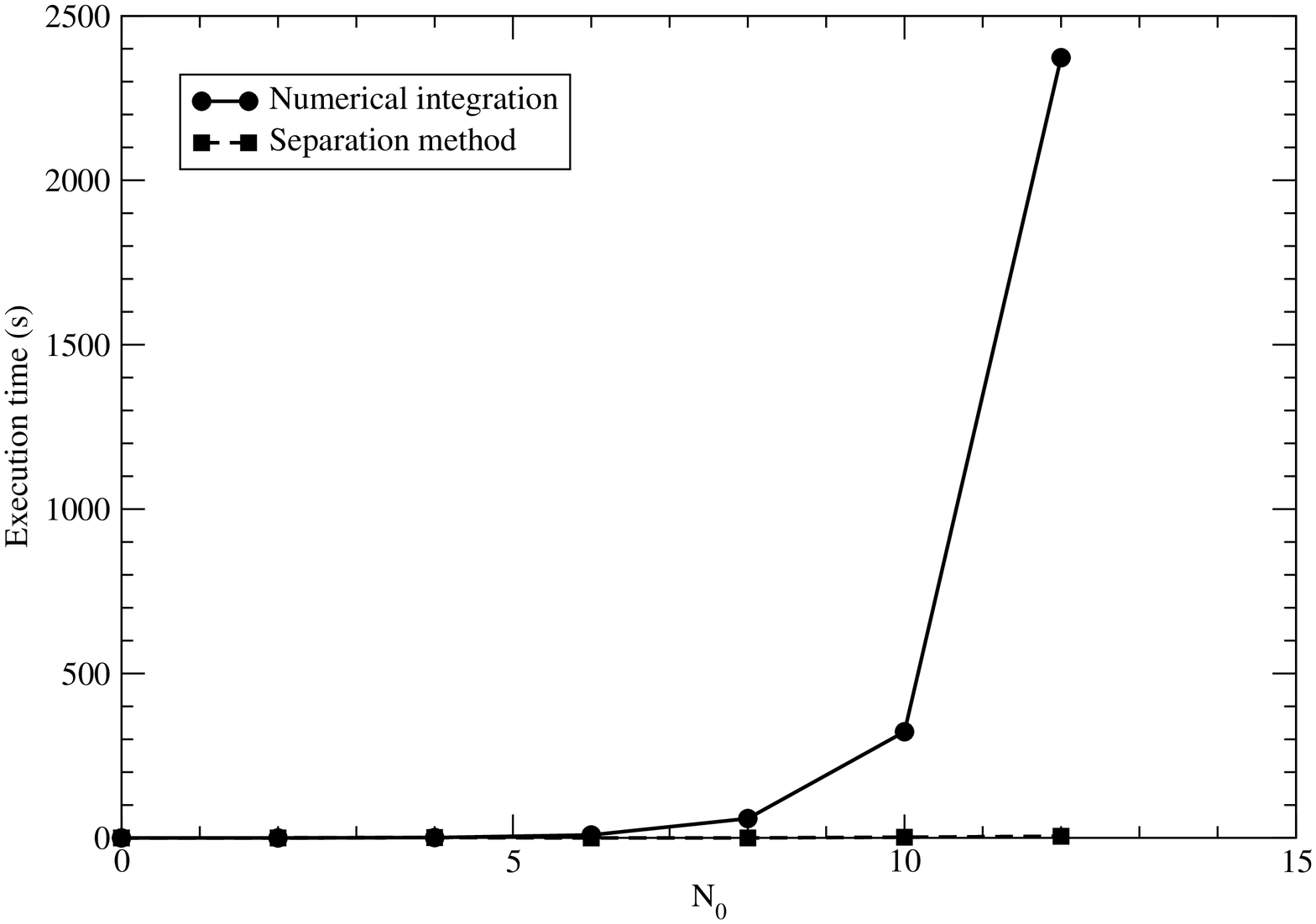}

\caption{\label{cap:rtime}Comparison of total execution times for the evaluation
of $V_{ijkl}^{\left(r\right)}$ by numerical integration and by the
separation method in Eq. (\ref{eq:vijkl-r}), as a function of basis
size $N_{0}$.}
\end{figure}

The calculations have been performed on a 2.13-GHz Pentium M processor
in double-precision mode. The execution times are plotted in Fig.
\ref{cap:ztime} for the z component of the matrix element, and in
Fig. \ref{cap:rtime} for the radial component. The times plotted
include the setup time needed to pre-calculate the harmonic-oscillator
function values and separation coefficients appropriate to each method.
The difference in execution times between the numerical and separation
methods become staggering with increasing basis size. For large-scale
computations requiring matrix-element calculations over a range of
values of the harmonic-oscillator parameters $b_{\bot}$ and $b_{z}$,
such as maps of fission shapes for a single nucleus or maps of nuclear
properties for large sets of nuclei, direct numerical integrations
rapidly become unfeasible without parallel machines. Even with parallel
processing, modern nuclear-physics problems (e.g., the microscopic
treatment of fission in a multidimensional collective-coordinate space)
will eventually overwhelm any given computational resource, and in
order to match the accuracy of the separation method, numerical integrals
will generally require an inordinate number of quadrature points.

\section{Conclusion}

We have derived explicit expressions for Gaussian matrix elements
in a cylindrical harmonic-oscillator basis, using the separation method.
These expressions have been tested against direct numerical integration
and found to be highly accurate and computationally efficient. These
characteristics make the separation method an invaluable tool for
computationally-intensive applications, such as the microscopic description
of fission. The work presented here has wider relevance than to the
Gaussian form, or to nuclear-physics problems alone. In particular,
the methodology used in the present derivations, which relies heavily
on generating functions, can be applied to other types of interactions
and a wider class of basis states to derive analytical, computationally-efficient
expressions for matrix elements. For example, in future publications,
we will apply the separation method to the Coulomb and Yukawa interactions,
and extend the formalism to bases of displaced and two-center deformed
harmonic oscillators. These planned extensions to the separation formalism
enlarge the range of applications of the method to many problems of
central importance in nuclear, atomic, and molecular systems.

We wish to thank D. Gogny for invaluable guidance in the development
of the formalism and preparation of this manuscript. This work was
performed under the auspices of the U.S. Department of Energy by Lawrence
Livermore National Laboratory under Contract DE-AC52-07NA27344.

\appendix

\section{Mapping between Cartesian and polar coordinates for harmonic-oscillator
functions\label{sec:Mapping-from-Cartesian}}

In this section, we derive an identity relating the harmonic-oscillator
functions expressed in two-dimensional Cartesian coordinates $\left(x,y\right)$
to those in polar coordinates $\left(\rho,\varphi\right)$ where\begin{eqnarray*}
\rho^{2} & = & x^{2}+y^{2}\\
\tan\varphi & = & \frac{y}{x}\end{eqnarray*}
To this end, we will first need to derive generating functions for
the harmonic-oscillator functions in the two coordinate systems.

\subsection{Generating function in Cartesian coordinates}

In this appendix, we derive the generating function\begin{equation}
\boxed{{e^{-t^{2}+2tx/b-x^{2}/\left(2b^{2}\right)}=\sqrt{b\sqrt{\pi}}\sum_{k=0}^{\infty}\frac{2^{k/2}}{\sqrt{k!}}t^{k}\Phi_{k}\left(x;b\right)}}\label{eq:carho-genf}\end{equation}
for the Cartesian harmonic-oscillator functions in Eq. (\ref{eq:phicar-def}).

We begin with the generating function for Hermite polynomials (Eq.
8.957(1), p. 1034 in \cite{gradshteyn79}), for arbitrary variables
$x$ and $t$,\begin{eqnarray*}
e^{-t^{2}+2tx} & = & \sum_{k=0}^{\infty}\frac{t^{k}}{k!}H_{k}\left(x\right)\end{eqnarray*}
making the substitution $x\rightarrow x/b$ in order to introduce
the harmonic-oscillator parameter $b$,\begin{eqnarray*}
e^{-t^{2}+2tx/b} & = & \sum_{k=0}^{\infty}\frac{t^{k}}{k!}H_{k}\left(\frac{x}{b}\right)\end{eqnarray*}
Next, we introduce the Gaussian and normalization factors appearing
in the definition of the harmonic oscillator function in Eq. (\ref{eq:phicar-def})\begin{eqnarray*}
e^{-x^{2}/\left(2b^{2}\right)}e^{-t^{2}+2tx/b} & = & \sum_{k=0}^{\infty}\frac{t^{k}}{k!\mathcal{N}_{k}}\mathcal{N}_{k}e^{-x^{2}/\left(2b^{2}\right)}H_{k}\left(\frac{x}{b}\right)\end{eqnarray*}
or, in terms of the harmonic-oscillator functions,\begin{eqnarray*}
e^{-t^{2}+2tx/b-x^{2}/\left(2b^{2}\right)} & = & \sqrt{b\sqrt{\pi}}\sum_{k=0}^{\infty}\frac{2^{k/2}}{\sqrt{k!}}t^{k}\Phi_{k}\left(x;b\right)\end{eqnarray*}

\subsection{Generating function in polar coordinates\label{sub:Generating-function-in}}

Here, we derive a generating function for the polar harmonic-oscillator
functions defined in Eq. (\ref{eq:phipol-def}),\begin{equation}
\boxed{{e^{-t^{2}+2\rho t\cos\varphi/b-\rho^{2}/\left(2b^{2}\right)}=b\sqrt{\pi}\sum_{k=-\infty}^{\infty}\sum_{n=0}^{\infty}\frac{\left(-1\right)^{n}t^{2n+\left|k\right|}}{\sqrt{n!\left(n+\left|k\right|\right)!}}\Phi_{n,k}\left(\rho,\varphi;b\right)}}\label{eq:polho-genf}\end{equation}
which we also cast in the form\begin{equation}
\boxed{{e^{-\vec{t}^{2}+2\vec{\rho}\cdot\vec{t}/b-\rho^{2}/\left(2b^{2}\right)}=b^{2}\sqrt{\frac{\pi}{2}}\sum_{k=-\infty}^{\infty}\sum_{n=0}^{\infty}\mathcal{N}_{n,\left|k\right|}\chi_{n,k}\left(\vec{t}\right)\Phi_{n,k}\left(\rho,\varphi;b\right)}}\label{eq:polho-genf-vec}\end{equation}
where the functions $\chi_{n,k}\left(\vec{t}\right)$ are defined
by Eq. (\ref{eq:chi-def}).

To derive a generating function for harmonic-oscillator functions
in polar coordinates, we begin with the generating function for Laguerre
polynomials (Eq. 8.975(3), p. 1038 in \cite{gradshteyn79}), for arbitrary
variables $x$ and $z$, and $\alpha>-1$\begin{eqnarray}
J_{\alpha}\left(2\sqrt{xz}\right)e^{z}\left(xz\right)^{-\alpha/2} & = & \sum_{n=0}^{\infty}\frac{z^{n}}{\Gamma\left(n+\alpha+1\right)}L_{n}^{\alpha}\left(x\right)\label{eq:laguerre-genf}\end{eqnarray}
In order to match the definition of the harmonic-oscillator function
in Eq. (\ref{eq:phirad-def}), we substitute $\sqrt{x}=\rho/b$, $\sqrt{z}=-it$,
and $\alpha=\left|k\right|$ where $k$ is an integer. Then, isolating
the Bessel function on the left-hand side, Eq. (\ref{eq:laguerre-genf})
takes the form\begin{eqnarray}
J_{\left|k\right|}\left(-2i\rho t/b\right) & = & e^{t^{2}}\left(-i\right)^{\left|k\right|}\left(\frac{\rho t}{b}\right)^{\left|k\right|}\sum_{n=0}^{\infty}\frac{\left(-1\right)^{n}t^{2n}}{\left(n+\left|k\right|\right)!}L_{n}^{\left|k\right|}\left(\frac{\rho^{2}}{b^{2}}\right)\label{eq:laguerre-genf1}\end{eqnarray}
On the other hand, the generating function for a Bessel function of
the first kind for arbitrary $z$ and $\varphi$ is (Eq. 8.511(4),
p. 973 in \cite{gradshteyn79})\begin{eqnarray}
e^{iz\cos\varphi} & = & \sum_{k=-\infty}^{\infty}i^{k}J_{k}\left(z\right)e^{ik\varphi}\nonumber \\
 & = & \sum_{k=-\infty}^{\infty}i^{\left|k\right|}J_{\left|k\right|}\left(z\right)e^{ik\varphi}\label{eq:bessel-genf}\end{eqnarray}
where the second line follows from Eq. 8.404(2) in \cite{gradshteyn79}.
Substituting $z=-2i\rho t/b^{2}$ into Eq. (\ref{eq:bessel-genf}),\begin{eqnarray}
e^{2\rho t\cos\varphi/b} & = & \sum_{k=-\infty}^{\infty}i^{\left|k\right|}J_{\left|k\right|}\left(-2i\frac{\rho t}{b}\right)e^{ik\varphi}\label{eq:bessel-genf1}\end{eqnarray}
Finally, plugging Eq. (\ref{eq:laguerre-genf1}) into Eq. (\ref{eq:bessel-genf1})
yields\begin{eqnarray*}
e^{2\rho t\cos\varphi/b} & = & e^{t^{2}}\sum_{k=-\infty}^{\infty}\left(-i\right)^{\left|k\right|}i^{\left|k\right|}\left(\frac{\rho t}{b}\right)^{\left|k\right|}\\
 &  & \times\sum_{n=0}^{\infty}\frac{\left(-1\right)^{n}t^{2n}}{\left(n+\left|k\right|\right)!}L_{n}^{\left|k\right|}\left(\frac{\rho^{2}}{b^{2}}\right)e^{ik\varphi}\end{eqnarray*}
where the right-hand side can be made to look more like the harmonic-oscillator
function definition in Eq. (\ref{eq:phirad-def}),\begin{eqnarray*}
e^{-t^{2}+2\rho t\cos\varphi/b} & = & \sum_{k=-\infty}^{\infty}\sum_{n=0}^{\infty}\left(-1\right)^{n}\frac{t^{2n+\left|k\right|}}{\left(n+\left|k\right|\right)!}\frac{\sqrt{2\pi}e^{\rho^{2}/\left(2b^{2}\right)}}{\mathcal{N}_{n,\left|k\right|}}\\
 &  & \times\left[\mathcal{N}_{n,\left|k\right|}\left(\frac{\rho}{b}\right)^{\left|k\right|}e^{-\rho^{2}/\left(2b^{2}\right)}L_{n}^{\left|k\right|}\left(\frac{\rho^{2}}{b^{2}}\right)\frac{e^{ik\varphi}}{\sqrt{2\pi}}\right]\end{eqnarray*}
or, after straightforward simplifications,\begin{eqnarray*}
e^{-t^{2}+2\rho t\cos\varphi/b-\rho^{2}/\left(2b^{2}\right)} & = & b\sqrt{\pi}\sum_{k=-\infty}^{\infty}\sum_{n=0}^{\infty}\frac{\left(-1\right)^{n}t^{2n+\left|k\right|}}{\sqrt{n!\left(n+\left|k\right|\right)!}}\Phi_{n,k}\left(\rho,\varphi;b\right)\end{eqnarray*}
Note that there is a potential ambiguity in the meaning of the angle
$\varphi$ in Eq. (\ref{eq:polho-genf}). In fact, Eq. (\ref{eq:polho-genf})
was derived for any arbitrary value of $\varphi$ but on left-hand
side, the term $\rho t\cos\varphi$ in the exponent suggests a dot
product $\vec{\rho}\cdot\vec{t}$ with $\varphi$ the angle between
the vectors, while on the right-hand side, writing the harmonic-oscillator
function $\Phi_{n,k}\left(\rho,\varphi;b\right)$ suggests that $\varphi$
is the polar angle of the vector $\vec{\rho}$. To lift this apparent
ambiguity, we introduce the polar angle $\varphi_{t}$ of vector $\vec{t}$
explicitly by noting that if $\theta$ is the angle between vectors
$\vec{\rho}$ and $\vec{t}$ with $\theta=\varphi-\varphi_{t}$, then
according to Eq. (\ref{eq:phipol-def})\begin{eqnarray*}
\Phi_{n,k}\left(\rho,\theta;b\right) & = & \Phi_{n,\left|k\right|}\left(\rho;b\right)\frac{e^{ik\theta}}{\sqrt{2\pi}}\end{eqnarray*}
and therefore\begin{eqnarray}
\Phi_{n,k}\left(\rho,\theta;b\right) & = & \Phi_{n,k}\left(\rho,\varphi;b\right)e^{-ik\varphi_{t}}\label{eq:pull-out-phit}\end{eqnarray}
Writing the left-hand side of Eq. (\ref{eq:polho-genf}) in vector
form, we now have\begin{eqnarray*}
e^{-\vec{t}^{2}+2\vec{\rho}\cdot\vec{t}/b-\rho^{2}/\left(2b^{2}\right)} & = & b\sqrt{\pi}\sum_{k=-\infty}^{\infty}\sum_{n=0}^{\infty}\frac{\left(-1\right)^{n}t^{2n+\left|k\right|}}{\sqrt{n!\left(n+\left|k\right|\right)!}}e^{-ik\varphi_{t}}\Phi_{n,k}\left(\rho,\varphi;b\right)\end{eqnarray*}
For convenience, we introduce the function\begin{eqnarray}
\chi_{n,k}\left(\vec{t}\right) & \equiv & \frac{\left(-1\right)^{n}}{n!}t^{2n+\left|k\right|}e^{-ik\varphi_{t}}\label{eq:chi-def}\end{eqnarray}
which allows us to write the generating function for polar harmonic-oscillator
functions as\begin{eqnarray*}
e^{-\vec{t}^{2}+2\vec{\rho}\cdot\vec{t}/b-\rho^{2}/\left(2b^{2}\right)} & = & b^{2}\sqrt{\frac{\pi}{2}}\sum_{k=-\infty}^{\infty}\sum_{n=0}^{\infty}\mathcal{N}_{n,\left|k\right|}\chi_{n,k}\left(\vec{t}\right)\Phi_{n,k}\left(\rho,\varphi;b\right)\end{eqnarray*}
This form will be convenient for some derivations, and we will obtain
useful properties of the function $\chi_{n,k}\left(\vec{t}\right)$
in section \ref{sub:Properties-of-the}.

\subsection{Polar-to-Cartesian mapping}

Having derived generating functions for the harmonic-oscillator functions
in both polar and Cartesian coordinates, we can now obtain a relation
between the two,\begin{equation}
\boxed{{\Phi_{n_{x}}\left(x;b\right)\Phi_{n_{y}}\left(y;b\right)=\sum_{k=-n_{x}-n_{y},2}^{n_{x}+n_{y}}C_{n,k}^{n_{x},n_{y}}\Phi_{\frac{n_{x}+n_{y}-\left|k\right|}{2},k}\left(\rho,\varphi;b\right)}}\label{eq:pol-to-car-map}\end{equation}
where the coefficients $C_{n,k}^{n_{x},n_{y}}$ are given by Eq. (\ref{eq:pol-to-car-coeff}).

In order to relate the polar and Cartesian harmonic-oscillator functions
we will use Eqs. (\ref{eq:carho-genf}) and (\ref{eq:polho-genf}).
We will assume axial symmetry and use the same parameter $b$ for
all the coordinates involved. Consider the arbitrary vectors $\vec{\rho}=x\hat{x}+y\hat{y}$
and $\vec{t}=t_{x}\hat{x}+t_{y}\hat{y}$ in the two-dimensional Cartesian
coordinate system, with $\vec{\rho}\cdot\vec{t}=\rho t\cos\theta$.
Note that we are using the symbol $\theta$ for the angle between
vectors $\vec{\rho}$ and $\vec{t}$. We can write\begin{eqnarray*}
e^{-t_{x}^{2}+2t_{x}x/b-x^{2}/\left(2b^{2}\right)}e^{-t_{y}^{2}+2t_{y}y-y^{2}/\left(2b^{2}\right)} & = & e^{-t^{2}+2\rho t\cos\theta/b-\rho^{2}/\left(2b^{2}\right)}\end{eqnarray*}
Using Eqs. (\ref{eq:carho-genf}) and (\ref{eq:polho-genf}), this
can also be written as\begin{eqnarray}
 &  & b\sqrt{\pi}\sum_{n_{x}=0}^{\infty}\sum_{n_{y}=0}^{\infty}\frac{2^{\left(n_{x}+n_{y}\right)/2}}{\sqrt{n_{x}!n_{y}!}}t_{x}^{n_{x}}t_{y}^{n_{y}}\Phi_{n_{x}}\left(x;b\right)\Phi_{n_{y}}\left(y;b\right)\nonumber \\
 & = & b\sqrt{\pi}\sum_{k=-\infty}^{\infty}\sum_{n=0}^{\infty}\frac{\left(-1\right)^{n}t^{2n+\left|k\right|}}{\sqrt{n!\left(n+\left|k\right|\right)!}}\Phi_{n,k}\left(\rho,\theta;b\right)\label{eq:car-eq-pol}\end{eqnarray}
We must now equate the terms on the left-hand side to those on the
right-hand side. We would like to introduce the polar angle $\varphi$
of the vector $\vec{\rho}$ instead of the angle $\theta$ between
$\vec{\rho}$ and $\vec{t}$ in these expressions, because the final
result should be completely independent of the choice of vector $\vec{t}$.
Using Eq. (\ref{eq:pull-out-phit}), Eq. (\ref{eq:car-eq-pol}) becomes\begin{eqnarray}
 &  & \sum_{n_{x}=0}^{\infty}\sum_{n_{y}=0}^{\infty}\frac{2^{\left(n_{x}+n_{y}\right)/2}}{\sqrt{n_{x}!n_{y}!}}\left(\frac{t_{x}}{b}\right)^{n_{x}}\left(\frac{t_{y}}{b}\right)^{n_{y}}\Phi_{n_{x}}\left(x;b\right)\Phi_{n_{y}}\left(y;b\right)\nonumber \\
 & = & \sum_{k=-\infty}^{\infty}\sum_{n=0}^{\infty}\frac{\left(-1\right)^{n}}{\sqrt{n!\left(n+\left|k\right|\right)!}}\left(\frac{t}{b}\right)^{2n+\left|k\right|}e^{-ik\varphi_{t}}\Phi_{n,k}\left(\rho,\varphi;b\right)\label{eq:car-eq-pol1}\end{eqnarray}
All we have to do now is identify terms on the left- and right-hand
sides. We can establish this correspondence by expressing $t$ and
$\varphi_{t}$ in terms of $t_{x}$ and $t_{y}$. To this end, we
write\begin{eqnarray*}
t^{2n+\left|k\right|}e^{-ik\varphi_{t}} & = & t^{2n}\left(te^{-is_{k}\varphi_{t}}\right)^{\left|k\right|}\end{eqnarray*}
where we have introduced the sign quantity\begin{eqnarray}
s_{k} & \equiv & \begin{cases}
1 & \quad k\geq0\\
-1 & \quad k<0\end{cases}\label{eq:sk-def}\end{eqnarray}
Note that we can write\begin{eqnarray*}
te^{-is_{k}\varphi_{t}} & = & t\cos\left(s_{k}\varphi_{t}\right)-it\sin\left(s_{k}\varphi_{t}\right)\\
 & = & t\cos\varphi_{t}-is_{k}t\sin\varphi_{t}\\
 & = & t_{x}-is_{k}t_{y}\end{eqnarray*}
where the second line follows because $s_{k}=\pm1$. Thus we have\begin{eqnarray*}
t^{2n+\left|k\right|}e^{-ik\varphi_{t}} & = & \left(t_{x}^{2}+t_{y}^{2}\right)^{n}\left(t_{x}-is_{k}t_{y}\right)^{\left|k\right|}\\
 & = & \sum_{p=0}^{n}\sum_{q=0}^{\left|k\right|}\left(\begin{array}{c}
n\\
p\end{array}\right)\left(\begin{array}{c}
\left|k\right|\\
q\end{array}\right)\left(-is_{k}\right)^{\left|k\right|-q}t_{x}^{2p+q}t_{y}^{2n+\left|k\right|-2p-q}\end{eqnarray*}
We substitute this result into the right-hand side of Eq. (\ref{eq:car-eq-pol1})
to get\begin{eqnarray}
RHS & = & \sum_{k=-\infty}^{\infty}\sum_{n=0}^{\infty}\frac{\left(-1\right)^{n}}{\sqrt{n!\left(n+\left|k\right|\right)!}}\left[\sum_{p=0}^{n}\sum_{q=0}^{\left|k\right|}\left(\begin{array}{c}
n\\
p\end{array}\right)\left(\begin{array}{c}
\left|k\right|\\
q\end{array}\right)\left(-is_{k}\right)^{\left|k\right|-q}\right.\nonumber \\
 &  & \left.\times t_{x}^{2p+q}t_{y}^{2n+\left|k\right|-2p-q}\right]\Phi_{n,k}\left(\rho,\varphi;b\right)\label{eq:car-eq-pol-rhs}\end{eqnarray}
Comparing with the left-hand side of Eq. (\ref{eq:car-eq-pol1}),
we see that we will need to make the identifications\begin{eqnarray*}
2p+q & = & n_{x}\\
2n+\left|k\right|-2p-q & = & n_{y}\end{eqnarray*}
which also implies the important relation\begin{eqnarray}
n_{x}+n_{y} & = & 2n+\left|k\right|\label{eq:nxny-nk}\end{eqnarray}
We wish to replace the sums in Eq. (\ref{eq:car-eq-pol-rhs}) over
$n$ and $p$ with sums over $n_{x}$ and $n_{y}$. Since $n_{x}=2p+q$,
it is clear that $n_{x}$ will span the full range of integers starting
with 0. Similarly, Eq. (\ref{eq:nxny-nk}) implies that $n_{y}=2n+\left|k\right|-n_{x}$
and for any $n_{x}$, there will always be a set of $n$ and $k$
values such that $n_{y}$ spans the full range of integers from 0,
independently of the value of index $n_{x}$. Thus we can make the
substitution\begin{eqnarray*}
\sum_{n=0}^{\infty}\sum_{p=0}^{n} & \rightarrow & \sum_{n_{x}=0}^{\infty}\sum_{n_{y}=0}^{\infty}\end{eqnarray*}
Next, we note that Eq. (\ref{eq:nxny-nk}) can also be written as
$2n=n_{x}+n_{y}-\left|k\right|$, and since $n\geq0$, we must therefore
have $\left|k\right|\leq n_{x}+n_{y}$. Finally, $2p=n_{x}-q$, and
since $p\geq0$, we conclude that $q\leq n_{x}$. Thus we can also
make the substitution\begin{eqnarray*}
\sum_{k=-\infty}^{\infty}\sum_{q=0}^{\left|k\right|} & \rightarrow & \sum_{k=-n_{x}-n_{y}}^{n_{x}+n_{y}}\sum_{q=0}^{\min\left(n_{x},\left|k\right|\right)}\end{eqnarray*}
and Eq. (\ref{eq:car-eq-pol-rhs}) becomes\begin{eqnarray}
RHS & = & \sum_{n_{x}=0}^{\infty}\sum_{n_{y}=0}^{\infty}\left[\sum_{k=-n_{x}-n_{y},2}^{n_{x}+n_{y}}\sum_{q=0}^{\min\left(n_{x},\left|k\right|\right)}\frac{\left(-1\right)^{\left(n_{x}+n_{y}-\left|k\right|\right)/2}\left(-is_{k}\right)^{\left|k\right|-q}}{\sqrt{\frac{n_{x}+n_{y}-\left|k\right|}{2}!\frac{n_{x}+n_{y}+\left|k\right|}{2}!}}\right.\nonumber \\
 &  & \left.\times\left(\begin{array}{c}
\frac{n_{x}+n_{y}-\left|k\right|}{2}\\
\frac{n_{x}-q}{2}\end{array}\right)\left(\begin{array}{c}
\left|k\right|\\
q\end{array}\right)\Phi_{\frac{n_{x}+n_{y}-\left|k\right|}{2},k}\left(\rho,\varphi;b\right)\right]t_{x}^{n_{x}}t_{y}^{n_{y}}\label{eq:car-eq-pol-rhs1}\end{eqnarray}
Note that in the sum over $k$, the index can be stepped by 2 units
at a time, because of the restrictions imposed by the factorials.
Comparing the left-hand side of Eq. (\ref{eq:car-eq-pol1}), and its
right-hand side given by Eq. (\ref{eq:car-eq-pol-rhs1}), we deduce\begin{eqnarray*}
\frac{2^{\left(n_{x}+n_{y}\right)/2}}{\sqrt{n_{x}!n_{y}!}}\Phi_{n_{x}}\left(x;b\right)\Phi_{n_{y}}\left(y;b\right) & = & \sum_{k=-n_{x}-n_{y},2}^{n_{x}+n_{y}}\sum_{q=0}^{\min\left(n_{x},\left|k\right|\right)}\frac{\left(-1\right)^{\left(n_{x}+n_{y}-\left|k\right|\right)/2}\left(-is_{k}\right)^{\left|k\right|-q}}{\sqrt{\frac{n_{x}+n_{y}-\left|k\right|}{2}!\frac{n_{x}+n_{y}+\left|k\right|}{2}!}}\\
 &  & \times\left(\begin{array}{c}
\frac{n_{x}+n_{y}-\left|k\right|}{2}\\
\frac{n_{x}-q}{2}\end{array}\right)\left(\begin{array}{c}
\left|k\right|\\
q\end{array}\right)\Phi_{\frac{n_{x}+n_{y}-\left|k\right|}{2},k}\left(\rho,\varphi;b\right)\end{eqnarray*}
or, in more compact notation,\begin{eqnarray*}
\Phi_{n_{x}}\left(x;b\right)\Phi_{n_{y}}\left(y;b\right) & = & \sum_{k=-n_{x}-n_{y},2}^{n_{x}+n_{y}}C_{n,k}^{n_{x},n_{y}}\Phi_{\frac{n_{x}+n_{y}-\left|k\right|}{2},k}\left(\rho,\varphi;b\right)\end{eqnarray*}
where\begin{equation}
\boxed{{C_{n,k}^{n_{x},n_{y}}\equiv\frac{\sqrt{n_{x}!n_{y}!}}{2^{\left(n_{x}+n_{y}\right)/2}}\frac{\left(-1\right)^{\left(n_{x}+n_{y}-\left|k\right|\right)/2}}{\sqrt{\frac{n_{x}+n_{y}-\left|k\right|}{2}!\frac{n_{x}+n_{y}+\left|k\right|}{2}!}}\sum_{q=0}^{\min\left(n_{x},\left|k\right|\right)}\left(-is_{k}\right)^{\left|k\right|-q}\left(\begin{array}{c}
\frac{n_{x}+n_{y}-\left|k\right|}{2}\\
\frac{n_{x}-q}{2}\end{array}\right)\left(\begin{array}{c}
\left|k\right|\\
q\end{array}\right)}}\label{eq:pol-to-car-coeff}\end{equation}
The appearance of the index $n$ in the symbol $C_{n,k}^{n_{x},n_{y}}$,
even though it is not explicitly used, serves as a reminder of the
implicit relation between the indices given by Eq. (\ref{eq:nxny-nk}).

\subsection{Cartesian-to-polar mapping}

In this section,we derive the inverse transformation corresponding
to Eq. (\ref{eq:pol-to-car-map}),\begin{equation}
\boxed{{\Phi_{n,k}\left(\rho,\varphi;b\right)=\sum_{n_{y}=0}^{2n+\left|k\right|}C_{n_{x},n_{y}}^{n,k}\Phi_{2n+\left|k\right|-n_{y}}\left(x;b\right)\Phi_{n_{y}}\left(y;b\right)}}\label{eq:car-to-pol-map}\end{equation}
which expresses the polar harmonic-oscillator functions in terms of
the Cartesian functions. The coefficients $C_{n_{x},n_{y}}^{n,k}$
are given by Eq. (\ref{eq:car-to-pol-coeff}).

We start again from Eq. (\ref{eq:car-eq-pol1}), but this time, we
express $t_{x}$ and $t_{y}$ on the left-hand side in terms of $t$
and $\varphi_{t}$. Consider then\begin{eqnarray*}
t_{x}^{n_{x}}t_{y}^{n_{y}} & = & \left(t\cos\varphi_{t}\right)^{n_{x}}\left(t\sin\varphi_{t}\right)^{n_{y}}\\
 & = & t^{n_{x}+n_{y}}\left(\frac{e^{i\varphi_{t}}+e^{-i\varphi_{t}}}{2}\right)^{n_{x}}\left(\frac{e^{i\varphi_{t}}-e^{-i\varphi_{t}}}{2i}\right)^{n_{y}}\end{eqnarray*}
Expanding the powers and grouping terms yields\begin{eqnarray*}
t_{x}^{n_{x}}t_{y}^{n_{y}} & = & \frac{t^{n_{x}+n_{y}}}{2^{n_{x}+n_{y}}i^{n_{y}}}\sum_{p=0}^{n_{x}}\sum_{q=0}^{n_{y}}\left(\begin{array}{c}
n_{x}\\
p\end{array}\right)\left(\begin{array}{c}
n_{y}\\
q\end{array}\right)\left(-1\right)^{n_{y}-q}e^{-i\left(n_{x}+n_{y}-2p-2q\right)\varphi_{t}}\end{eqnarray*}
Substituting into the left-hand side of Eq. (\ref{eq:car-eq-pol1})
produces\begin{eqnarray*}
LHS & = & \sum_{n_{x}=0}^{\infty}\sum_{n_{y}=0}^{\infty}\frac{t^{n_{x}+n_{y}}}{\sqrt{n_{x}!n_{y}!}2^{\left(n_{x}+n_{y}\right)/2}i^{n_{y}}}\sum_{p=0}^{n_{x}}\sum_{q=0}^{n_{y}}\left(\begin{array}{c}
n_{x}\\
p\end{array}\right)\left(\begin{array}{c}
n_{y}\\
q\end{array}\right)\left(-1\right)^{n_{y}-q}\\
 &  & \times e^{-i\left(n_{x}+n_{y}-2p-2q\right)\varphi_{t}}\Phi_{n_{x}}\left(x;b\right)\Phi_{n_{y}}\left(y;b\right)\end{eqnarray*}
Comparing with the right-hand side of Eq. (\ref{eq:car-eq-pol1})
we see that we need to make the identifications\begin{eqnarray}
n_{x}+n_{y} & = & 2n+\left|k\right|\label{eq:car-to-pol-ida}\\
n_{x}+n_{y}-2p-2q & = & k\label{eq:car-to-pol-id}\end{eqnarray}
we therefore introduce a summation over $n$ and $k$ with the help
of Kronecker-delta functions,\begin{eqnarray*}
LHS & = & \sum_{n=0}^{\infty}\sum_{k=-\infty}^{\infty}t^{2n+\left|k\right|}e^{-ik\varphi_{t}}2^{-\left(n_{x}+n_{y}\right)/2}\sum_{n_{x}=0}^{\infty}\sum_{n_{y}=0}^{\infty}\delta_{n_{x}+n_{y},2n+\left|k\right|}\\
 &  & \times\frac{\Phi_{n_{x}}\left(x;b\right)\Phi_{n_{y}}\left(y;b\right)}{\sqrt{n_{x}!n_{y}!}i^{n_{y}}}\sum_{p=0}^{n_{x}}\sum_{q=0}^{n_{y}}\delta_{2p+2q,n_{x}+n_{y}-k}\left(\begin{array}{c}
n_{x}\\
p\end{array}\right)\left(\begin{array}{c}
n_{y}\\
q\end{array}\right)\left(-1\right)^{n_{y}-q}\end{eqnarray*}
where the Kronecker-delta functions collect those terms in the remaining
summations needed to satisfy Eqs. (\ref{eq:car-to-pol-ida}) and (\ref{eq:car-to-pol-id}).
The restrictions imposed by the Kronecker-delta functions can be used
to eliminate the summations over $n_{x}$ and $p$\begin{eqnarray*}
LHS & = & \sum_{n=0}^{\infty}\sum_{k=-\infty}^{\infty}t^{2n+\left|k\right|}e^{-ik\varphi_{t}}2^{-n-\left|k\right|/2}\sum_{n_{y}=0}^{2n+\left|k\right|}\frac{\Phi_{2n+\left|k\right|-n_{y}}\left(x;b\right)\Phi_{n_{y}}\left(y;b\right)}{\sqrt{\left(2n+\left|k\right|-n_{y}\right)!n_{y}!}i^{n_{y}}}\\
 &  & \sum_{q=0}^{\min\left(n_{y},n-q+\left(\left|k\right|-k\right)/2\right)}\left(\begin{array}{c}
2n+\left|k\right|-n_{y}\\
n-q+\frac{\left|k\right|-k}{2}\end{array}\right)\left(\begin{array}{c}
n_{y}\\
q\end{array}\right)\left(-1\right)^{n_{y}-q}\end{eqnarray*}
Comparing with the right-hand side of Eq. (\ref{eq:car-eq-pol1})
we deduce the relation\begin{eqnarray*}
2^{-n-\left|k\right|/2}\sum_{n_{y}=0}^{2n+\left|k\right|}\frac{\Phi_{2n+\left|k\right|-n_{y}}\left(x;b\right)\Phi_{n_{y}}\left(y;b\right)}{\sqrt{\left(2n+\left|k\right|-n_{y}\right)!n_{y}!}i^{n_{y}}}\\
\times\sum_{q=0}^{q_{max}}\left(\begin{array}{c}
2n+\left|k\right|-n_{y}\\
n-q+\frac{\left|k\right|-k}{2}\end{array}\right)\left(\begin{array}{c}
n_{y}\\
q\end{array}\right)\left(-1\right)^{n_{y}-q} & = & \frac{\left(-1\right)^{n}}{\sqrt{n!\left(n+\left|k\right|\right)!}}\Phi_{n,k}\left(\rho,\varphi;b\right)\end{eqnarray*}
where\begin{eqnarray*}
q_{max} & \equiv & \min\left(n_{y},n+\left(\left|k\right|-k\right)/2\right)\end{eqnarray*}
which we write as\begin{eqnarray*}
\Phi_{n,k}\left(\rho,\varphi;b\right) & = & \sum_{n_{y}=0}^{2n+\left|k\right|}C_{n_{x},n_{y}}^{n,k}\Phi_{2n+\left|k\right|-n_{y}}\left(x;b\right)\Phi_{n_{y}}\left(y;b\right)\end{eqnarray*}
with\begin{equation}
\boxed{{C_{n_{x},n_{y}}^{n,k}=\frac{2^{-n-\left|k\right|/2}\left(-1\right)^{n}\sqrt{n!\left(n+\left|k\right|\right)!}}{\sqrt{\left(2n+\left|k\right|-n_{y}\right)!n_{y}!}i^{n_{y}}}\sum_{q=0}^{q_{max}}\left(\begin{array}{c}
2n+\left|k\right|-n_{y}\\
n-q+\frac{\left|k\right|-k}{2}\end{array}\right)\left(\begin{array}{c}
n_{y}\\
q\end{array}\right)\left(-1\right)^{n_{y}-q}}}\label{eq:car-to-pol-coeff}\end{equation}
The appearance of the index $n_{x}$ in the symbol $C_{n_{x},n_{y}}^{n,k}$,
even though it is not explicitly used, serves as a reminder of the
implicit relation between the indices given by Eq. (\ref{eq:nxny-nk}).

\subsection{Properties of the function $\chi_{n,k}\left(\vec{t}\right)$\label{sub:Properties-of-the}}

In section \ref{sub:Generating-function-in} we introduced the function
$\chi_{n,k}\left(\vec{t}\right)$ which was used to obtain a generating
function for harmonic-oscillator functions in polar coordinates. This
function has many useful properties which we will exploit in further
derivations. In this section, we obtain some important properties
of $\chi_{n,k}\left(\vec{t}\right)$. From the definition of the $\chi_{n,k}\left(\vec{t}\right)$
function in Eq. (\ref{eq:chi-def}),\begin{eqnarray*}
\chi_{n,k}\left(\vec{t}\right) & \equiv & \frac{\left(-1\right)^{n}}{n!}t^{2n+\left|k\right|}e^{-ik\varphi_{t}}\end{eqnarray*}
 we can easily show that

\begin{equation}
\boxed{{t^{2m}\chi_{n,k}\left(\vec{t}\right)=\left(-1\right)^{m}\frac{\left(n+m\right)!}{n!}\chi_{n+m,k}\left(\vec{t}\right)}}\label{eq:chiprops-transl}\end{equation}
As a corollary, we can use Eq. (\ref{eq:chiprops-transl}) to show\begin{equation}
\boxed{{e^{at^{2}}\chi_{n,k}\left(\vec{t}\right)=\sum_{m=0}^{\infty}\frac{\left(-a\right)^{m}\left(n+m\right)!}{m!n!}\chi_{n+m,k}\left(\vec{t}\right)}}\label{eq:chiprops-exptransl}\end{equation}
The complex conjugate of $\chi_{n,k}\left(\vec{t}\right)$ is also
readily expressed as\begin{equation}
\boxed{{\chi_{n,k}^{*}\left(\vec{t}\right)=\chi_{n,-k}\left(\vec{t}\right)}}\label{eq:chiprops-complex}\end{equation}
and a scale factor can be factored out,\begin{equation}
\boxed{{\chi_{n,k}\left(a\vec{t}\right)=a^{2n+\left|k\right|}\chi_{n,k}\left(\vec{t}\right)}}\label{eq:chiprops-scale}\end{equation}

Next, We will use the function $\chi_{n,k}$, to expand the expression
$\exp\left(2\vec{t}_{1}\cdot\vec{t}_{2}\right)$. Starting with the
generating function for Bessel functions of the first kind, Eq. (\ref{eq:bessel-genf})
with $z=-2it_{1}t_{2}$ and $\varphi=\varphi_{1}-\varphi_{2}$,\begin{eqnarray}
e^{2\vec{t}_{1}\cdot\vec{t}_{2}} & = & \sum_{k=-\infty}^{\infty}i^{\left|k\right|}J_{\left|k\right|}\left(-2it_{1}t_{2}\right)e^{ik\left(\varphi_{1}-\varphi_{2}\right)}\label{eq:chiprops-expt1t2-1}\end{eqnarray}
Next, we use the series expansion for Bessel functions (Eq. 8.440
in \cite{gradshteyn79}),\begin{eqnarray*}
J_{\nu}\left(z\right) & = & \left(\frac{z}{2}\right)^{\nu}\sum_{k=0}^{\infty}\frac{\left(-1\right)^{k}}{k!\left(\nu+k\right)!}\left(\frac{z}{2}\right)^{2k}\end{eqnarray*}
to write Eq. (\ref{eq:chiprops-expt1t2-1}) as\begin{eqnarray*}
e^{2\vec{t}_{1}\cdot\vec{t}_{2}} & = & \sum_{k=-\infty}^{\infty}i^{\left|k\right|}e^{ik\left(\varphi_{1}-\varphi_{2}\right)}\left(-it_{1}t_{2}\right)^{\left|k\right|}\sum_{n=0}^{\infty}\frac{\left(-1\right)^{n}\left(-it_{1}t_{2}\right)^{2n}}{n!\left(\left|k\right|+n\right)!}\\
 & = & \sum_{n=0}^{\infty}\sum_{k=-\infty}^{\infty}\frac{1}{n!\left(\left|k\right|+n\right)!}\left(t_{1}t_{2}\right)^{2n+\left|k\right|}e^{ik\left(\varphi_{1}-\varphi_{2}\right)}\end{eqnarray*}
or,\begin{equation}
\boxed{{e^{2\vec{t}_{1}\cdot\vec{t}_{2}}=\frac{b^{2}}{2}\sum_{n=0}^{\infty}\sum_{k=-\infty}^{\infty}\mathcal{N}_{n,\left|k\right|}^{2}\left(b\right)\chi_{n,k}^{*}\left(\vec{t}_{1}\right)\chi_{n,k}\left(\vec{t}_{2}\right)}}\label{eq:chiprops-expt1t2}\end{equation}
where $\mathcal{N}_{n_{r},\left|\Lambda\right|}\left(b\right)$ is
given by Eq. (\ref{eq:normpol-def}), and the oscillator parameter
$b$ cancels out in the right-hand side. Next, we derive an expression
for the product of two $\chi_{n,k}$ functions, using the definition
in Eq. (\ref{eq:chi-def})\begin{eqnarray}
\chi_{n_{1},k_{1}}\left(\vec{t}\right)\chi_{n_{2},k_{2}}\left(\vec{t}\right) & = & \frac{\left(-1\right)^{n_{1}+n_{2}}}{n_{1}!n_{2}!}t^{2n_{1}+2n_{2}+\left|k_{1}\right|+\left|k_{2}\right|}e^{-i\left(k_{1}+k_{2}\right)\varphi_{t}}\label{eq:chiprod1}\end{eqnarray}
at this point, it is convenient to define the quantities\begin{eqnarray}
n_{1,2} & \equiv & n_{1}+n_{2}+\frac{\left|k_{1}\right|+\left|k_{2}\right|-\left|k_{1}+k_{2}\right|}{2}\label{eq:n12-def}\\
k_{1,2} & \equiv & \frac{\left|k_{1}\right|+\left|k_{2}\right|-\left|k_{1}+k_{2}\right|}{2}\label{eq:k12-def}\end{eqnarray}
which recur throughout the paper. Then Eq. (\ref{eq:chiprod1}) becomes\begin{eqnarray*}
\chi_{n_{1},k_{1}}\left(\vec{t}\right)\chi_{n_{2},k_{2}}\left(\vec{t}\right) & = & \left(-1\right)^{-k_{1,2}}\frac{n_{1,2}!}{n_{1}!n_{2}!}\frac{\left(-1\right)^{n_{1,2}}}{n_{1,2}!}t^{2n_{1,2}+\left|k_{1}+k_{2}\right|}e^{-i\left(k_{1}+k_{2}\right)\varphi_{t}}\end{eqnarray*}
or,\begin{equation}
\boxed{{\chi_{n_{1},k_{1}}\left(\vec{t}\right)\chi_{n_{2},k_{2}}\left(\vec{t}\right)=\left(-1\right)^{k_{1,2}}\frac{n_{1,2}!}{n_{1}!n_{2}!}\chi_{n_{1,2},k_{1}+k_{2}}\left(\vec{t}\right)}}\label{eq:chiprops-chiprod}\end{equation}
Next, we obtain an expression for the function $\chi_{n,k}\left(\vec{t}_{1}+\vec{t}_{2}\right)$
of a sum of vectors. We write for an arbitrary vector $\vec{t}$\begin{eqnarray}
e^{2\left(\vec{t}_{1}+\vec{t}_{2}\right)\cdot\vec{t}} & = & e^{2\vec{t}_{1}\cdot\vec{t}}e^{2\vec{t}_{2}\cdot\vec{t}}\label{eq:exp-vec-sum}\end{eqnarray}
Using Eq. (\ref{eq:chiprops-expt1t2}), the left-hand side is\begin{eqnarray}
LHS & = & \frac{b^{2}}{2}\sum_{n=0}^{\infty}\sum_{k=-\infty}^{\infty}\mathcal{N}_{n,\left|k\right|}^{2}\left(b\right)\chi_{n,k}^{*}\left(\vec{t}_{1}+\vec{t}_{2}\right)\chi_{n,k}\left(\vec{t}\right)\label{eq:exp-vec-sum-lhs}\end{eqnarray}
while the right-hand side of Eq. (\ref{eq:exp-vec-sum}) is\begin{eqnarray*}
RHS & = & \frac{b^{4}}{4}\sum_{n_{1}=0}^{\infty}\sum_{k_{1}=-\infty}^{\infty}\sum_{n_{2}=0}^{\infty}\sum_{k_{2}=-\infty}^{\infty}\mathcal{N}_{n_{1},\left|k_{1}\right|}^{2}\left(b\right)\mathcal{N}_{n_{2},\left|k_{2}\right|}^{2}\left(b\right)\\
 &  & \times\chi_{n_{1},k_{1}}^{*}\left(\vec{t}_{1}\right)\chi_{n_{2},k_{2}}^{*}\left(\vec{t}_{2}\right)\chi_{n_{1},k_{1}}\left(\vec{t}\right)\chi_{n_{2},k_{2}}\left(\vec{t}\right)\end{eqnarray*}
Using Eq. (\ref{eq:chiprops-chiprod}), this reduces to\begin{eqnarray*}
RHS & = & \frac{b^{4}}{4}\sum_{n_{1}=0}^{\infty}\sum_{k_{1}=-\infty}^{\infty}\sum_{n_{2}=0}^{\infty}\sum_{k_{2}=-\infty}^{\infty}\mathcal{N}_{n_{1},\left|k_{1}\right|}^{2}\left(b\right)\mathcal{N}_{n_{2},\left|k_{2}\right|}^{2}\left(b\right)\left(-1\right)^{k_{1,2}}\frac{n_{1,2}!}{n_{1}!n_{2}!}\\
 &  & \times\chi_{n_{1},k_{1}}^{*}\left(\vec{t}_{1}\right)\chi_{n_{2},k_{2}}^{*}\left(\vec{t}_{2}\right)\chi_{n_{1,2},k_{1}+k_{2}}\left(\vec{t}\right)\end{eqnarray*}
In order to compare with Eq. (\ref{eq:exp-vec-sum-lhs}), we introduce
summations over the indices $n$ and $k$ with the help of Kronecker-delta
functions,\begin{eqnarray}
RHS & = & \frac{b^{4}}{4}\sum_{n=0}^{\infty}\sum_{k=-\infty}^{\infty}\sum_{n_{1}=0}^{\infty}\sum_{k_{1}=-\infty}^{\infty}\sum_{n_{2}=0}^{\infty}\sum_{k_{2}=-\infty}^{\infty}\mathcal{N}_{n_{1},\left|k_{1}\right|}^{2}\left(b\right)\mathcal{N}_{n_{2},\left|k_{2}\right|}^{2}\left(b\right)\left(-1\right)^{k_{1,2}}\frac{n_{1,2}!}{n_{1}!n_{2}!}\nonumber \\
 &  & \times\chi_{n_{1},k_{1}}^{*}\left(\vec{t}_{1}\right)\chi_{n_{2},k_{2}}^{*}\left(\vec{t}_{2}\right)\delta_{n,n_{1,2}}\delta_{k,k_{1}+k_{2}}\chi_{n,k}\left(\vec{t}\right)\label{eq:exp-vec-sum-rhs}\end{eqnarray}
Comparing Eqs. (\ref{eq:exp-vec-sum-lhs}) and (\ref{eq:exp-vec-sum-rhs})
for an arbitrary vector $\vec{t}$, and taking the complex conjugate,
we are lead to write\begin{equation}
\boxed{{\chi_{n,k}\left(\vec{t}_{1}+\vec{t}_{2}\right)=\sum_{n_{1}=0}^{\infty}\sum_{k_{1}=-\infty}^{\infty}\sum_{n_{2}=0}^{\infty}\sum_{k_{2}=-\infty}^{\infty}D_{n_{1},k_{1};n_{2},k_{2}}^{n,k}\chi_{n_{1},k_{1}}\left(\vec{t}_{1}\right)\chi_{n_{2},k_{2}}\left(\vec{t}_{2}\right)}}\label{eq:chiprops-chisum-expr}\end{equation}
where\begin{equation}
\boxed{{D_{n_{1},k_{1};n_{2},k_{2}}^{n,k}=\left(-1\right)^{n_{1}+n_{2}-n}\frac{\left(n+\left|k\right|\right)!}{\left(n_{1}+\left|k_{1}\right|\right)!\left(n_{2}+\left|k_{2}\right|\right)!}\delta_{n,n_{1,2}}\delta_{k,k_{1}+k_{2}}}}\label{eq:chiprops-chisum-coeff}\end{equation}
Note that we have used the condition imposed by the Kronecker-delta
function $\delta_{n,n_{1,2}}$ and the definition of $n_{1,2}$ in
Eq. (\ref{eq:n12-def}) to write\begin{eqnarray*}
\left(-1\right)^{k_{1,2}} & = & \left(-1\right)^{n_{1}+n_{2}-n}\end{eqnarray*}

Finally, we derive an expansion for the product $\exp\left(2\vec{t}_{1}\cdot\vec{t}_{2}\right)\chi_{n,k}\left(\vec{t}_{1}+\vec{t}_{2}\right)$.
Though it is tempting to use Eq. (\ref{eq:chiprops-expt1t2}) for
this, we will adopt a different approach which will yield a simpler
expression in the end. We write\begin{eqnarray*}
e^{2\vec{t}_{1}\cdot\vec{t}_{2}/b^{2}}\chi_{n,k}\left(\vec{t}_{1}+\vec{t}_{2}\right) & = & e^{\left(\vec{t}_{1}+\vec{t}_{2}\right)^{2}}e^{-\left(t_{1}^{2}+t_{2}^{2}\right)}\chi_{n,k}\left(\vec{t}_{1}+\vec{t}_{2}\right)\end{eqnarray*}
We treat the first exponential on the right-hand side using Eq. (\ref{eq:chiprops-exptransl}),
so that\begin{eqnarray*}
e^{2\vec{t}_{1}\cdot\vec{t}_{2}}\chi_{n,k}\left(\vec{t}_{1}+\vec{t}_{2}\right) & = & e^{-\left(t_{1}^{2}+t_{2}^{2}\right)}\sum_{m=0}^{\infty}\frac{\left(-1\right)^{m}\left(n+m\right)!}{m!n!}\chi_{n+m,k}\left(\vec{t}_{1}+\vec{t}_{2}\right)\end{eqnarray*}
Next, we use Eq. (\ref{eq:chiprops-chisum-expr}) to expand the $\chi_{n+m,k}\left(\vec{t}_{1}+\vec{t}_{2}\right)$
function\begin{eqnarray*}
e^{2\vec{t}_{1}\cdot\vec{t}_{2}}\chi_{n,k}\left(\vec{t}_{1}+\vec{t}_{2}\right) & = & e^{-\left(t_{1}^{2}+t_{2}^{2}\right)}\sum_{m=0}^{\infty}\frac{\left(-1\right)^{m}\left(n+m\right)!}{m!n!}\\
 &  & \times\sum_{n_{1}=0}^{\infty}\sum_{k_{1}=-\infty}^{\infty}\sum_{n_{2}=0}^{\infty}\sum_{k_{2}=-\infty}^{\infty}D_{n_{1},k_{1};n_{2},k_{2}}^{n+m,k}\\
 &  & \times\chi_{n_{1},k_{1}}\left(\vec{t}_{1}\right)\chi_{n_{2},k_{2}}\left(\vec{t}_{2}\right)\end{eqnarray*}
and use Eq. (\ref{eq:chiprops-exptransl}) again to eliminate the
remaining exponential on the right-hand side\begin{eqnarray}
e^{2\vec{t}_{1}\cdot\vec{t}_{2}}\chi_{n,k}\left(\vec{t}_{1}+\vec{t}_{2}\right) & = & \sum_{m_{1}=0}^{\infty}\sum_{m_{2}=0}^{\infty}\sum_{n_{1}=0}^{\infty}\sum_{k_{1}=-\infty}^{\infty}\sum_{n_{2}=0}^{\infty}\sum_{k_{2}=-\infty}^{\infty}D_{n_{1},k_{1},m_{1};n_{2},k_{2},m_{2}}^{n,k}\nonumber \\
 &  & \times\chi_{n_{1}+m_{1},k_{1}}\left(\vec{t}_{1}\right)\chi_{n_{2}+m_{2},k_{2}}\left(\vec{t}_{2}\right)\label{eq:chiprops-expchi-expr}\end{eqnarray}
where we have defined\begin{eqnarray*}
D_{n_{1},k_{1},m_{1};n_{2},k_{2},m_{2}}^{n,k} & \equiv & \sum_{m=0}^{\infty}\frac{\left(-1\right)^{m}\left(n+m\right)!\left(n_{1}+m_{1}\right)!\left(n_{2}+m_{2}\right)!}{m!n!m_{1}!n_{1}!m_{2}!n_{2}!}D_{n_{1},k_{1};n_{2},k_{2}}^{n+m,k}\\
 & = & \sum_{m=0}^{\infty}\frac{\left(-1\right)^{n_{1}+n_{2}-n}\left(n+m\right)!\left(n_{1}+m_{1}\right)!\left(n_{2}+m_{2}\right)!}{m!n!m_{1}!n_{1}!m_{2}!n_{2}!}\\
 &  & \times\frac{\left(n+m+\left|k\right|\right)!}{\left(n_{1}+\left|k_{1}\right|\right)!\left(n_{2}+\left|k_{2}\right|\right)!}\delta_{n+m,n_{1,2}}\delta_{k,k_{1}+k_{2}}\end{eqnarray*}
which simplifies to\begin{eqnarray}
D_{n_{1},k_{1},m_{1};n_{2},k_{2},m_{2}}^{n,k} & \equiv & \frac{n_{1,2}!\left(n_{1}+m_{1}\right)!\left(n_{2}+m_{2}\right)!\left(n_{1,2}+\left|k_{1}+k_{2}\right|\right)!}{\left(n_{1,2}-n\right)!n!m_{1}!n_{1}!m_{2}!n_{2}!\left(n_{1}+\left|k_{1}\right|\right)!\left(n_{2}+\left|k_{2}\right|\right)!}\nonumber \\
 &  & \times\left(-1\right)^{n_{1}+n_{2}-n}\delta_{n\leq n_{1,2}}\delta_{k,k_{1}+k_{2}}\label{eq:chiprops-expchi-coeff}\end{eqnarray}
Note the disappearance of the infinite sum over $m$ in favor of the
Kronecker-delta function $\delta_{n\leq n_{1,2}}$.

\section{Decomposition of two-body Gaussian form\label{sec:Decomposition-of-two-body}}

Consider the two-body Gaussian potential function in cylindrical coordinates\begin{eqnarray*}
V\left(\vec{r}_{1},\vec{r}_{2}\right) & = & e^{-\left(\vec{r}_{1}-\vec{r}_{2}\right)^{2}/\mu^{2}}\\
 & = & e^{-\left(\vec{\rho}_{1}-\vec{\rho}_{2}\right)^{2}/\mu^{2}}e^{-\left(z_{1}-z_{2}\right)^{2}/\mu^{2}}\end{eqnarray*}
The critical first step in the separation method for harmonic-oscillator
matrix elements is to write the potential itself in a form where the
dependence on the coordinates $\vec{r}_{1}$ and $\vec{r}_{2}$ has
been explicitly separated. We will therefore write this two-body function
as a sum of one-body functions in the two coordinates. Note that the
resulting sum will contain and infinite number of terms, while the
matrix elements of the potential will be limited to a finite sum,
thanks to properties of the harmonic-oscillator functions.

\subsection{Cartesian component}

The radial and Cartesian components of the potential can be expanded
independently. We begin with the Cartesian term and postulate\begin{eqnarray*}
V\left(z_{1},z_{2}\right) & = & e^{-\left(z_{1}-z_{2}\right)^{2}/\mu^{2}}\\
 & \equiv & \sum_{n_{z}=0}^{\infty}f_{n_{z}}\left(z_{1};b_{z}\right)\hat{\Phi}_{n_{z}}\left(z_{2};b_{z}\right)\end{eqnarray*}
choosing for the expansion the functions\begin{equation}
\boxed{{\hat{\Phi}_{n_{z}}\left(z;b_{z}\right)\equiv e^{z^{2}/\left(2b_{z}^{2}\right)}\Phi_{n_{z}}\left(z;b_{z}\right)}}\label{eq:phihatcar}\end{equation}
We will now show that\begin{equation}
\boxed{{f_{n_{z}}\left(z_{1};b_{z}\right)=K_{z}^{1/2}\lambda_{n_{z}}e^{-z_{1}^{2}/\left(2G_{z}b_{z}^{2}\right)}\Phi_{n_{z}}\left(z_{1};G_{z}^{1/2}b_{z}\right)}}\label{eq:fcar}\end{equation}
where the coefficients $K_{z}$ and $\lambda_{n_{z}}$ are given by
Eqs. (\ref{eq:Kz-def}) and (\ref{eq:Lz-def}), respectively.

The exponential function in $z^{2}$ in front of the harmonic-oscillator
function on the left-hand side has been added for computational convenience,
as we shall see. Then, by orthogonality of the harmonic-oscillator
functions, we have\begin{eqnarray*}
\int_{-\infty}^{\infty}dz_{2}e^{-z_{2}^{2}/\left(2b_{z}^{2}\right)}\Phi_{n_{z}}\left(z_{2};b_{z}\right)V\left(z_{1},z_{2}\right) & = & \int_{-\infty}^{\infty}dz_{2}e^{-z_{2}^{2}/\left(2b_{z}^{2}\right)}\Phi_{n_{z}}\left(z_{2};b_{z}\right)\\
 &  & \times\left[\sum_{n_{z}^{\prime}=0}^{\infty}f_{n_{z}^{\prime}}\left(z_{1};b_{z}\right)\hat{\Phi}_{n_{z}^{\prime}}\left(z_{2};b_{z}\right)\right]\\
 & = & f_{n_{z}}\left(z_{1};b_{z}\right)\end{eqnarray*}
from which we obtain an explicit expression for the weight function
$f_{n_{z}}\left(z_{1};b_{z}\right)$,\begin{eqnarray}
f_{n_{z}}\left(z_{1};b_{z}\right) & = & \int_{-\infty}^{\infty}dz_{2}e^{-z_{2}^{2}/\left(2b_{z}^{2}\right)}\Phi_{n_{z}}\left(z_{2};b_{z}\right)V\left(z_{1},z_{2}\right)\nonumber \\
 & = & \mathcal{N}_{n_{z}}\int_{-\infty}^{\infty}dz_{2}e^{-z_{2}^{2}/b_{z}^{2}}e^{-\left(z_{1}-z_{2}\right)^{2}/\mu^{2}}H_{n_{z}}\left(\frac{z_{2}}{b_{z}}\right)\label{eq:fcar-integral}\end{eqnarray}
Completing the square, we write\begin{eqnarray*}
-\frac{z_{2}^{2}}{b_{z}^{2}}-\frac{\left(z_{1}-z_{2}\right)^{2}}{\mu^{2}} & = & -\left[G_{z}^{1/2}\frac{z_{2}}{\mu}-G_{z}^{-1/2}\frac{z_{1}}{\mu}\right]^{2}-\left(1-\frac{1}{G_{z}}\right)\left(\frac{z_{1}}{\mu}\right)^{2}\end{eqnarray*}
where we have defined\begin{eqnarray}
G_{z} & \equiv & 1+\frac{\mu^{2}}{b_{z}^{2}}\label{eq:Gz-def}\end{eqnarray}
and the integral becomes\begin{eqnarray*}
f_{n_{z}}\left(z_{1};b_{z}\right) & = & \mathcal{N}_{n_{z}}\exp\left[-\left(1-\frac{1}{G_{z}}\right)\left(\frac{z_{1}}{\mu}\right)^{2}\right]\\
 &  & \times\int_{-\infty}^{\infty}dz_{2}\exp\left[-\left(G_{z}^{1/2}\frac{z_{2}}{\mu}-G_{z}^{-1/2}\frac{z_{1}}{\mu}\right)^{2}\right]H_{n_{z}}\left(\frac{z_{2}}{b_{z}}\right)\end{eqnarray*}
Making the substitutions $x\equiv G_{z}^{1/2}z_{2}/\mu$, $y\equiv G_{z}^{-1/2}z_{1}/\mu$,
$\alpha\equiv G_{z}^{-1/2}\mu/b_{z}$, the remaining integral can
be evaluated using Eq. 7.374(8), p. 837 in \cite{gradshteyn79},\begin{eqnarray*}
f_{n_{z}}\left(z_{1};b_{z}\right) & = & \mu G_{z}^{-1/2}\mathcal{N}_{n_{z}}\pi^{1/2}\left(1-\alpha^{2}\right)^{n_{z}/2}\exp\left[-\left(G_{z}-1\right)y^{2}\right]\\
 &  & \times H_{n_{z}}\left(\frac{\alpha y}{\sqrt{1-\alpha^{2}}}\right)\end{eqnarray*}
After some straightforward algebra and re-grouping of terms, this
can be written as\begin{eqnarray*}
f_{n_{z}}\left(z_{1};b_{z}\right) & = & \pi^{1/2}\mu G_{z}^{-1/2}G_{z}^{-n_{z}/2}e^{-z_{1}^{2}/\left(2G_{z}b_{z}^{2}\right)}G_{z}^{1/4}\\
 &  & \times\left[\frac{1}{G_{z}^{1/4}}\mathcal{N}_{n_{z}}e^{-z_{1}^{2}/\left(2G_{z}b_{z}^{2}\right)}H_{n_{z}}\left(\frac{z_{1}}{G_{z}^{1/2}b_{z}}\right)\right]\end{eqnarray*}
or, identifying the term in the square brackets with a harmonic-oscillator
function with parameter $G_{z}^{1/2}b_{z}$ (note the extra factor
$G_{z}^{1/4}$ needed to get the proper normalization constant $\mathcal{N}_{n_{z}}\left(G_{z}^{1/2}b_{z}\right)$),\begin{eqnarray*}
f_{n_{z}}\left(z_{1};b_{z}\right) & = & K_{z}^{1/2}\lambda_{n_{z}}e^{-z_{1}^{2}/\left(2G_{z}b_{z}^{2}\right)}\Phi_{n_{z}}\left(z_{1};G_{z}^{1/2}b_{z}\right)\end{eqnarray*}
where\begin{eqnarray}
K_{z} & \equiv & \frac{\pi\mu^{2}}{G_{z}^{1/2}}\label{eq:Kz-def}\\
\lambda_{n_{z}} & \equiv & G_{z}^{-n_{z}/2}\label{eq:Lz-def}\end{eqnarray}

\subsection{Radial component}

For the radial component of the Gaussian potential, we write\begin{eqnarray*}
V\left(\vec{\rho}_{1},\vec{\rho}_{2}\right) & = & e^{-\left(\vec{\rho}_{1}-\vec{\rho}_{2}\right)^{2}/\mu^{2}}\\
 & \equiv & \sum_{n_{r}=0}^{\infty}\sum_{\Lambda=-\infty}^{\infty}f_{n_{r},\Lambda}\left(\rho_{1},\varphi_{1};b_{\bot}\right)\hat{\Phi}_{n_{r},\Lambda}\left(\rho_{2},\varphi_{2};b_{\bot}\right)\end{eqnarray*}
where we have chosen\begin{equation}
\boxed{{\hat{\Phi}_{n_{r},\Lambda}\left(\rho,\varphi;b_{\bot}\right)\equiv e^{\frac{\rho^{2}}{2b_{\bot}^{2}}}\Phi_{n_{r},\Lambda}\left(\rho,\varphi;b_{\perp}\right)}}\label{eq:phihatpol}\end{equation}
We will then show that\begin{equation}
\boxed{{f_{n_{r},\Lambda}\left(\rho_{1},\varphi_{1};b_{\bot}\right)=K_{\bot}\lambda_{2n_{r}+\left|\Lambda\right|}e^{-\rho_{1}^{2}/\left(2G_{\bot}b_{\bot}^{2}\right)}\Phi_{n_{r},\Lambda}\left(\rho_{1},\varphi_{1};G_{\bot}^{1/2}b_{\perp}\right)}}\label{eq:fpol}\end{equation}
where the coefficients $K_{\bot}$ and $\lambda_{2n_{r}+\left|\Lambda\right|}$
are given by Eqs. (\ref{eq:Kp-def}) and (\ref{eq:Lp-def}), respectively.

By orthogonality of the harmonic-oscillator function we then have\begin{eqnarray*}
f_{n_{r},\Lambda}\left(\rho_{1},\varphi_{1};b_{\bot}\right) & = & \int_{0}^{\infty}\rho_{2}d\rho_{2}\int_{0}^{2\pi}d\varphi_{2}e^{-\frac{\rho_{2}^{2}}{2b_{\bot}^{2}}}e^{-\left(\vec{\rho}_{1}-\vec{\rho}_{2}\right)^{2}/\mu^{2}}\Phi_{n_{r},\Lambda}\left(\rho_{2},\varphi_{2};b_{\perp}\right)\end{eqnarray*}
This integral can be evaluated in a straightforward way by transforming
to a Cartesian coordinate system, and using Eq. (\ref{eq:car-to-pol-map}),\begin{eqnarray*}
f_{n_{r},\Lambda}\left(\rho_{1},\varphi_{1};b_{\bot}\right) & = & \int_{-\infty}^{\infty}dx_{2}\int_{-\infty}^{\infty}dy_{2}e^{-\frac{x_{2}^{2}+y_{2}^{2}}{2b_{\bot}^{2}}}e^{-\left(x_{1}-x_{2}\right)^{2}/\mu^{2}-\left(y_{1}-y_{2}\right)^{2}/\mu^{2}}\\
 &  & \times\sum_{n_{y}=0}^{2n_{r}+\left|\Lambda\right|}C_{n_{x},n_{y}}^{n_{r},\Lambda}\Phi_{2n_{r}+\left|\Lambda\right|-n_{y}}\left(x_{2};b_{\perp}\right)\Phi_{n_{y}}\left(y_{2};b_{\perp}\right)\\
 & = & \sum_{n_{y}=0}^{2n_{r}+\left|\Lambda\right|}C_{n_{x},n_{y}}^{n_{r},\Lambda}\left[\int_{-\infty}^{\infty}dx_{2}e^{-\frac{x_{2}^{2}}{2b_{\bot}^{2}}}e^{-\left(x_{1}-x_{2}\right)^{2}/\mu^{2}}\Phi_{2n_{r}+\left|\Lambda\right|-n_{y}}\left(x_{2};b_{\perp}\right)\right]\\
 &  & \times\left[\int_{-\infty}^{\infty}dy_{2}e^{-\frac{y_{2}^{2}}{2b_{\bot}^{2}}}e^{-\left(y_{1}-y_{2}\right)^{2}/\mu^{2}}\Phi_{n_{y}}\left(y_{2};b_{\perp}\right)\right]\end{eqnarray*}
The integrals in the square brackets are precisely those appearing
in Eq. (\ref{eq:fcar-integral}), and they are given by Eq. (\ref{eq:fcar})\begin{eqnarray}
f_{n_{r},\Lambda}\left(\rho_{1},\varphi_{1};b_{\bot}\right) & = & \sum_{n_{y}=0}^{2n_{r}+\left|\Lambda\right|}C_{n_{x},n_{y}}^{n_{r},\Lambda}\left[K_{\bot}^{1/2}\lambda_{2n_{r}+\left|\Lambda\right|-n_{y}}e^{-x_{1}^{2}/\left(2G_{\bot}b_{\bot}^{2}\right)}\right.\nonumber \\
 &  & \left.\Phi_{2n_{r}+\left|\Lambda\right|-n_{y}}\left(x_{1};G_{\bot}^{1/2}b_{\bot}\right)\right]\nonumber \\
 &  & \left[K_{\bot}^{1/2}\lambda_{n_{y}}e^{-y_{1}^{2}/\left(2G_{\bot}b_{\bot}^{2}\right)}\Phi_{n_{y}}\left(y_{1};G_{\bot}^{1/2}b_{\bot}\right)\right]\label{eq:fcyl1}\end{eqnarray}
where\begin{eqnarray}
G_{\bot} & \equiv & 1+\frac{\mu^{2}}{b_{\bot}^{2}}\label{eq:Gp-def}\\
K_{\bot} & \equiv & \frac{\pi\mu^{2}}{G_{\bot}^{1/2}}\label{eq:Kp-def}\\
\lambda_{n} & \equiv & G_{\bot}^{-n/2}\label{eq:Lp-def}\end{eqnarray}
and Eq. (\ref{eq:fcyl1}) can be further reduced to\begin{eqnarray*}
f_{n_{r},\Lambda}\left(\rho_{1},\varphi_{1};b_{\bot}\right) & = & K_{\bot}\lambda_{2n_{r}+\left|\Lambda\right|}e^{-\rho_{1}^{2}/\left(2G_{\bot}b_{\bot}^{2}\right)}\\
 &  & \sum_{n_{y}=0}^{2n_{r}+\left|\Lambda\right|}C_{n_{x},n_{y}}^{n_{r},\Lambda}\Phi_{2n_{r}+\left|\Lambda\right|-n_{y}}\left(x_{1};G_{\bot}^{1/2}b_{\bot}\right)\Phi_{n_{y}}\left(y_{1};G_{\bot}^{1/2}b_{\bot}\right)\end{eqnarray*}
Finally, using Eq. (\ref{eq:car-to-pol-map}) again to return to polar
coordinates, we get\begin{eqnarray*}
f_{n_{r},\Lambda}\left(\rho_{1},\varphi_{1};b_{\bot}\right) & = & K_{\bot}\lambda_{2n_{r}+\left|\Lambda\right|}e^{-\rho_{1}^{2}/\left(2G_{\bot}b_{\bot}^{2}\right)}\Phi_{n_{r},\Lambda}\left(\rho_{1},\varphi_{1};G_{\bot}^{1/2}b_{\perp}\right)\end{eqnarray*}

\section{Product of harmonic-oscillator functions\label{sec:Product-of-harmonic-oscillator}}

In this section, we will express the product of two harmonic-oscillator
functions in terms of a sum of single oscillator functions. These
results will be particularly useful in evaluating integrals where
the integrand includes products of harmonic-oscillator functions.

\subsection{Product of Cartesian harmonic-oscillator functions}

In this section, we derive the form\begin{equation}
\boxed{{\Phi_{k_{1}}\left(x;b\right)\Phi_{k_{2}}\left(x;b\right)=\frac{e^{-x^{2}/\left(2b^{2}\right)}}{\sqrt{b\sqrt{\pi}}}\sum_{k=\left|k_{1}-k_{2}\right|,2}^{k_{1}+k_{2}}T_{k_{1},k_{2}}^{k}\Phi_{k}\left(x;b\right)}}\label{eq:carprod-expr}\end{equation}
for the Cartesian harmonic-oscillator functions of Eq. (\ref{eq:phicar-def}),
with the coefficients $T_{k_{1},k_{2}}^{k}$ given by Eq. (\ref{eq:carprod-coeff}).

Using the generating function in Eq. (\ref{eq:carho-genf}), we write
for any arbitrary variables $t_{1}$ and $t_{2}$,\begin{eqnarray}
 &  & e^{-t_{1}^{2}+2t_{1}x/b-x^{2}/\left(2b^{2}\right)}e^{-t_{2}^{2}+2t_{2}x/b-x^{2}/\left(2b^{2}\right)}\nonumber \\
 & = & \left[\sqrt{b\sqrt{\pi}}\sum_{k_{1}=0}^{\infty}\frac{2^{k_{1}/2}}{\sqrt{k_{1}!}}t_{1}^{k_{1}}\Phi_{k_{1}}\left(x;b\right)\right]\nonumber \\
 &  & \times\left[\sqrt{b\sqrt{\pi}}\sum_{k_{2}=0}^{\infty}\frac{2^{k_{2}/2}}{\sqrt{k_{2}!}}t_{2}^{k_{2}}\Phi_{k_{2}}\left(x;b\right)\right]\label{eq:carprod-genf}\end{eqnarray}
With the intent of manipulating the left-hand side of this equation
into a form similar to the left-hand side of Eq. (\ref{eq:carho-genf}),
we write\begin{eqnarray*}
LHS & = & e^{-t_{1}^{2}-t_{2}^{2}+2\left(t_{1}+t_{2}\right)x/b-x^{2}/b^{2}}\\
 & = & e^{-\left(t_{1}+t_{2}\right)^{2}+2\left(t_{1}+t_{2}\right)x/b-x^{2}/\left(2b^{2}\right)}e^{2t_{1}t_{2}-x^{2}/\left(2b^{2}\right)}\end{eqnarray*}
Using Eq. (\ref{eq:carho-genf}), this becomes\begin{eqnarray}
LHS & = & e^{2t_{1}t_{2}-x^{2}/\left(2b^{2}\right)}\sqrt{b\sqrt{\pi}}\sum_{k=0}^{\infty}\frac{2^{k/2}}{\sqrt{k!}}\left(t_{1}+t_{2}\right)^{k}\Phi_{k}\left(x;b\right)\nonumber \\
 & = & \sqrt{b\sqrt{\pi}}e^{-x^{2}/\left(2b^{2}\right)}\sum_{k=0}^{\infty}\frac{2^{k/2}}{\sqrt{k!}}\Phi_{k}\left(x;b\right)\sum_{p=0}^{\infty}\frac{\left(2t_{1}t_{2}\right)^{p}}{p!}\nonumber \\
 &  & \times\sum_{q=0}^{k}\left(\begin{array}{c}
k\\
q\end{array}\right)t_{1}^{q}t_{2}^{k-q}\nonumber \\
 & = & \sqrt{b\sqrt{\pi}}e^{-x^{2}/\left(2b^{2}\right)}\sum_{k=0}^{\infty}\frac{2^{k/2}}{\sqrt{k!}}\Phi_{k}\left(x;b\right)\nonumber \\
 &  & \times\sum_{q=0}^{k}\left(\begin{array}{c}
k\\
q\end{array}\right)\sum_{p=0}^{\infty}\frac{2^{p}}{p!}t_{1}^{q+p}t_{2}^{k+p-q}\label{eq:carprod-lhs}\end{eqnarray}
We can also group the terms in the right-hand side of Eq. (\ref{eq:carprod-genf}),\begin{eqnarray}
RHS & = & b\sqrt{\pi}\sum_{k_{1}=0}^{\infty}\sum_{k_{2}=0}^{\infty}\frac{2^{\left(k_{1}+k_{2}\right)/2}}{\sqrt{k_{1}!k_{2}!}}t_{1}^{k_{1}}t_{2}^{k_{2}}\Phi_{k_{1}}\left(x;b\right)\Phi_{k_{2}}\left(x;b\right)\label{eq:carprod-rhs}\end{eqnarray}
Now we equate powers of $t_{1}$ and $t_{2}$ between Eqs. (\ref{eq:carprod-lhs})
and (\ref{eq:carprod-rhs}). We find that we must make the identifications\begin{eqnarray*}
q+p & = & k_{1}\\
k+p-q & = & k_{2}\end{eqnarray*}
which lead to\begin{eqnarray*}
p & = & \left(k_{1}+k_{2}-k\right)/2\\
q & = & \left(k_{1}-k_{2}+k\right)/2\end{eqnarray*}
so that Eq. (\ref{eq:carprod-lhs}) can be written\begin{eqnarray}
LHS & = & \sqrt{b\sqrt{\pi}}e^{-x^{2}/\left(2b^{2}\right)}\sum_{k=\left|k_{1}-k_{2}\right|,2}^{k_{1}+k_{2}}\frac{2^{k/2}}{\sqrt{k!}}\Phi_{k}\left(x;b\right)\nonumber \\
 &  & \times\sum_{k_{1}=0}^{\infty}\sum_{k_{2}=0}^{\infty}\left(\begin{array}{c}
k\\
\frac{k_{1}-k_{2}+k}{2}\end{array}\right)\frac{2^{\left(k_{1}+k_{2}-k\right)/2}}{\left(\frac{k_{1}+k_{2}-k}{2}\right)!}t_{1}^{k_{1}}t_{2}^{k_{2}}\label{eq:carprod-lhs1}\end{eqnarray}
Note that the limits and step size for the summation over $k$ are
dictated by the need to keep the arguments of the factorials non-negative.
In particular, the {}``2'' appearing in the lower limit of the sum
over $k$ indicates that the index should be incremented by steps
of 2. Direct comparison of Eqs. (\ref{eq:carprod-rhs}) and (\ref{eq:carprod-lhs1})
now yields\begin{eqnarray*}
 &  & \sqrt{b\sqrt{\pi}}e^{-x^{2}/\left(2b^{2}\right)}\sum_{k=\left|k_{1}-k_{2}\right|,2}^{k_{1}+k_{2}}\frac{2^{\left(k_{1}+k_{2}\right)/2}\sqrt{k!}}{\left(\frac{k_{1}-k_{2}+k}{2}\right)!\left(\frac{k_{2}-k_{1}+k}{2}\right)!\left(\frac{k_{1}+k_{2}-k}{2}\right)!}\Phi_{k}\left(x;b\right)\\
 & = & b\sqrt{\pi}\frac{2^{\left(k_{1}+k_{2}\right)/2}}{\sqrt{k_{1}!k_{2}!}}\Phi_{k_{1}}\left(x;b\right)\Phi_{k_{2}}\left(x;b\right)\end{eqnarray*}
which leads to\begin{eqnarray*}
\Phi_{k_{1}}\left(x;b\right)\Phi_{k_{2}}\left(x;b\right) & = & \frac{e^{-x^{2}/\left(2b^{2}\right)}}{\sqrt{b\sqrt{\pi}}}\sum_{k=\left|k_{1}-k_{2}\right|,2}^{k_{1}+k_{2}}T_{k_{1},k_{2}}^{k}\Phi_{k}\left(x;b\right)\end{eqnarray*}
where\begin{equation}
\boxed{{T_{k_{1},k_{2}}^{k}\equiv\frac{\sqrt{k_{1}!k_{2}!k!}}{\left(\frac{k_{1}-k_{2}+k}{2}\right)!\left(\frac{k_{2}-k_{1}+k}{2}\right)!\left(\frac{k_{1}+k_{2}-k}{2}\right)!}}}\label{eq:carprod-coeff}\end{equation}

\subsection{Product of radial harmonic-oscillator functions}

Here, we obtain the relation\begin{equation}
\boxed{{\Phi_{n_{1},k_{1}}\left(\rho,\varphi;b\right)\Phi_{n_{2},k_{2}}\left(\rho,\varphi;b\right)=\frac{e^{-\rho^{2}/\left(2b^{2}\right)}}{\sqrt{\pi}b}\sum_{n=0}^{n_{1,2}}T_{n_{1},k_{1};n_{2},k_{2}}^{n,k_{1}+k_{2}}\Phi_{n,k_{1}+k_{2}}\left(\rho,\varphi;b\right)}}\label{eq:polprod-expr}\end{equation}
between the harmonic-oscillator functions in polar coordinates defined
in Eq. (\ref{eq:phipol-def}). The expansion coefficients $T_{n_{1},k_{1};n_{2},k_{2}}^{n,k_{1}+k_{2}}$
are defined by Eq. (\ref{eq:polprod-coeff1}).

Starting from the generating function in Eq. (\ref{eq:polho-genf-vec}),
and for arbitrary vectors $\vec{t}_{1}$ and $\vec{t}_{2}$\begin{eqnarray}
 &  & e^{-\vec{t}_{1}^{2}+2\vec{\rho}\cdot\vec{t}_{1}/b-\rho^{2}/\left(2b^{2}\right)}e^{-\vec{t}_{2}^{2}+2\vec{\rho}\cdot\vec{t}_{2}/b-\rho^{2}/\left(2b^{2}\right)}\nonumber \\
 & = & \left[b^{2}\sqrt{\frac{\pi}{2}}\sum_{k_{1}=-\infty}^{\infty}\sum_{n_{1}=0}^{\infty}\mathcal{N}_{n_{1},\left|k_{1}\right|}\chi_{n_{1},k_{1}}\left(\vec{t}_{1}\right)\Phi_{n_{1},k_{1}}\left(\rho,\varphi;b\right)\right]\nonumber \\
 &  & \times\left[b^{2}\sqrt{\frac{\pi}{2}}\sum_{k_{2}=-\infty}^{\infty}\sum_{n_{2}=0}^{\infty}\mathcal{N}_{n_{2},\left|k_{2}\right|}\chi_{n_{2},k_{2}}\left(\vec{t}_{2}\right)\Phi_{n_{2},k_{2}}\left(\rho,\varphi;b\right)\right]\label{eq:polprod-genf}\end{eqnarray}
The left-hand side can be written\begin{eqnarray*}
LHS & = & e^{-\left(\vec{t}_{1}+\vec{t}_{2}\right)^{2}+2\vec{\rho}\cdot\left(\vec{t}_{1}+\vec{t}_{2}\right)/b-\rho^{2}/\left(2b^{2}\right)}e^{2\vec{t}_{1}\cdot\vec{t}_{2}-\rho^{2}/\left(2b^{2}\right)}\end{eqnarray*}
Using Eq. (\ref{eq:polho-genf-vec}) again to expand the first exponential,
we get\begin{eqnarray*}
LHS & = & e^{2\vec{t}_{1}\cdot\vec{t}_{2}-\rho^{2}/\left(2b^{2}\right)}b^{2}\sqrt{\frac{\pi}{2}}\sum_{k=-\infty}^{\infty}\sum_{n=0}^{\infty}\mathcal{N}_{n,\left|k\right|}\chi_{n,k}\left(\vec{t}_{1}+\vec{t}_{2}\right)\Phi_{n,k}\left(\rho,\varphi;b\right)\end{eqnarray*}
and using Eq. (\ref{eq:chiprops-expchi-expr}) to absorb the remaining
exponential,\begin{eqnarray*}
LHS & = & b^{2}\sqrt{\frac{\pi}{2}}e^{-\rho^{2}/\left(2b^{2}\right)}\sum_{k=-\infty}^{\infty}\sum_{n=0}^{\infty}\mathcal{N}_{n,\left|k\right|}\Phi_{n,k}\left(\rho,\varphi;b\right)\\
 &  & \times\sum_{m_{1}=0}^{\infty}\sum_{m_{2}=0}^{\infty}\sum_{p_{1}=0}^{\infty}\sum_{k_{1}=-\infty}^{\infty}\sum_{p_{2}=0}^{\infty}\sum_{k_{2}=-\infty}^{\infty}D_{p_{1},k_{1},m_{1};p_{2},k_{2},m_{2}}^{n,k}\\
 &  & \times\chi_{p_{1}+m_{1},k_{1}}\left(\vec{t}_{1}\right)\chi_{p_{2}+m_{2},k_{2}}\left(\vec{t}_{2}\right)\end{eqnarray*}
Comparing with the right-hand side of Eq. (\ref{eq:polprod-genf})
for arbitrary vectors $\vec{t}_{1}$ and $\vec{t}_{2}$, we make the
identifications \begin{eqnarray*}
p_{1}+m_{1} & = & n_{1}\\
p_{2}+m_{2} & = & n_{2}\end{eqnarray*}
and write the left-hand side as\begin{eqnarray*}
LHS & = & b^{2}\sqrt{\frac{\pi}{2}}e^{-\rho^{2}/\left(2b^{2}\right)}\sum_{k=-\infty}^{\infty}\sum_{n=0}^{\infty}\mathcal{N}_{n,\left|k\right|}\Phi_{n,k}\left(\rho,\varphi;b\right)\\
 &  & \times\sum_{m_{1}=0}^{\infty}\sum_{m_{2}=0}^{\infty}\sum_{n_{1}=0}^{\infty}\sum_{k_{1}=-\infty}^{\infty}\sum_{n_{2}=0}^{\infty}\sum_{k_{2}=-\infty}^{\infty}D_{n_{1}-m_{1},k_{1},m_{1};n_{2}-m_{2},k_{2},m_{2}}^{n,k}\\
 &  & \times\chi_{n_{1},k_{1}}\left(\vec{t}_{1};b\right)\chi_{n_{2},k_{2}}\left(\vec{t}_{2};b\right)\end{eqnarray*}
Comparing again with the right-hand side of Eq. (\ref{eq:polprod-genf}),
we readily deduce\begin{eqnarray*}
 &  & b^{4}\frac{\pi}{2}\mathcal{N}_{n_{1},\left|k_{1}\right|}\mathcal{N}_{n_{2},\left|k_{2}\right|}\Phi_{n_{1},k_{1}}\left(\rho,\varphi;b\right)\Phi_{n_{2},k_{2}}\left(\rho,\varphi;b\right)\\
 & = & b^{2}\sqrt{\frac{\pi}{2}}e^{-\rho^{2}/\left(2b^{2}\right)}\sum_{k=-\infty}^{\infty}\sum_{n=0}^{\infty}\mathcal{N}_{n,\left|k\right|}\Phi_{n,k}\left(\rho,\varphi;b\right)\\
 &  & \times\sum_{m_{1}=0}^{\infty}\sum_{m_{2}=0}^{\infty}D_{n_{1}-m_{1},k_{1},m_{1};n_{2}-m_{2},k_{2},m_{2}}^{n,k}\end{eqnarray*}
The sum over $k$ disappears because of the Kronecker-delta function
inside the $D$ coefficient in Eq. (\ref{eq:chiprops-expchi-coeff})
restricting the value of $k$ to $k_{1}+k_{2}$, and the sum over
$n$ is cut off at $n=n_{1,2}$, because of the other Kronecker-delta
function in Eq. (\ref{eq:chiprops-expchi-coeff}) restricting its
value. Therefore,\begin{eqnarray*}
\Phi_{n_{1},k_{1}}\left(\rho,\varphi;b\right)\Phi_{n_{2},k_{2}}\left(\rho,\varphi;b\right) & = & \frac{e^{-\rho^{2}/\left(2b^{2}\right)}}{b^{2}}\sqrt{\frac{2}{\pi}}\sum_{n=0}^{n_{1,2}}\frac{\mathcal{N}_{n,\left|k_{1}+k_{2}\right|}}{\mathcal{N}_{n_{1},\left|k_{1}\right|}\mathcal{N}_{n_{2},\left|k_{2}\right|}}\\
 &  & \times\left[\sum_{m_{1}=0}^{\infty}\sum_{m_{2}=0}^{\infty}D_{n_{1}-m_{1},k_{1},m_{1};n_{2}-m_{2},k_{2},m_{2}}^{n,k_{1}+k_{2}}\right]\\
 &  & \times\Phi_{n,k_{1}+k_{2}}\left(\rho,\varphi;b\right)\end{eqnarray*}
which we write as\begin{eqnarray*}
\Phi_{n_{1},k_{1}}\left(\rho,\varphi;b\right)\Phi_{n_{2},k_{2}}\left(\rho,\varphi;b\right) & = & \frac{e^{-\rho^{2}/\left(2b^{2}\right)}}{\sqrt{\pi}b}\sum_{n=0}^{n_{1,2}}T_{n_{1},k_{1};n_{2},k_{2}}^{n,k_{1}+k_{2}}\Phi_{n,k_{1}+k_{2}}\left(\rho,\varphi;b\right)\end{eqnarray*}
The coefficients $T_{n_{1},k_{1};n_{2},k_{2}}^{n,k_{1}+k_{2}}$ are
obtained from Eq. (\ref{eq:chiprops-expchi-coeff}), being careful
to make the substitutions $n_{1}\rightarrow n_{1}-m_{1}$ and $n_{2}\rightarrow n_{2}-m_{2}$
(and therefore, according to Eq. (\ref{eq:n12-def}), $n_{1,2}\rightarrow n_{1,2}-m_{1}-m_{2}$
as well). Then,\begin{eqnarray*}
T_{n_{1},k_{1};n_{2},k_{2}}^{n,k_{1}+k_{2}} & = & \left(-1\right)^{n_{1}+n_{2}-n}\sqrt{\frac{n!\left(n_{1}+\left|k_{1}\right|\right)!\left(n_{2}+\left|k_{2}\right|\right)!}{n_{1}!n_{2}!\left(n+\left|k_{1}+k_{2}\right|\right)!}}\sum_{m_{1}=0}^{n_{1}}\sum_{m_{2}=0}^{n_{2}}\left(-1\right)^{m_{1}+m_{2}}\\
 &  & \times\left(\begin{array}{c}
n_{1}\\
m_{1}\end{array}\right)\left(\begin{array}{c}
n_{2}\\
m_{2}\end{array}\right)\left(\begin{array}{c}
n_{1,2}-m_{1}-m_{2}\\
n\end{array}\right)\\
 &  & \times\frac{\left(n_{1,2}+\left|k_{1}+k_{2}\right|-m_{1}-m_{2}\right)!}{\left(n_{1}+\left|k_{1}\right|-m_{1}\right)!\left(n_{2}+\left|k_{2}\right|-m_{2}\right)!}\delta_{n\leq n_{1,2}-m_{1}-m_{2}}\end{eqnarray*}
or, in more compact notation,\begin{eqnarray}
T_{n_{1},k_{1};n_{2},k_{2}}^{n,k_{1}+k_{2}} & = & \left(-1\right)^{n_{1}+n_{2}-n}\sqrt{\frac{n!\left(n_{1}+\left|k_{1}\right|\right)!\left(n_{2}+\left|k_{2}\right|\right)!}{n_{1}!n_{2}!\left(n+\left|k_{1}+k_{2}\right|\right)!}}\nonumber \\
 &  & \times\sum_{m_{1}=0}^{n_{1}}\sum_{m_{2}=0}^{n_{2}}\delta_{n\leq n_{1,2}-m_{1}-m_{2}}C_{n_{1},k_{1},m_{1};n_{2},k_{2},m_{2}}^{n,k_{1}+k_{2}}\label{eq:polprod-coeff1}\end{eqnarray}
with\begin{eqnarray}
C_{n_{1},k_{1},m_{1};n_{2},k_{2},m_{2}}^{n,k_{1}+k_{2}} & \equiv & \left(-1\right)^{m_{1}+m_{2}}\left(\begin{array}{c}
n_{1}\\
m_{1}\end{array}\right)\left(\begin{array}{c}
n_{2}\\
m_{2}\end{array}\right)\left(\begin{array}{c}
n_{1,2}-m_{1}-m_{2}\\
n\end{array}\right)\nonumber \\
 &  & \times\frac{\left(n_{1,2}+\left|k_{1}+k_{2}\right|-m_{1}-m_{2}\right)!}{\left(n_{1}+\left|k_{1}\right|-m_{1}\right)!\left(n_{2}+\left|k_{2}\right|-m_{2}\right)!}\label{eq:polprod-coeff2}\end{eqnarray}
Note again that the Kronecker-delta function $\delta_{n\leq n_{1,2}-m_{1}-m_{2}}$
ensures that we always have $n\leq n_{1,2}$, which we used to limit
the sum over $n$ in Eq. (\ref{eq:polprod-expr}).

\section{Formalism for large oscillator shell number\label{sec:Formalism-for-large}}

In this section, we derive the result in \cite{egido97}, \begin{eqnarray}
\left\langle n_{1}\left|f_{n}\right|n_{2}\right\rangle  & = & \frac{\mu b^{-1/2}}{\sqrt{2\pi^{5/2}}}\frac{\Gamma\left(\xi-n_{1}\right)\Gamma\left(\xi-n_{2}\right)\Gamma\left(\xi-n\right)}{z^{\xi}\sqrt{n!n_{1}!n_{2}!}}\nonumber \\
 &  & \times\,_{2}F_{1}\left(-n_{1},-n_{2};-\xi+n+1;1-z\right)\label{eq:fcarn-12}\end{eqnarray}
with $\xi$ given by Eq. (\ref{eq:xi-def}) and $z$ by Eq. (\ref{eq:z-def}),
for the numerically accurate calculation of the matrix element $\left\langle n_{1}\left|f_{n}\right|n_{2}\right\rangle $
in Eq. (\ref{eq:fcar-ik-def}) when large oscillator-shell numbers
are involved. Note that our result differs slightly from \cite{egido97}
in that a {}``$b^{-1/2}$'' factors appears in Eq. (\ref{eq:fcarn-12})
instead of {}``$b^{1/2}$'' (see discussion at the end of this section).
The formula in Eq. (\ref{eq:fcarn-12}) is preferred to the one in
Eq. (\ref{eq:fcar-ik}) for large oscillator-shell numbers, because
the latter requires the evaluation of a sum of products of large ($T$)
and small ($\bar{I}$) coefficients, which can be numerically unstable.
We also obtain the corresponding matrix elements in Eq. (\ref{eq:Vijkl-z})\begin{eqnarray}
V_{ijkl}^{\left(z\right)} & = & \frac{\mu}{\sqrt{2\pi^{3}}b_{z}}\sum_{n_{z}=\left|n_{z}^{(j)}-n_{z}^{(l)}\right|,2}^{n_{z}^{(j)}+n_{z}^{(l)}}T_{n_{z}^{(j)},n_{z}^{(l)}}^{n_{z}}\bar{F}_{n_{z}^{(i)},n_{z}^{(k)}}^{n_{z}}\label{eq:vijkl-z-big}\end{eqnarray}
where the coefficients $\bar{F}_{n_{z}^{(i)},n_{z}^{(k)}}^{n_{z}}$
are defined by Eq. (\ref{eq:fbar-def}).

Starting from the definition,\begin{eqnarray*}
\left\langle n_{1}\left|f_{n}\right|n_{2}\right\rangle  & = & K_{z}^{1/2}\lambda_{n}\int_{-\infty}^{\infty}dz\,\Phi_{n_{1}}\left(z;b\right)e^{-z^{2}/\left(2Gb^{2}\right)}\Phi_{n}\left(z;G^{1/2}b\right)\Phi_{n_{2}}\left(z;b\right)\end{eqnarray*}
we use the generating function, Eq. (\ref{eq:carho-genf}), to integrate
the product of three harmonic-oscillator functions with the Gaussian
factor. This produces\begin{eqnarray}
 &  & e^{-t_{1}^{2}-t_{2}^{2}-t^{2}}\int_{-\infty}^{\infty}dz\, e^{2\left(t_{1}+t_{2}\right)z/b+2tz/B-\nu z^{2}}\nonumber \\
 & = & \sum_{n_{1}=0}^{\infty}\sum_{n_{2}=0}^{\infty}\sum_{n=0}^{\infty}C_{n_{1},n_{2},n}t_{1}^{n_{1}}t_{2}^{n_{2}}t^{n}\left\langle n_{1}\left|f_{n}\right|n_{2}\right\rangle \label{eq:fcar-genf}\end{eqnarray}
where\begin{eqnarray*}
B & \equiv & G^{1/2}b\end{eqnarray*}
\begin{eqnarray*}
\nu & \equiv & \frac{1}{b^{2}}+\frac{1}{B^{2}}\end{eqnarray*}
\begin{eqnarray*}
C_{n_{1},n_{2},n} & \equiv & \frac{b\sqrt{\pi}\sqrt{B\sqrt{\pi}}}{K_{z}^{1/2}\lambda_{n}}\frac{2^{\left(n_{1}+n_{2}+n\right)/2}}{\sqrt{n_{1}!}\sqrt{n_{2}!}\sqrt{n!}}\end{eqnarray*}
The integral in the left-hand side of Eq. (\ref{eq:fcar-genf}) is
easily evaluated by completing the square in the exponential, giving\begin{eqnarray}
LHS & = & \sqrt{\frac{\pi}{\nu}}e^{-t_{1}^{2}-t_{2}^{2}-t^{2}+\tau^{2}/\nu}\label{eq:fcar-genf-lhs}\end{eqnarray}
where\begin{eqnarray*}
\tau & \equiv & \frac{t_{1}+t_{2}}{b}+\frac{t}{B}\end{eqnarray*}
After some simplification, Eq. (\ref{eq:fcar-genf-lhs}) takes the
form\begin{eqnarray*}
LHS & = & \sqrt{\frac{\pi}{\nu}}\exp\left\{ \left[\alpha\left(t_{1}+t_{2}\right)-t\right]^{2}\zeta+2t_{1}t_{2}\right\} \end{eqnarray*}
with\begin{eqnarray*}
\alpha & \equiv & G^{-1/2}\end{eqnarray*}
\begin{eqnarray*}
\zeta & \equiv & -\frac{G}{G+1}\end{eqnarray*}
which we expand as a series\begin{eqnarray*}
LHS & = & \sqrt{\frac{\pi}{\nu}}\sum_{i=0}^{\infty}\frac{\left(2t_{1}t_{2}\right)^{i}}{i!}\sum_{p=0}^{\infty}\frac{1}{p!}\left[\alpha\left(t_{1}+t_{2}\right)-t\right]^{2p}\zeta^{p}\\
 & = & \sqrt{\frac{\pi}{\nu}}\sum_{i=0}^{\infty}\sum_{p=0}^{\infty}\sum_{q=0}^{2p}\sum_{s=0}^{2p-q}\frac{2^{i}}{p!i!}\left(\begin{array}{c}
2p\\
q\end{array}\right)\left(\begin{array}{c}
2p-q\\
s\end{array}\right)\left(-\alpha\right)^{2p-q}\zeta^{p}t_{1}^{s+i}t_{2}^{2p-q-s+i}t^{q}\end{eqnarray*}
comparing with the right-hand side of Eq. (\ref{eq:fcar-genf}), we
make the identifications\begin{eqnarray*}
s+i & = & n_{1}\Rightarrow s=n_{1}-i\\
2p-q-s+i & = & n_{2}\Rightarrow p=\frac{n_{1}+n_{2}+q}{2}-i\\
q & = & n\end{eqnarray*}
 Note that this implies $n_{1}+n_{2}+n$ must be even, and the summation
over $i$ terminates after a finite number of terms, although we will
let it run up to $\infty$ for notational convenience, letting the
factorial terms implicitly truncate the sum. Then we have\begin{eqnarray}
 &  & C_{n_{1},n_{2},n}\left\langle n_{1}\left|f_{n}\right|n_{2}\right\rangle \nonumber \\
 & = & \sqrt{\frac{\pi}{\nu}}\left(-\alpha\right)^{n_{1}+n_{2}}\zeta^{\left(n_{1}+n_{2}+n\right)/2}\nonumber \\
 &  & \times\sum_{i=0}^{\infty}\frac{\left(\begin{array}{c}
n_{1}+n_{2}+n-2i\\
n\end{array}\right)\left(\begin{array}{c}
n_{1}+n_{2}-2i\\
n_{1}-i\end{array}\right)}{\left(\frac{n_{1}+n_{2}+n}{2}-i\right)!i!}\left(\frac{2}{\alpha^{2}\zeta}\right)^{i}\nonumber \\
 & = & \sqrt{\frac{\pi}{\nu}}\left(-\alpha\right)^{n_{1}+n_{2}}\zeta^{\left(n_{1}+n_{2}+n\right)/2}\nonumber \\
 &  & \times\sum_{i=0}^{\infty}\frac{\left(n_{1}+n_{2}+n-2i\right)!}{n!\left(n_{1}-i\right)!\left(n_{2}-i\right)!\left(\frac{n_{1}+n_{2}+n}{2}-i\right)!i!}\left(\frac{2}{\alpha^{2}\zeta}\right)^{i}\label{eq:cfcar-series}\end{eqnarray}
Next, we simplify the ratio of factorials\begin{eqnarray*}
\frac{\left(2p\right)!}{p!} & = & \frac{\left(n_{1}+n_{2}+n-2i\right)!}{\left(\frac{n_{1}+n_{2}+n}{2}-i\right)!}\end{eqnarray*}
using the doubling formula for the Gamma function (Eq. 8.335(1) in
\cite{gradshteyn79}),\begin{eqnarray*}
\frac{\left(2p\right)!}{p!} & = & \frac{2^{2p}}{\sqrt{\pi}}\Gamma\left(p+\frac{1}{2}\right)\end{eqnarray*}
For convenience, we define\begin{eqnarray}
\xi & \equiv & \frac{n_{1}+n_{2}+n+1}{2}\label{eq:xi-def}\end{eqnarray}
which is a half-integer since we have already noted that $n_{1}+n_{2}+n$
is even. Then $p=\xi-i-1/2$, and\begin{eqnarray}
\frac{\left(2p\right)!}{p!} & = & \frac{2^{2p}}{\sqrt{\pi}}\Gamma\left(\xi-i\right)\label{eq:2p-p-factorial-ratio}\end{eqnarray}
In order to simplify this further, we derive the following useful
identity\begin{eqnarray*}
\Gamma\left(1-\xi+i\right) & = & \left(i-\xi\right)\Gamma\left(i-\xi\right)\\
 & = & \left(i-\xi\right)\left(i-\xi-1\right)\cdots\left(1-\xi\right)\Gamma\left(1-\xi\right)\\
 & = & \left(-1\right)^{i}\left(\xi-1\right)\cdots\left(\xi-\left(i-1\right)\right)\left(\xi-i\right)\Gamma\left(1-\xi\right)\end{eqnarray*}
Similarly, we can write\begin{eqnarray*}
\Gamma\left(\xi\right) & = & \left(\xi-1\right)\cdots\left(\xi-\left(i-1\right)\right)\left(\xi-i\right)\Gamma\left(\xi-i\right)\end{eqnarray*}
Therefore,\begin{eqnarray}
\Gamma\left(1-\xi+i\right) & = & \left(-1\right)^{i}\frac{\Gamma\left(\xi\right)\Gamma\left(1-\xi\right)}{\Gamma\left(\xi-i\right)}\label{eq:gamma-identity1}\end{eqnarray}
and, equivalently,\begin{eqnarray}
\Gamma\left(\xi-i\right) & = & \left(-1\right)^{i}\frac{\Gamma\left(\xi\right)\Gamma\left(1-\xi\right)}{\Gamma\left(1-\xi+i\right)}\label{eq:gamma-identity2}\end{eqnarray}
Thus, Eq. (\ref{eq:2p-p-factorial-ratio}) becomes\begin{eqnarray*}
\frac{\left(2p\right)!}{p!} & = & \frac{2^{2p}}{\sqrt{\pi}}\left(-1\right)^{i}\frac{\Gamma\left(\xi\right)\Gamma\left(1-\xi\right)}{\Gamma\left(1-\xi+i\right)}\\
 & = & \frac{2^{2p}}{\sqrt{\pi}}\left(-1\right)^{i}\frac{\Gamma\left(\xi\right)}{\left(1-\xi\right)_{i}}\end{eqnarray*}
where we have used the Pochhammer symbol\begin{eqnarray*}
\left(x\right)_{n} & \equiv & \frac{\Gamma\left(x+n\right)}{\Gamma\left(x\right)}=x\left(x+1\right)\cdots\left(x+n-1\right)\end{eqnarray*}
Returning to Eq. (\ref{eq:cfcar-series}), we replace the $\left(n_{1}-i\right)!$
and $\left(n_{2}-i\right)!$ terms with Pochhammer symbols as well
using Eq. (\ref{eq:gamma-identity2}) with $\xi\rightarrow n_{1}+1$
to write\begin{eqnarray*}
\left(n_{1}-i\right)! & = & \Gamma\left(n_{1}-i+1\right)\\
 & = & \left(-1\right)^{i}\frac{\Gamma\left(n_{1}+1\right)\Gamma\left(-n_{1}\right)}{\Gamma\left(-n_{1}+i\right)}\\
 & = & \left(-1\right)^{i}\frac{n_{1}!}{\left(-n_{1}\right)_{i}}\end{eqnarray*}
and similarly for $\left(n_{2}-i\right)!$. Then, Eq. (\ref{eq:cfcar-series})
yields\begin{eqnarray*}
\left\langle n_{1}\left|f_{n}\right|n_{2}\right\rangle  & = & \frac{2^{2\xi-1}\Gamma\left(\xi\right)\left(-\alpha\right)^{n_{1}+n_{2}}\zeta^{\left(n_{1}+n_{2}+n\right)/2}}{\sqrt{\nu}C_{n_{1},n_{2},n}n!n_{1}!n_{2}!}\sum_{i=0}^{\infty}\frac{\left(-n_{1}\right)_{i}\left(-n_{2}\right)_{i}}{\left(1-\xi\right)_{i}i!}\left(-\frac{1}{2\alpha^{2}\zeta}\right)^{i}\end{eqnarray*}
which we express as a hypergeometric function, as defined in \cite{gradshteyn79}
Eq. 9.100 (see also section 9.14(2) in \cite{gradshteyn79} for the
notation in terms of a generalized hypergeometric function),\begin{eqnarray*}
\left\langle n_{1}\left|f_{n}\right|n_{2}\right\rangle  & = & \frac{2^{\xi-1/2}\mu\Gamma\left(\xi\right)\left(-\alpha\right)^{n_{1}+n_{2}}\zeta^{\left(n_{1}+n_{2}+n\right)/2}}{\sqrt{\nu}b\sqrt{B\sqrt{\pi}}\sqrt{n!n_{1}!n_{2}!}G^{1/4}G^{n/2}}\,_{2}F_{1}\left(-n_{1},-n_{2};1-\xi;z\right)\end{eqnarray*}
where\begin{eqnarray}
z & \equiv & -\frac{1}{2\alpha^{2}\zeta}=1+\frac{\mu^{2}}{2b^{2}}\label{eq:z-def}\end{eqnarray}
Simplifying further, we find\begin{eqnarray}
\left\langle n_{1}\left|f_{n}\right|n_{2}\right\rangle  & = & \left(-1\right)^{\left(n_{1}+n_{2}-n\right)/2}\frac{\mu}{\sqrt{2b\sqrt{\pi}}}\frac{\Gamma\left(\xi\right)}{\sqrt{n!n_{1}!n_{2}!}}z^{-\xi}\nonumber \\
 &  & \times\,_{2}F_{1}\left(-n_{1},-n_{2};1-\xi;z\right)\label{eq:fcar-hypgeo}\end{eqnarray}
Comparing with Eq. (3) in \cite{egido97}, we note that the hypergeometric
function is evaluated at $1-z$ rather than $z$ in that paper. In
order to make a direct comparison with \cite{egido97}, we use Eq.
9.131(2) in \cite{gradshteyn79},\begin{eqnarray*}
 &  & \,_{2}F_{1}\left(-n_{1},-n_{2};1-\xi;z\right)\\
 & = & \frac{\Gamma\left(1-\xi\right)\Gamma\left(1-\xi+n_{1}+n_{2}\right)}{\Gamma\left(1-\xi+n_{1}\right)\Gamma\left(1-\xi+n_{2}\right)}\,_{2}F_{1}\left(-n_{1},-n_{2};-n_{1}-n_{2}+\xi;1-z\right)\\
 &  & +\left(1-z\right)^{1-\xi+n_{1}+n_{2}}\frac{\Gamma\left(1-\xi\right)\Gamma\left(-n_{1}-n_{2}+\xi-1\right)}{\Gamma\left(-n_{1}\right)\Gamma\left(-n_{2}\right)}\\
 &  & \times\,_{2}F_{1}\left(1-\xi+n_{1},1-\xi+n_{2};2-\xi+n_{1}+n_{2};1-z\right)\end{eqnarray*}
The second term vanishes because of the Gamma functions with negative-integer
(or zero) arguments in the denominator. We can simplify the third
argument of the hypergeometric function in the first term to\begin{eqnarray*}
-n_{1}-n_{2}+\xi & = & \frac{-n_{1}-n_{2}+n+1}{2}\\
 & = & -\xi+n+1\end{eqnarray*}
Thus,\begin{eqnarray*}
\,_{2}F_{1}\left(-n_{1},-n_{2};1-\xi;z\right) & = & \frac{\Gamma\left(1-\xi\right)\Gamma\left(\xi-n\right)}{\Gamma\left(1-\xi+n_{1}\right)\Gamma\left(1-\xi+n_{2}\right)}\\
 &  & \times\,_{2}F_{1}\left(-n_{1},-n_{2};-\xi+n+1;1-z\right)\end{eqnarray*}
Next, we use Eq. (\ref{eq:gamma-identity1}) to re-write the Gamma
functions in the denominator,\begin{eqnarray*}
\,_{2}F_{1}\left(-n_{1},-n_{2};1-\xi;z\right) & = & \left(-1\right)^{n_{1}+n_{2}}\frac{\Gamma\left(\xi-n_{1}\right)\Gamma\left(\xi-n_{2}\right)\Gamma\left(\xi-n\right)}{\Gamma\left(1-\xi\right)\Gamma\left(\xi\right)\Gamma\left(\xi\right)}\\
 &  & \times\,_{2}F_{1}\left(-n_{1},-n_{2};-\xi+n+1;1-z\right)\end{eqnarray*}
Substituting this expression into Eq. (\ref{eq:fcar-hypgeo}) gives\begin{eqnarray*}
\left\langle n_{1}\left|f_{n}\right|n_{2}\right\rangle  & = & \left(-1\right)^{\left(-n_{1}-n_{2}-n\right)/2}\frac{\mu}{\sqrt{2b\sqrt{\pi}}}\frac{\Gamma\left(\xi-n_{1}\right)\Gamma\left(\xi-n_{2}\right)\Gamma\left(\xi-n\right)}{\sqrt{n!n_{1}!n_{2}!}\Gamma\left(1-\xi\right)\Gamma\left(\xi\right)}z^{-\xi}\\
 &  & \times\,_{2}F_{1}\left(-n_{1},-n_{2};-\xi+n+1;1-z\right)\end{eqnarray*}
Finally, we use Eq. 8.334(3) in \cite{gradshteyn79} to write\begin{eqnarray*}
\left\langle n_{1}\left|f_{n}\right|n_{2}\right\rangle  & = & \left(-1\right)^{\left(-n_{1}-n_{2}-n\right)/2}\frac{\mu}{\sqrt{2b\sqrt{\pi}}}\frac{\Gamma\left(\xi-n_{1}\right)\Gamma\left(\xi-n_{2}\right)\Gamma\left(\xi-n\right)}{\sqrt{n!n_{1}!n_{2}!}\pi\left(-1\right)^{\left(n_{1}+n_{2}+n\right)/2}}z^{-\xi}\\
 &  & \times\,_{2}F_{1}\left(-n_{1},-n_{2};-\xi+n+1;1-z\right)\\
 & = & \frac{\mu b^{-1/2}}{\sqrt{2\pi^{5/2}}}\frac{\Gamma\left(\xi-n_{1}\right)\Gamma\left(\xi-n_{2}\right)\Gamma\left(\xi-n\right)}{z^{\xi}\sqrt{n!n_{1}!n_{2}!}}\\
 &  & \times\,_{2}F_{1}\left(-n_{1},-n_{2};-\xi+n+1;1-z\right)\end{eqnarray*}
This result is nearly identical to Eq. (3) in \cite{egido97}, after
properly adjusting for the choice of variable names, the only minor
difference being the oscillator parameter which appears as $b^{-1/2}$
in the present work, and $b^{1/2}$ in \cite{egido97}. However, dimensional
analysis favors the $b^{-1/2}$ form, as the matrix element $\left\langle n_{1}\left|f_{n}\right|n_{2}\right\rangle $
must carry dimensions of length to the $1/2$ power, according to
its definition in Eq. (\ref{eq:fcar-ik-def}). In closing, we use
Eq. (\ref{eq:fcarn-12}) to write the expression for the two-body
matrix element (corresponding to Eq. (\ref{eq:Vijkl-z}) in the large
oscillator-shell limit),\begin{eqnarray*}
V_{ijkl}^{\left(z\right)} & = & \frac{\mu}{\sqrt{2\pi^{3}}b_{z}}\sum_{n_{z}=\left|n_{z}^{(j)}-n_{z}^{(l)}\right|,2}^{n_{z}^{(j)}+n_{z}^{(l)}}T_{n_{z}^{(j)},n_{z}^{(l)}}^{n_{z}}\bar{F}_{n_{z}^{(i)},n_{z}^{(k)}}^{n_{z}}\end{eqnarray*}
where\begin{eqnarray}
\bar{F}_{n_{z}^{(i)},n_{z}^{(k)}}^{n_{z}} & \equiv & \frac{\Gamma\left(\xi-n_{z}^{(i)}\right)\Gamma\left(\xi-n_{z}^{(k)}\right)\Gamma\left(\xi-n_{z}\right)}{z^{\xi}\sqrt{n_{z}!n_{z}^{(i)}!n_{z}^{(k)}!}}\nonumber \\
 &  & \times\,_{2}F_{1}\left(-n_{z}^{(i)},-n_{z}^{(k)};-\xi+n_{z}+1;1-z\right)\label{eq:fbar-def}\end{eqnarray}

\section{Angular integral\label{sec:Angular-integral}}

We wish to evaluate the radial part of the matrix-element integral
\begin{eqnarray*}
V_{ijkl}^{\left(r\right)} & \equiv & \int_{0}^{\infty}\rho_{1}d\rho_{1}\int_{0}^{2\pi}d\varphi_{1}\int_{0}^{\infty}\rho_{2}d\rho_{2}\int_{0}^{2\pi}d\varphi_{2}\\
 &  & \times\Phi_{n_{r}^{(i)},\Lambda^{(i)}}^{*}\left(\rho_{1},\varphi_{1};b_{\perp}\right)\Phi_{n_{r}^{(j)},\Lambda^{(j)}}^{*}\left(\rho_{2},\varphi_{2};b_{\perp}\right)\\
 &  & \times e^{-\left(\vec{\rho}_{1}-\vec{\rho}_{2}\right)^{2}/\mu^{2}}\Phi_{n_{r}^{(k)},\Lambda^{(k)}}\left(\rho_{1},\varphi_{1};b_{\perp}\right)\Phi_{n_{r}^{(l)},\Lambda^{(l)}}\left(\rho_{2},\varphi_{2};b_{\perp}\right)\end{eqnarray*}
numerically, where the harmonic-oscillator functions are defined in
Eq. (\ref{eq:phipol-def}). By rotational invariance of the Gaussian
potential, we have\begin{eqnarray*}
-\Lambda^{(i)}-\Lambda^{(j)}+\Lambda^{(k)}+\Lambda^{(l)} & = & 0\end{eqnarray*}
The angular integrals over $\varphi_{1}$ and $\varphi_{2}$ are particularly
problematic because of their oscillatory nature. Therefore, we focus
on those integrals and introduce the function\begin{eqnarray}
\Theta_{k}\left(x\right) & \equiv & \frac{1}{\left(2\pi\right)^{2}}\int_{0}^{2\pi}d\varphi_{1}\int_{0}^{2\pi}d\varphi_{2}e^{ik\left(\varphi_{1}-\varphi_{2}\right)}e^{2x\cos\left(\varphi_{1}-\varphi_{2}\right)}\label{eq:bigtheta-def}\end{eqnarray}
so that we may write\begin{eqnarray}
V_{ijkl}^{\left(r\right)} & = & \int_{0}^{\infty}\rho_{1}d\rho_{1}\int_{0}^{\infty}\rho_{2}d\rho_{2}e^{-\left(\rho_{1}^{2}+\rho_{2}^{2}\right)/\mu^{2}}\Phi_{n_{r}^{(i)},\left|\Lambda^{(i)}\right|}\left(\rho_{1};b_{\perp}\right)\Phi_{n_{r}^{(j)},\left|\Lambda^{(j)}\right|}\left(\rho_{2};b_{\perp}\right)\nonumber \\
 &  & \Phi_{n_{r}^{(k)},\left|\Lambda^{(k)}\right|}\left(\rho_{1};b_{\perp}\right)\Phi_{n_{r}^{(l)},\left|\Lambda^{(l)}\right|}\left(\rho_{2};b_{\perp}\right)\Theta_{-\Lambda^{(i)}+\Lambda^{(k)}}\left(\frac{\rho_{1}\rho_{2}}{\mu^{2}}\right)\label{eq:vijkl-r-integral}\end{eqnarray}
We simplify Eq. (\ref{eq:bigtheta-def}) using the generating function
for the Bessel function, given in Eq. (\ref{eq:bessel-genf}), with
$z=-2ix$ and $\varphi=\varphi_{1}-\varphi_{2}$,\begin{eqnarray*}
e^{2x\cos\left(\varphi_{1}-\varphi_{2}\right)} & = & \sum_{n=-\infty}^{\infty}i^{\left|n\right|}J_{\left|n\right|}\left(-2ix\right)e^{in\left(\varphi_{1}-\varphi_{2}\right)}\end{eqnarray*}
from which the integral in Eq. (\ref{eq:bigtheta-def}) yields\begin{eqnarray*}
\Theta_{k}\left(x\right) & = & i^{\left|k\right|}J_{\left|k\right|}\left(-2ix\right)\end{eqnarray*}
From the series expansion of the modified Bessel function of the first
kind, Eq. 8.445 in \cite{gradshteyn79}, we get\begin{eqnarray}
\Theta_{k}\left(x\right) & = & \left(-1\right)^{\left|k\right|}I_{\left|k\right|}\left(-2x\right)\nonumber \\
 & = & \sum_{n=0}^{\infty}\frac{x^{2n}}{n!\left(n+\left|k\right|\right)!}\label{eq:bigtheta-series}\end{eqnarray}
We find that the series in Eq. (\ref{eq:bigtheta-series}) is extremely
well converged if we include terms up to $m$ such that\begin{eqnarray*}
\left|\frac{x^{2m}}{m!\left(m+\left|k\right|\right)!}\right| & < & \epsilon\end{eqnarray*}
where $\epsilon=10^{-2N_{0}-N_{\textrm{quad}}/8}$ for a calculation
in up to $N_{0}$ oscillator shells and $N_{\textrm{quad}}$ quadrature
points. The remaining integrals over $\rho_{1}$ and $\rho_{2}$ in
Eq. (\ref{eq:vijkl-r-integral}) were evaluated by Gauss-Laguerre
quadrature.


\begin{thebibliography}{10}
\bibitem{brink67}D. M. Brink, E. Boeker, Nucl. Phys. \textbf{A91}, (1967) 1.
\bibitem{decharge80}J. Dechargé, D. Gogny, Phys. Rev. C \textbf{21}, (1980) 1568.
\bibitem{gogny70}D. Gogny, Phys. Lett. \textbf{B32}, (1970) 591.
\bibitem{gogny75}D. Gogny, Nucl. Phys. \textbf{A237}, (1975) 399.
\bibitem{berger84}J. F. Berger, M. Girod, D. Gogny, Nucl. Phys. \textbf{A428}, (1984)
23.
\bibitem{goutte05}H. Goutte, J. F. Berger, P. Casoli, D. Gogny, Phys. Rev. C \textbf{71},
(2005) 024316.
\bibitem{dubray08}N. Dubray, H. Goutte, and J.-P. Delaroche, Phys. Rev. C \textbf{77},
(2008) 014310.
\bibitem{gradshteyn79}I. S. Gradshteyn, I. M. Ryzhik, Tables of Integrals, Series, and Products
(Academic Press Inc., San Diego CA 1979).
\bibitem{egido97}J. L. Egido, L. M. Robledo, R. R. Chasman, Phys. Lett. B 393, (1997)
13.
\bibitem{wolfram03}S. Wolfram, The Mathematica book, $5^{\textrm{th}}$ ed., (Wolfram
Media, 2003).
\bibitem{younes07}W. Younes, D. Gogny, LLNL Tech. Rep. (2007) UCRL-TR-234682.
\bibitem{warda02}M. Warda, J. L. Egido, L. M. Robledo, and K. Pomorski, Phys. Rev.
C \textbf{66}, 014310 (2002).
\end{thebibliography}
\end{document}